\documentclass[]{aastex631}

\usepackage{epsfig}

\usepackage{xcolor}
\usepackage{soul}

\newcommand{\RK}[2]{{}{{#2}}}
\newcommand{\HG}[2]{{}{{#2}}}

\newcommand{\FIG}[1]{#1}

\shorttitle{Super-Alfv\'enic Rotational Instability}
\shortauthors{Goedbloed and Keppens}

\def\BEQ {\begin{equation}} 
\def\EEQ {\end{equation}}
\def\BEQN {\begin{displaymath}} 
\def\EEQN {\end{displaymath}}
\def\BEQAR {\begin{eqnarray}}
\def\EEQAR {\end{eqnarray}}
\def\non {\nonumber}

\def\En#1{\label{Eq.#1}}
\def\E#1{(\ref{Eq.#1})}

\def\R#1{\citep{#1}}
\def\Ron#1{\citet{#1}}
\def\Fn#1{\label{Fig.#1}}
\def\F#1{\ref{Fig.#1}}
\def\Sn#1{\label{Sec.#1}}
\def\S#1{\ref{Sec.#1}}

\def\hor #1{{\hskip #1 mm}}

\def\hs {{\hskip 0.2mm}}

\def\bfe {{\bf e}}

\def\bfj {{\bf j}}

\def\bfr {{\bf r}}
\def\bfv {{\bf v}}
\def\bfB {{\bf B}}
\def\bfF {{\bf F}}
\def\bfG {{\bf G}}

\def\bfxi{\mbox{\boldmath $\xi$}}  

\def\div #1{\nabla\cdot #1}
\def\curl #1{\nabla\times #1}

\def\IT #1{{\it #1\/}}

\def\smalltext#1#2{{\medskip\small 
   \noindent{{\bf $\triangleright$ \hskip.3mm #1\hskip.3mm} #2 \hfill {\bf
   $\triangleleft$}}\par\medskip}}

\def\half {{\textstyle\frac{1}{2}}}
\def\onefourth {{\textstyle\frac{1}{4}}}
\def\fivefourth {{\textstyle\frac{5}{4}}}
\def\fiveeighth {{\textstyle\frac{5}{8}}}
\def\onefifth {{\textstyle\frac{1}{5}}}
\def\onesixth {{\textstyle\frac{1}{6}}}
\def\onethird {{\textstyle\frac{1}{3}}}
\def\fivehalf {{\textstyle\frac{5}{2}}}

\def\doublel {\lbrack\!\lbrack}
\def\doubler {\rbrack\!\rbrack}

\begin{document}

\title{The Super-Alfv\'enic Rotational Instability\\[2mm]
in accretion disks about black holes}

\author{Hans Goedbloed}
\affiliation{DIFFER -- Dutch Institute for Fundamental Energy Research, De Zaale 20, 
5612 AJ Eindhoven, the Netherlands}
\author{Rony Keppens}
\affiliation{Centre for mathematical Plasma-Astrophysics, Department of Mathematics, 
KU Leuven, Celestijnenlaan 200B, 3001 Leuven, Belgium}

\date{\today}

\begin{abstract}
 
The theory of instability of accretion disks about black holes, neutron stars or 
proto\-planets, is revi\-sited by means of the recent method of the Spectral Web. The 
cylindrical accretion disk differential equation is shown to be governed by the forward 
and backward Doppler-shifted continuous Alfv\'en spectra $\Omega_{\rm A}^\pm \equiv 
m \Omega \pm \omega_{\rm A}$, where $\omega_{\rm A}$ is the static Alfv\'en 
frequency. It is crucial to take non-axisymmetry ($m \ne 0$) and super-Alfv\'enic rotation 
of the Doppler frames ($|m\Omega| \gg |\omega_{\rm A}|$) into account. The continua 
$\Omega_{\rm A}^+$ and $\Omega_{\rm A}^-$ then overlap, ejecting a plethora of 
{\em Super-Alfv\'enic Rotational Instabilities} (SARIs). In-depth analysis for small 
inhomogeneity shows that the two Alfv\'en singularities reduce the extent of the modes to 
sizes much smaller than the width of the accretion disk. Generalization for large 
inhomogeneity leads to the completely unprecedented result that, for mode numbers $|k| 
\gg |m|$, any complex $\omega$ in a wide neighborhood of the real axis is an 
approximate `eigenvalue'. The difference with genuine eigenmodes is that the amount of 
complementary energy to excite the modes is tiny, $|W_{\rm com}| \le c\hs$, with $c$ 
the machine accuracy of the computation. This yields a multitude of two-dimensional 
continua of quasi-discrete modes: {\em quasi-continuum SARIs}. We conjecture that the 
onset of 3D turbulence in magnetized accretion disks is governed, not by the excitation of discrete 
axisymmetric Magneto-Rotational Instabilities, but by the excitation of modes from these 
two-dimensional continua of quasi-discrete non-axisymmetric Super-Alfv\'enic 
Rotational Instabilities. 

\end{abstract}

\section{Introduction}{\Sn{I}}

\subsection{Accretion and the Magneto-Rotational Instabilities}{\Sn{IA}}

Accretion processes in disks are of central importance in astrophysics, as disks arise in 
the vicinity of black holes at both galactic or stellar scales (in Active Galactic Nuclei -- 
AGN -- or X-ray binary systems), but they also play a crucial role in the context of 
protoplanet formation. The accretion disk radiative properties relate to the accretion 
rate~\R{ShakSun73}, which is in turn connected with angular momentum transport. 
\RK{Modern views distinguish many accretion disk types}{The literature distinguishes 
many accretion disk types, see e.g.~\Ron{Ar2011,Turner2014} for reviews on 
protoplanetary disk physics, or \Ron{AF2013,FN2014} for reviews of accretion onto 
black holes. Disks may be} varying in geometry between thick versus thin disks, varying 
in radiative aspects being hot versus cold (optically thin versus thick), or whether 
advection, gas dynamical or radiation effects are dominant. \RK{The various disk 
types}{Many disk structures} can be described in terms of essentially one-dimensional 
(height-integrated) radially self-similar accretion solutions, as pioneered 
by~\Ron{ShakSun73}. These self-similar profiles involve an angular frequency 
$\Omega=v_\theta(r)/r \sim r^{-p}$, which can deviate from perfect Keplerian flow 
(where $p=1.5$) due to e.g.\ gas pressure or magnetic field effects, and ultimately relate 
the radially inward accretion velocity $v_r \sim -\alpha r^{-a}$ with the celebrated 
$\alpha$-parameter collecting ``turbulent viscous'' processes. In this $v_r(r)$ profile, the 
accretion inflow power index $a$ is taken according to prevailing physics, and e.g., 
\Ron{SMIS87} argue for radial profiles where $v_r \sim -\alpha r^{-0.5}$ while the 
sound speed varies as $r^{-0.5}$, a property shared by the viscous advection-dominated 
accretion flows (ADAFs) discussed in~\Ron{NY94}. These purely hydrodynamic 
prescriptions can be generalized in various ways, e.g.\ by incorporating convection 
effects, yielding convection-dominated accretion flows  (CDAFs)~\R{NIA2000}, or by 
allowing for a magnetic field at a fixed plasma beta~\R{YWB2012}. The latter really 
requires the use of magnetohydrodynamics (MHD), which has become a central working 
tool for accretion physics as a whole. Indeed, in \RK{}{various} reviews on accretion 
disk theory (see e.g.~\Ron{BH98}), the role of MHD turbulence in the disk is 
acknowledged as essential for accretion to proceed, as the fully developed turbulent state 
would realize the corresponding $\alpha$-values to agree with observed disk properties. 
The ultimate trigger for disk turbulence is \RK{nowadays fully ascribed}{in the ideal 
MHD regime invariably ascribed} to a linear MHD instability, the magneto-rotational 
instability (MRI). \RK{}{Naturally, extensions beyond ideal MHD have rightfully gained 
attention, especially in the protoplanetary disk context \R{Turner2014}, where resistive, 
ambipolar and Hall MHD effects came into focus, introducing MRI-active versus `dead' 
zones, see e.g. \Ron{Lesur2014}. Cartoon views on protoplanetary disks identify where 
ideal MHD, MRI-like turbulence is still applicable, or where only weak turbulence is 
expected,  and how the $\alpha$-parameter affects the growth of coupled gas-particle 
streaming instabilities~\R{Umurhan2020}.  For both protoplanetary and black hole disks, 
the way the observationally inferred $\alpha$-parameter (or the related central accretion 
rate, see~\Ron{Rafikov2017}) maps to turbulence-induced angular momentum transport 
is in reality also complicated by (magnetized) winds and/or jets in the accreting object + 
disk environment, which are known to influence the overall angular momentum budget, 
see e.g.~\Ron{CK2002,CK2004}.}

The MRI was introduced as a ``powerful local shear instability in weakly magnetized 
disks'' in the seminal work by~\Ron{BH91a}. Their work presented dispersion relations 
\RK{}{from linear ideal MHD} for axisymmetric, linear perturbations about a weakly 
magnetized disk of finite vertical extent, and all it requires for instability is a radially 
decreasing angular velocity law where $\Omega'<0$ and a weak poloidal $B_z$ 
magnetic field component. Their linear MHD treatment of a disk, with added (but weak) 
toroidal $B_\theta(r,z)$ and poloidal $B_z(r)$ components, assumed a Boussinesq 
approximation and analyzed $\exp[{\rm i}(k_r r+k_z z- \omega t)]$ perturbations with 
radial and vertical mode numbers $k_r$ and $k_z$. In an accompanying 
paper~\R{HB91b}, 2.5D axisymmetric nonlinear MHD simulations were performed of a 
magnetized Keplerian, cylindrical Couette flow, where the presence of a weak magnetic 
field was confirmed as essential to get growing linear mode behavior, and where its role 
in triggering turbulence and angular momentum redistribution was demonstrated for the 
first time. Further 2.5D nonlinear MHD simulations, employing the  `shearing sheet' 
approximation as appropriate for disks, provided additional support~\R{HB92a}. The 
MRI, whose essential ingredients were analyzed earlier by \Ron{Vel59} and 
\Ron{Chandra60}, has ever since been \RK{recognized as}{} paramount to understand 
and explain accretion disk behavior. \RK{}{State-of-the-art research invokes MRI in 
countless studies. For example, in protoplanetary disks, \Ron{Xu2022} investigate how 
MRI-turbulence in the outer disk zones interplays with gas-coupled particle prescriptions 
(dust), where gas-dust velocity differences affect dust clumping by streaming 
instabilities. The `gas' is then described by standard MHD equations, extended with 
ambipolar diffusion to account for the influence of partial ionization. In modern studies 
of accretion flows about black holes, general relativistic MHD (GRMHD) simulations 
serve to interpret observations by~\Ron{EHT2019}, or observed multi-wavelength 
variability in lightcurves~\R{Chatterjee2021}, whereas they employ `MRI-quality 
factors'  that dictate the needed numerical resolutions to `resolve MRI turbulence'. 
General relativistic, resistive MHD simulations nowadays address how plasmoids may 
form within current sheets in black hole accretion disks~\R{Ripperda2020}, clearly 
differing for `Standard And Normal Evolution' (SANE) versus `Magnetically Arrested 
Disks' (MAD) states, whereby SANE disks are particularly prone to MRI turbulence 
throughout. Also in Newtonian settings, modern 3D resistive MHD runs address 
plasmoid reconnection aspects generated by `primary MRI' \R{Rosenberg2021}. 
Numerous authors performed MRI studies in shearing boxes, with 
e.g.~\Ron{Simon2009} addressing visco-resistive modifications, or 
recently~\Ron{Held2022} studying the effect of parameterized cooling prescriptions, 
where MRI interacts with thermal instability cycles.}

The original axisymmetric MRI treatment~\R{BH91a} presented a purely local stability 
analysis, akin to a WKB treatment, and various later studies highlighted important novel 
aspects. For example, \Ron{Knobloch1992} revisited the axisymmetric MRI using linear 
incompressible MHD in a more rigorous way, pointing out the role of boundary 
conditions when performing a local stability analysis of shear flows, and the fact that any 
azimuthal field will turn the purely growing mode into an overstable oscillation. This was 
further elaborated on by~\Ron{BSKG05}, where axisymmetric modes in a stratified disk, 
as analyzed using the WKB approach, were confirmed by a complete numerical treatment 
of the linear compressible MHD equations. In terms of the traditional plasma beta 
parameter $\beta \equiv 2p/B^2$, this study identified how $m=0$ MRIs could persist up 
to equipartion $\beta\approx 1$ conditions, when dominant toroidal magnetic field 
components are present.

Although the initial 2.5D nonlinear MHD simulations~\R{HB91b,HB92a} gave clear 
hints for efficient angular momentum transport due to MRI, the assumption of 
axisymmetry precluded definitive statements on its importance for sustaining 3D MHD 
turbulence, as needed for e.g.\ disk dynamo processes. Purely hydrodynamic 3D 
simulations of accretion tori did show non-axisymmetric (low azimuthal mode number 
$m$) spirals forming~\R{Hawley1991}, but no hints of actual turbulence. That linearly 
unstable $m=1$ modes indeed exist for hydrodynamic tori with constant specific angular 
momentum was predicted analytically by~\Ron{PP1984}. To make the case for the MRI 
as relevant for triggering 3D MHD turbulence, non-axisymmetric modes for a thin disk 
were addressed in~\Ron{BH92b}, where all radial structure was ignored except for the 
shear flow resulting from $\Omega(r)$. The analysis used time-dependent radial mode 
numbers where $dk_r/dt=-m\Omega'$, and demonstrated transient amplification of 
magnetic field perturbations, up to many orders of magnitude. Further strengthened by 
full 3D nonlinear MHD simulations, originally in the local `shearing box' 
approach~\R{HGB95}, but later \R{Hawley2000} in global 3D MHD simulations of 
accretion tori, all these findings essentially appeared to close the case on MRIs, as 
relevant (1) to explain turbulence in disks, (2) for providing efficient outer angular 
momentum fluxes due to magnetic stresses to get accretion going, and (3) for their role in 
disk magnetic field amplification. 

However, as this paper intends to show, many crucial aspects of especially 
non-axisymmetric linear MHD instabilities in realistic accretion disks have been 
completely ignored thus far. This is mainly because -- with few exceptions, 
e.g.~\Ron{Ogilvie1998} -- the astrophysical literature on MHD disk eigenmodes does 
not exploit the full power of MHD spectroscopy\RK{}{, where a key role is played by 
the continuous parts in the MHD spectrum}. This `art' of quantifying the complete 
spectral eigenstructure of a given MHD equilibrium state has matured in laboratory 
fusion context, and has been transferred to astrophysical settings as documented in our 
textbook~\R{GKP2019}. The variation of all MHD equilibrium quantities introduces 
both continuous and discrete sequences of eigenmodes, which have been shown to 
organize and link all known stable and unstable (overstable/damped) modes in intriguing 
ways. By using a fully general treatment of MHD eigenmodes in the cylindrical disk 
approach (see section~\S{Ib}), we therefore aim to show (1)~that the \IT{axisymmetric 
MRIs represent only a finite amount of actually unstable (i.e., overstable) modes}, within 
an otherwise infinite clustering sequence of mostly stable modes accumulating towards 
the continuous spectra; (2)~a novel type of non-axisymmetric modes exists, which we 
denote as Super-Alfv\'enic Rotational Instabilities (SARIs), as a result of overlapping 
continuum frequency ranges, and these are \IT{truly infinite sequences consisting of 
unstable modes} only; (3)~the existence of `virtual walls' (due to skin currents at 
near-singularities) in the localization of the SARI eigenfunctions, which makes them 
insensitive to adopted boundary conditions, unlike the $m=0$ MRIs; (4)~the completely 
uncharted role of extremely localized quasi-continuum, non-axisymmetric instabilities, 
which seemingly occupy complete regions in the complex eigenfrequency plane. 
Together, these findings open up an entirely new window on linear MHD modes, far 
beyond the celebrated MRI mode, as likely triggers for MHD turbulence, accretion and 
dynamo activity in disks. 

\RK{}{Of course, even in purely hydrodynamical settings, the important role of 
non-axisymmetric perturbations on (thin, viscous) disks is well established, with 
possibilities for transient growth as in \Ron{Rebusco2009}, or $m\neq 0$ Rossby Wave 
Instabilities induced by local overdensities, as in~\Ron{Meheut2012}, as well as various 
other pure hydro effects discussed in protoplanetary context by~\Ron{Turner2014}. We 
will focus here on magnetized disks, extending the ideal MHD theory way beyond the 
original $m=0$ MRI, but point out that our findings have potential implications for 
purely hydro settings as well, as discussed in our Outlook, Sec.~\S{VIB}.} Various 
previous studies investigated non-axisymmetric modes in magnetized sheared flows. 
\Ron{OP1996} presented a treatment in a cylindrical setting where gravity was ignored, 
the density was homogeneous and a purely toroidal, potential $B_\theta \sim r^{-1}$ 
field was incorporated. That study did admit the role of forward and backward Alfv\'en 
continua, and found discrete non-axisymmetric modes whose limit points could be 
obtained for large axial wavenumber $k_z\rightarrow \infty$. \RK{}{Non-axisymmetric 
MRI modes, called azimuthal MRIs or AMRIs, were also found in a Taylor--Couette 
setting in visco-resistive MHD with a purely toroidal imposed field~\R{Rudiger2007a, 
Rudiger2007b,Hollerbach2010}. The latter works extended earlier 
findings~\R{Hollerbach2005} on $m=0$ modes in helical magnetic fields in the same 
Taylor--Couette context, and this `helical MRI'  is still a topic of concern for modern 
liquid sodium experiments~\R{Mishra2021}. How a uniform, axial magnetic field can 
open the way to non-axisymmetric modes in dissipative disks is studied 
in~\Ron{Kitchatinov2010}.} Truly 2D disks with purely toroidal $B_\theta(r,z)$ 
magnetic fields were analyzed as well in \Ron{TP1996}, pointing out that such discs are 
always unstable.  \Ron{Curry1996} studied non-axisymmetric modes in incompressible 
cylinder settings, and thereby generalized the hydrodynamic Papaloizou--Pringle 
\R{PP1984} mode to magnetic regimes,  while finding additional pure Alfv\'en modes 
coupled to the local rotation frequency of a cylindrical shell. Our treatment will 
generalize and augment these findings considerably, since we will allow for arbitrary 
poloidal $B_z(r)$ and toroidal $B_\theta(r)$, incorporating all subtleties due to the MHD 
continuous spectra\RK{}{, while we intentionally address the ideal MHD regime only, 
which we believe to be central to any further modification by non-ideal effects}. This will 
involve completely new spectral structures as organized in the {\em Spectral Web,} 
which allows us to reveal the surprising spread of modes throughout the complex 
eigenfrequency plane in unprecedented ways.

It must be emphasized that the countless nonlinear MHD simulations of turbulent disks, 
nowadays extended from Newtonian to fully general relativistic settings, covering aspects 
from magnetized protoplanetary disks, to disks in X-ray binaries, all the way up to 
galactic disks and AGNs, do indeed show clearly that MHD turbulence is virtually 
inevitable. In that respect, our general linear MHD treatment serves to seriously question 
the present attitude throughout the astrophysical literature, ascribing all turbulence in 
disks to MRI activity. In 3D shearing box simulations, it is clear that precisely the high 
$m$ modes are the ones of interest. In global disk settings, the 3D time-dependent (ideal) 
MHD state obtained at any individual moment may well be analyzed spectrally, and it 
has been shown \R{KD2016} that the same mathematical operators (a generalized force 
operator and the Doppler-Coriolis operator) built up from these instantaneous (ideal) 
MHD profiles govern the state's stability at any time in its nonlinear evolution. It would 
be a major oversimplification to not discriminate mode properties depending on the state 
at hand, which may transit in its evolution from weak to strong field, has disks in thin to 
thick settings, with locally different partitions between rotational, magnetic, thermal and 
radiative energies. In that respect, there is a clear dichotomy in the astrophysical 
literature, where the nonlinear, height-averaged equilibrium accretion flow of 
\RK{}{magnetized} disks is being categorized in various states known as \RK{}{SANE, 
MAD, }ADAFs, CDAFs, thin, thick, slim, radiatively efficient or inefficient, \ldots, 
while all turbulent activity is simply denoted as a direct consequence of the linear MRI.

\subsection{Cylindrical versus toroidal analysis}{\Sn{Ib}}

In the established approach to diagnose \IT{all waves and instabilities about a stationary 
(i.e.\ time-independent) ideal MHD state} (see section \S{IIB}), one must always specify 
the fully force-balanced equilibrium state up front. In the context of accretion disks, this 
equilibrium can be taken to be a magnetized, axisymmetric accretion torus, where the 
combination of pressure gradients, external gravity due to the central object, centrifugal 
and inertial effects, and Lorentz forces balance. This in essence renders the background 
equilibrium 2D in nature, with profiles $\rho(r,z)$, $p(r,z)$ and magnetic and flow vector 
quantities that in turn determine the linear eigenmode distribution. In laboratory fusion 
context, magnetic flux surfaces forming nested tori are realized within tokamak 
configurations, and the same magnetic topology can be adopted for initializing 
magnetized accretion tori equilibrium states. Rigorous analysis of the MHD spectrum of 
such 2D equilibria has been undertaken, and has already identified two new (i.e.\ distinct 
from MRI) likely routes to turbulence, related to the fact that purely flux surface 
localized modes, that define the continuous parts of the MHD spectrum, can actually be 
driven unstable. This invariably involves intricate coupling schemes between multiple 
linear modes of different poloidal mode number: the 2D poloidal $(r,z)$ variation of the 
equilibrium itself now causes Fourier modes of different poloidal angular variation to 
couple. In \Ron{GBHK2004a}, as well as in Chap.~18 of \Ron{GKP2019}, a new class 
of local instabilities called the Trans-Slow Alfv\'en Continuum (TSAC) modes was 
identified, whenever the poloidal Alfv\'en Mach number $M$ (found from $M^2 \equiv 
\rho v_p^2/B_p^2$ using the poloidal vector components) of the equilibrium would 
exceed a critical value. In a later study, \Ron{BKG2007} identified unstable continuous 
spectra related to convective modes, aptly called Convective Continuum Instabilities 
(CCIs), that even persist in strongly magnetized disks (a property known to suppress the 
standard MRI). The same is true for the TSAC modes, since in the strong field limit 
$\beta\ll 1$ of magnetically dominated thick accretion tori, this continuum instability 
switches on whenever the squared poloidal Alfv\'en Mach number exceeds a value of 
about $\half\gamma\beta$, and this for all poloidal-toroidal mode numbers pairs. At the 
same time, \Ron{Haverkort2012} clearly showed that in such 2D equilibrium settings as 
valid for accretion tori, a radially decreasing toroidal rotation frequency can act 
stabilizing for non-axisymmetric MHD modes, by a Coriolis-pressure effect. This finding 
is relevant for both accretion disks and for rotating, toroidally confined, laboratory 
plasmas (which are known to feature transport barriers).

Vertical stratification of the disk also features in the analyses of \Ron{MPJH2013} and 
\Ron{GX1994}. In both papers the main thrust is on the non-linear phase of the 
perturbations, either grown from a linearly stable state (like the hydrodynamic Keplerian 
disk analyzed in the first paper) or from a linear instability (like the MRI in magnetized 
accretion disks analyzed in the second paper). The paper by \Ron{MPJH2013} considers 
a purely hydrodynamic vertical flow field, sheared in the horizontal direction. While 
linearly stable, it generates sequences of vortices that advect in the cross-stream 
directions. It is conjectured that this mechanism of finite amplitude instability may 
destabilize hydrodynamic Keplerian rotations in protoplanetary disks with a negligible 
magnetic field. In the paper by \Ron{GX1994}, on the other hand, a magnetized plasma 
disk with small magnetic fields subject to MRIs is considered. The problem posed is 
whether these instabilities survive at finite amplitudes. It is indicated that the nonlinear 
development of secondary parasitic instabilities will stop the growth of the primary 
MRIs. However, an `exact' finite amplitude MRI is shown to occur for magnetic field 
perturbations of even equipartition strength! As admitted by the authors, this analysis is 
quite restricted in the number of degrees of freedom that can be handled. It gives 
additional significance to the MRI picture of the initial phase of turbulence in disks 
though, which might equally well apply to the non-axisymmetric instabilities discussed in 
the present paper.

For the time being, we will stick to the consideration of magnetic fields in the overall 
picture of accretion and the ensuing formation of jets. Also, we restrict the analysis to 
linear perturbations, awaiting a nonlinear approach of the non-axisymmetric instabilities 
which may show either saturation at low amplitude levels of the instabilities or further 
growth to large amplitudes, with the possible generation of parasitic instabilities, 
sequences of vortices, etc. As the history of the MRI illustrates, such analyses can only 
come after the linear version of the dynamics has been sufficiently documented. That is 
the implicit goal of the present paper.

Hence, we will rather adopt a purely 1D `cylindrical disk' model, and analyze the full 
consequences of having a radially stratified $\Omega(r)$, density $\rho(r)$, pressure 
$p(r)$ and further arbitrary $B_z(r)$ and $B_\theta(r)$. This cylindrical disk approach 
has the distinct advantage that the poloidal couplings mentioned as crucial to the TSAC 
and CCI unstable continuum mode types do not occur at all: a linear analysis can focus 
attention on a single mode $\exp[{\rm i}(m\theta+k z-\omega t)]$ at a time, with 
prechosen mode numbers $m$ and $k$. The radial variation of the eigenmodes must 
always be computed as part of the eigenmode quantification, unless some WKB-type 
assumption can be justified. The basic governing equations for such cylindrical disk 
eigenmodes in ideal MHD have been known for some time, and were presented in 
\Ron{KCG02}, building on all previous efforts that excluded the influence of a 
(cylindrical) gravitational potential. Due to the 1D nature of the equilibrium, the 
continuous -- and always stable -- parts of the MHD spectrum are known for each given 
set of mode numbers $(m,k)$. In the mentioned paper, the axisymmetric MRI modes of a 
weakly magnetized disk were computed, showing their relation to the continua, and clear 
indications of a richer non-axisymmetric eigenmode structure were given. The present 
paper revives this uncharted field of magneto-seismology of accretion disks, by using the 
powerful Spectral Web technique to locate all the eigenmodes. In this approach, we can 
choose the equilibrium properties according to the disk model believed to prevail, e.g.\ 
take profiles in accord with the ADAF, CDAF or whichever accretion flow solution one 
realizes (e.g.\ using instantaneously height-averaged profiles from a fully 3D MHD 
simulation), and then show how different unstable modes appear due to the intricate 
interplay of the various restoring forces that are present.

\section{Basic theory}{\Sn{II}}

\subsection{Cylindrical accretion disk equilibrium}{\Sn{IIA}}

Consider the cylindrical slice model of an accretion disk rotating about a compact or 
protostellar object of mass $M_{\textstyle*}$ (Fig.~\F{1}, adapted from 
\Ron{GKP2019}). The vertical variation of the plasma equilibrium is neglected so that 
we obtain an annular cylindrically symmetric slice of height $\Delta z$ and located 
within the radial range $r_1 \le r \le r_2$. The disk is confined by the gravitational field 
of the central object,
\BEQ \Phi_{\rm gr} = - \frac{G M_{\textstyle*}}{\sqrt{r^2 + z^2}} \approx - \frac{G 
M_{\textstyle*}}{r} \quad\Rightarrow\quad g = \Phi_{\rm gr}' = G M_{\textstyle*}/r^2 
\qquad \big(\,' \equiv d/dr \,\big)\,,\En{1} \EEQ
where the approximation  on the RHS is justified by the assumption of short wavelengths 
in the vertical direction, $k \Delta z \gg 1$. The Newtonian potential exploited could be 
replaced by the Paczy\'nski--Wiita potential~\R{PW1980}, to approximate general 
relativistic effects, but this complication is avoided here assuming that $r_1$ is far 
outside the Schwarzschild radius. Also, accretion is supposed to take place on a time 
scale much longer than that of the instabilities and ensuing turbulence that cause it, so 
that the radial velocity $v_r$ is negligible with respect to the rotation velocity $v_\theta$. 
The equilibrium is then described by rather arbitrary radial distributions of the following 
quantities. The disk has a density $\rho(r)$, it exerts a pressure  $p(r)$, it is rotating with 
angular frequency $\Omega \equiv v_\theta(r)/r$, and immersed in a helical magnetic 
field with a toroidal component $B_\theta(r)$ and a vertical component $B_z(r)$. The 
only restriction on these five profiles is that they should satisfy the radial equilibrium 
constraint
\BEQ \rho r(\Omega^2 -  G M_{\textstyle*}/r^3) = (p + \half B^2)' + B_\theta^2/r \,, 
\En{2}\EEQ
where the RHS pressure and magnetic field contributions produce deviations from the 
LHS Keplerian rotation. A vertical flow field $v_z(r)$ would not modify this 
equilibrium, and it has also been incorporated in the general analysis of Section 13.3 of 
\Ron{GKP2019} on which this paper is based, but it is put to zero in the present paper. 
The kinetic pressure is assumed large with respect to the magnetic pressure, $\beta \equiv 
2p/B^2 \gg 1$, so that the perturbations of this equilibrium are assumed to be 
incompressible in the explicit analyses of Sections~\S{III} -- \S{V}. The numerical 
results from the program ROC~\R{Goed2018a, Goed2018b} presented throughout the 
paper are not restricted by this condition; the program exploits the fully general spectral 
differential equation solver.

\begin{figure}[ht]
\FIG{\begin{center} 
\includegraphics*[height=3.8cm]{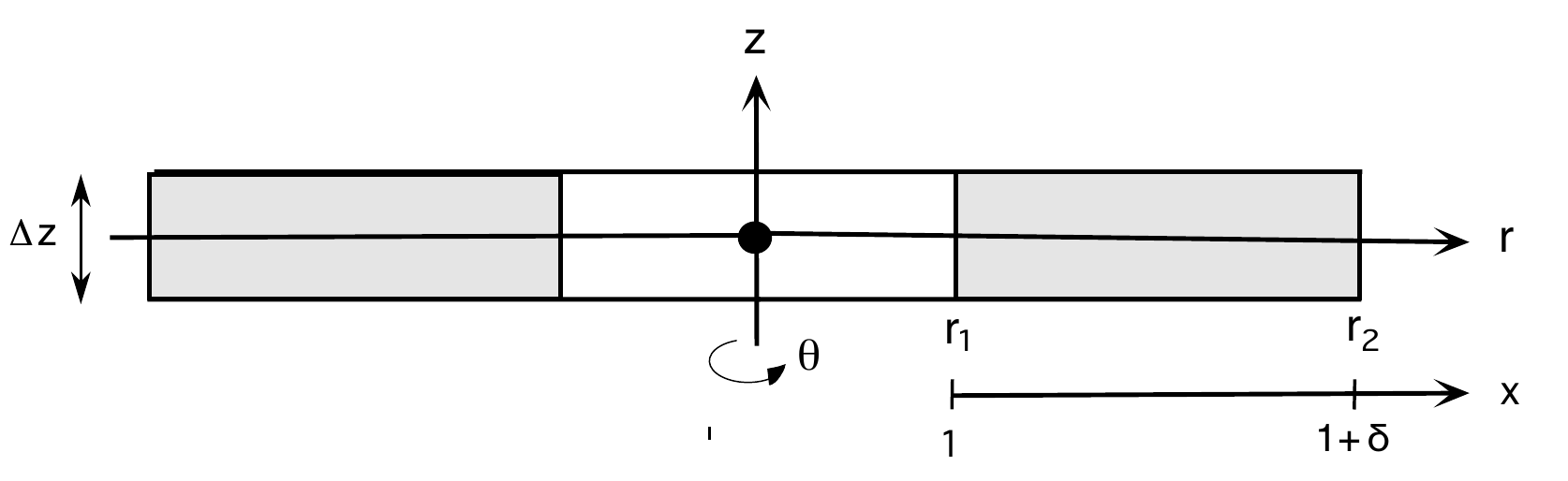}
\end{center}} 
\caption{Cylindrical slice model (annulus $r_1 \le r \le r_2$ of height $\Delta z$) of an 
accretion disk rotating about a compact object in the origin. The dimensionless radial 
coordinate $x \equiv r/r_1$ is used for the equilibria and eigenfunctions depicted in this 
paper.}{\Fn{1}}
\end{figure}

\begin{figure}[ht]
\FIG{\begin{center} 
\includegraphics*[height=10.8cm]{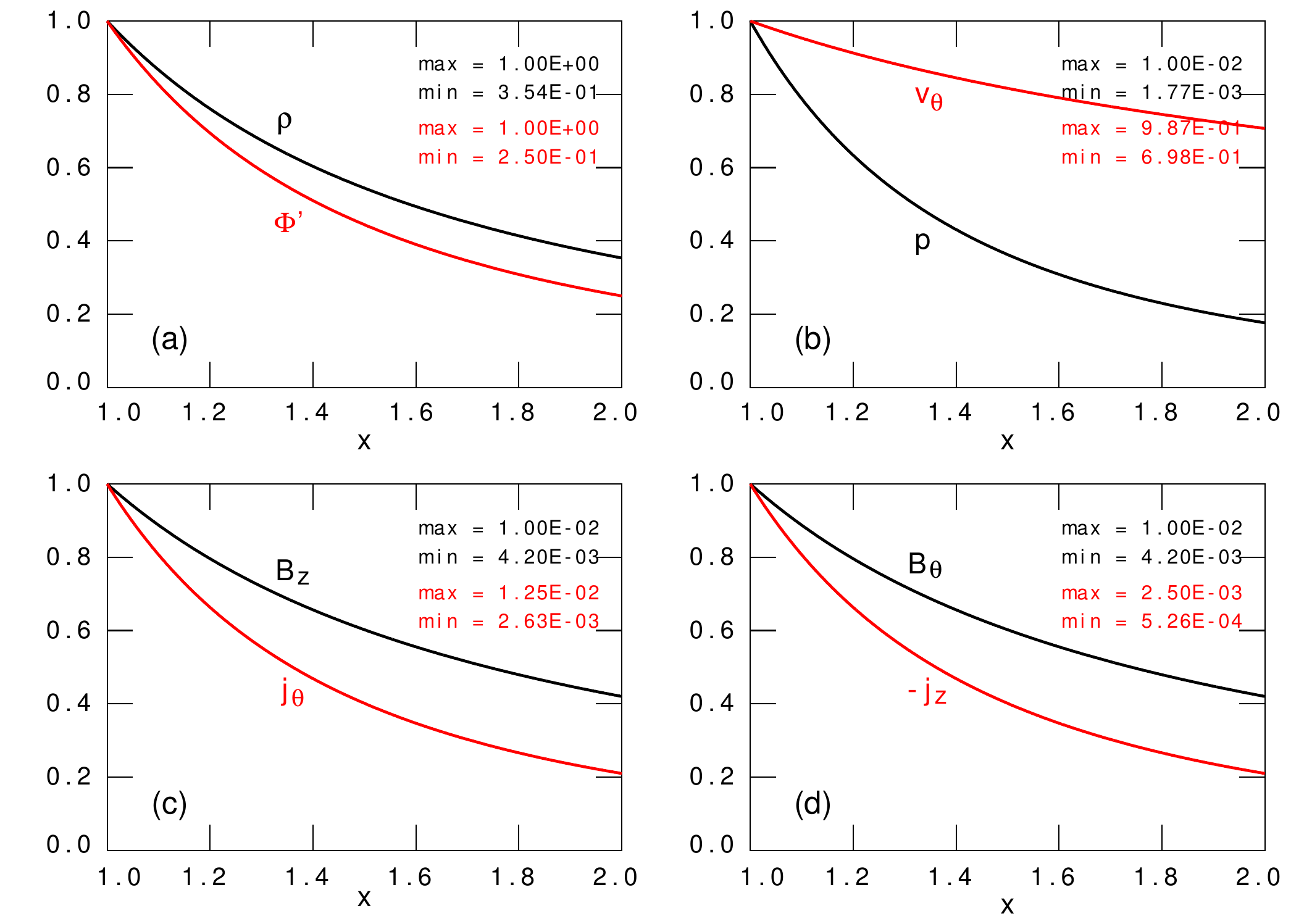}
\end{center}}
\caption{Radial equilibrium distributions (in black and red) for $\epsilon \equiv B_1 = 
0.0141$, $\mu_1 \equiv B_{\theta 1}/B_{z1}  =1$, $\beta \equiv 2p_1/B_1^2 = 100$, 
$\delta \equiv r_2/r_1 - 1 = 1$, of (a)~{density and derivative of the gravitational 
potential}, (b)~{pressure and azimuthal velocity}; (c)~{longitudinal magnetic field and 
associated current density}, (d)~{azimuthal magnetic field and associated current 
density}.}{\Fn{2}}
\end{figure}

We will utilize \IT{the scale independence} of the ideal MHD equations (see Section 
4.1.2 of \Ron{GKP2019}) by exploiting units of length, mass and time, effectively 
provided by the distance from the origin to the reference position $r = r_1$, the density 
$\rho_1$ and the Keplerian velocity $v_{\rm{K\hs}1} \equiv \sqrt{G 
M_{\textstyle*}/r_1}$ at that position. This yields dimensionless quantities $\bar{r} 
\equiv r/r_1$, $\bar{\rho} \equiv \rho/\rho_1$, $\bar{\Omega} \equiv 
r_1\Omega/v_{\rm{K\hs}1}$, $\bar{B}_{z,\theta} \equiv B_{z,\theta}/(\sqrt{\mu_0 
\rho_1}{\hs}v_{\rm{K\hs}1})$, etc. (In the plots, the coordinate $\bar{r}$ will be 
indicated by $x$, as shown in Fig.~\F{1}.) From now on, we get rid of these \IT{trivial 
scaling parameters} by dropping the bars and exploiting the resulting dimensionless 
quantities. This rescaling also gets us rid of the awkward constant $\mu_0$ of the 
magnetic variables, as anticipated in Eq.~\E{2}. In contrast, four \IT{essential 
parameters} will describe all our numerical examples, viz.
\BEQ \epsilon \equiv B_1 \equiv (B_{z{\hs}1}^2 + B_{\theta{\hs}1}^2)^{1/2} \,,\quad 
\beta \equiv 2 p_1/B_1^2 \,,\quad \mu_1 \equiv B_{\theta{\hs}1}/B_{z{\hs}1} 
\,,\quad\delta \equiv (r_2 - r_1)/r_1 \,,\En{3}\EEQ
where we have replaced the previous definitions of $\epsilon$ and $\delta$ used in 
\Ron{GKP2019} and \Ron{Goed2018b} by the present, more relevant, ones. In 
particular, the parameter $\epsilon$ now directly measures the magnitude of the magnetic 
field, which will be argued below to be the main dynamical quantity (next to rotation of 
course). The subscript on the constant $\mu_1$ is essential since the inverse pitch of the 
field lines is not constant: $\mu \equiv B_\theta/(rB_z) = \mu_1 r^{-1}$. Adopting the 
standard self-similar radial dependence of accretion disks introduced by~\Ron{SMIS87}, 
we then obtain the following explicit equilibrium:
\BEQ \rho = r^{-3/2} \,,\quad \Omega = \Omega_1 r^{-3/2} \,,\quad B_z = B_{z1} r^{-
5/4} \,,\quad B_\theta = \mu_1 B_{z1} r^{-5/4}\,,\quad p = \half \beta \epsilon^2 r^{-
5/2} \,, \En{4}\EEQ
where the angular rotation frequency $\Omega_1$  follows from the equilibrium 
equation~\E{2}:
\BEQ B_{z1} =  \frac{\epsilon}{(1 + \mu_1^2)^{1/2}} \quad\Rightarrow\quad 
\Omega_1 =  \Big\{1 - \fivefourth \epsilon^2 \Big[ \beta + \frac{1 + \onefifth 
\mu_1^2}{1 + \mu_1^2}\Big] \Big\}^{1/2} \approx 1 - \fiveeighth \epsilon^2 \beta 
\,.\En{5}\EEQ
Approximating $\Omega_1 \approx 1$ (i.e.\ Keplerian rotation at $r_1$ and, hence, 
everywhere) would result from the assumptions $\epsilon \ll 1$ and $\epsilon^2 \beta \ll 
1$ (but $\beta \sim \epsilon^{-1} \gg 1$) that are made for the examples in this paper. 
These assumptions also imply that the equilibrium rotation velocities considered will be 
\IT{extremely super-Alfv\'enic:} $v_\theta \sim 1 \gg v_{\rm A} \sim \epsilon$. It is to 
be noted that the parameter $\beta$, even though large, disappears from the spectral 
analysis since it leads to incompressibility of the modes. Hence, somewhat paradoxically, 
the characteristic ordering of importance for these accretion disk problems is: 
(1)~dominant rotation (quantified by the Keplerian velocity $v_{\rm{K\hs}1}$), 
(2)~small magnetic field magnitude (quantified by the parameter~$\epsilon$) and, 
finally, (3)~larger pressure effects (quantified by the parameter $\beta \epsilon^2$) that 
are of no importance though for the dynamics of the modes described in this paper. The 
radial distribution of the equilibrium variables for a typical cylindrical accretion disk is 
shown in Fig.~\F{2}.

Many more self-similar equilibria could be considered. In particular, the functions 
$\rho(r)$, $B_\theta^2(r)$, $B_z^2(r)$ and $p(r)$ could be multiplied with an arbitrary 
power of $r$. For definiteness and comparison with the existing literature~\R{SMIS87, 
ShakSun73}, we have set that power equal to $1$ in the expressions~\E{4}. Again, the 
numerical tools ROC~\R{Goed2018a, Goed2018b}, exploited in this paper, or 
Legolas~\R{CDK2021}, exploited in a forthcoming paper, are not restricted to these 
equilibria. 

\subsection{The Frieman--Rotenberg spectral problem}{\Sn{IIB}}

The most effective starting point for the study of waves and instabilities of stationary 
equilibria, such as the present one, is the seminal paper of~\Ron{FrieRo60}. For modes 
exponentiating as $\exp(-{\rm i} \omega t)$, their spectral differential equation may be 
written as
\BEQ \bfG(\bfxi) - 2 \hs\omega{\hs}U\bfxi + \rho \hs\omega^2 \bfxi = 0 \,,\En{6}\EEQ
where $\bfG(\bfxi) \equiv \bfF(\bfxi) + \div{\big(\hs \bfxi \hs\rho \bfv \cdot \nabla \bfv)} 
- \rho (\bfv \cdot \nabla)^2 \bfxi$ is the force operator expression, generalizing the more 
well-known expression $\bfF(\bfxi)$ for static equilibria~\R{BFKK58}, and $U \equiv - 
{\rm i} \hs\rho \bfv \cdot \nabla$ is the gradient operator projected onto the velocity field. 
With suitable boundary conditions (BCs) on $\bfxi$, Eq.~\E{6} becomes a quadratic 
eigenvalue problem in terms of the two operators $\bfG$ and $U$, which are both 
self-adjoint. Hence, the two associated quadratic forms
\BEQ W \equiv - \half \int \bfxi^* \cdot \bfG(\bfxi) \,dV \,,\quad\hbox{and}\quad\; V 
\equiv \half \int \bfxi^* \cdot U \bfxi \,dV \En{7}\EEQ
are real for solutions satisfying the BCs. The meaning of the ambiguous symbol $V$, 
referring to both the integral and the plasma volume, will be evident from the context.  
Introducing the normalization $I \equiv \half \int \rho |\bfxi|^2 \,dV$, Eqs.~\E{6} and 
\E{7} yield a quadratic equation for the complex variable $\omega \equiv \sigma + {\rm 
i}\hs\nu$:
\BEQ \omega^2 - 2 \hs\overline{V} \omega - \overline{W} = 0 \,,\qquad \overline{V} 
\equiv {V}/{I} \,,\qquad \overline{W} \equiv {W}/{I} \,, \En{8}\EEQ
where $\overline{V}$ and $\overline{W}$ are the solution averages of the 
Doppler--Coriolis shift and of the potential energy of the perturbations. The solutions of 
this quadratic equation for instabilities, with real frequency $\sigma$ and growth rate 
$\nu$, may be written as
\BEQ \sigma = \overline{V}, \qquad \nu = \pm \big(-\overline{W} -
\overline{V}^2\hs\big)^{1/2}. \En{9}\EEQ
It would look like: problem solved! However, {\em these expressions cannot be evaluated 
a priori\,}. In the standard approach, they can only be computed a posteriori when the 
final solution of the spectral equation \E{6}, consisting of the eigenvalue--eigenfunction 
pair $\{\omega, \bfxi(\bfr;\omega)\}$, is known. This obstacle is overcome by the new 
method of the {\em Spectral Web} described in Section~\S{IID}.

\subsection{Reduction to the cylindrical spectral differential equation}{\Sn{IIC}}

The complex eigenvalues of the waves and instabilities of the stationary cylindrical 
equilibria described in Section~\S{IIA} are determined by the solutions of an ordinary 
differential equation (ODE) for the radial component $\chi \equiv r \xi_r$ of the plasma 
displacement, that may be obtained from the Frieman--Rotenberg spectral 
equation~\E{6} by straightforward reduction. We skip over all special cases of this ODE 
that have been exploited over the past sixty years and, for reference, just indicate the 
general forms for different assumptions of the underlying equilibrium. For static 
isothermal ($\gamma = 1$) cylindrical equilibria, the relevant second order ODE was 
derived by \Ron{HL58}, and generalized to adiabatic equilibria by \Ron{Goed71}. The 
stationary counterpart was derived by \Ron{Hameiri81} and numerically investigated by 
\Ron{BIB87}. These studies were concerned with stability of magnetically confined 
laboratory plasmas for thermonuclear purposes. For astrophysical applications, the 
extension to the \IT{completely general} cylindrical ODE with a gravitational field, 
applied to the various instabilities of an accretion disk rotating about a massive central 
object, e.g.\ the Magneto-Rotational Instability or MRI~\R{Vel59, Chandra60, BH91a, 
BH98}, was obtained by \Ron{KCG02}. 

We exploit the latter form of the cylindrical spectral equation, as given in Section~3.3.2 
of \Ron{GKP2019} for normal modes of the form $\chi(r) \exp[\hs{\rm i} (m \theta + k z 
- \omega t)]$:
\BEQ \frac{d}{dr} \bigg(\, \frac{N}{D} \,\frac{d\chi}{dr} \,\bigg) + \bigg[\,A + 
\frac{B}{D} + \bigg(\frac{C}{D}\bigg)' \,\bigg] \, \chi = 0 \,, \En{10}\EEQ
where the modes are supposed to be localized within `rigid' boundaries at $r = r_1$ and 
$r = r_2$ represented by the boundary conditions
\BEQ \chi(r_1) = 0 \;\;\hbox{(left)}\,,\qquad \chi(r_2) = 0 \;\;\hbox{(right)} 
\,.\En{11}\EEQ
Of course, the actual boundaries of an accretion disk will not comply with the assumption 
of an annular cylindrical slice, so that we have to convince ourselves that, in the end, the 
eigenfunctions computed will be well within the range $r_1 \le r \le r_2$. For numerical 
integration it is expedient~\R{AGV74} to transform the ODE~\E{10} into a pair of 
complex first order ODEs in terms of $\chi$ and the total pressure perturbation, $\Pi 
\equiv p_1 + B_0 B_1 \equiv -(N/D) \chi' - (C/D) \chi$:
\BEQ
N \frac{d}{dr} \Bigg( \!\begin{array}{c}
\chi \\[3mm] 
\Pi
\end{array} \!\Bigg)
+ \Bigg( \! \begin{array}{cc}
C & D \\[2mm]
E & - C
\end{array} \!\Bigg)
\Bigg(\! \begin{array}{c}
\chi \\[3mm]
\Pi
\end{array} \!\Bigg) = 0 \,, \En{12}\EEQ
where the coefficient $E$ is defined in terms of the other coefficients, 
\BEQ E \equiv - N \big( A + {B}/{D} \big) - {C^2}/{D} \,.\En{13}\EEQ
Of course, all physics now resides in the expressions for the coefficients $N$, $D$, $A$, 
$B$ and $C$, which are functions of the radial position $r$ \IT{and of the 
Doppler-shifted complex frequency} $\widetilde{\omega}$ which is also a function of 
$r$:
\BEQ \widetilde{\omega} \equiv \omega - \Omega_0 ,\qquad \Omega_0 \equiv m 
\Omega + k v_z \,. \En{14}\EEQ
For generality, we have kept the vertical velocity $v_z$ though it will be neglected in the 
rest of this paper. In contrast to the usual analysis of axisymmetric MRIs ($m = 0$), 
where $\widetilde{\omega} = \omega$ is constant, the radial profile $\Omega_0(r)$ 
plays a central role in the description of the non-axisymmetric modes ($m \ne 0$), so that 
$\widetilde{\omega}$ is not constant. This significantly complicates the analysis but, of 
course, also greatly enlarges the possible dynamics of accretion disks about compact 
objects. In particular, the ODEs~\E{10} and \E{12} now become a fourth order system 
for the determination of the four components $\chi_1$, $\chi_2$, $\Pi_1$ and $\Pi_2$ of 
the essentially complex variables $\chi \equiv \chi_1 + {\rm i}\hs\chi_2$ and $\Pi \equiv 
\Pi_1 + {\rm i}\hs\Pi_2$. 

Recall that the collection of real frequencies $\omega = \{\Omega_0(r)|r_1 \le r \le r_2\}$ 
forms the flow continuous spectrum in hydrodynamics~\R{Case60}. In magnetized 
plasmas, these frequencies still play an important role, as we will see, but their role as a 
continuum is superseded by the \IT{forward and backward Doppler-shifted Alfv\'en and 
slow magneto-sonic continuum frequencies}
\BEQ 
\Omega_{\rm A}^\pm \equiv \Omega_0(r) \pm \omega_{\rm A}(r) \,,\qquad 
\Omega_{\rm S}^\pm \equiv \Omega_0(r) \pm \omega_{\rm S}(r) \,, \En{15} \EEQ
where $\omega_{\rm A} \equiv F/\sqrt{\rho}$, $F \equiv mB_\theta/r + k B_z$, and 
$\omega_{\rm S} \equiv [\gamma p/(\gamma p + B^2)]^{1/2} \hs\omega_{\rm A}$ are 
the static expressions. The latter square root factor $\approx 1 - (4\gamma\beta)^{-1} 
\approx 1$ for $\beta \gg1$, so that the modes are essentially  incompressible and the 
difference between the slow and Alfv\'en frequencies disappears. Note that {\em these 
frequencies are real!} The definitions of the singularity coefficients $N$ and $D$ can 
then be written as 
\BEQAR N &\equiv& \widetilde{A}\,\widetilde{S}/r \,,\qquad D \,\equiv\, \rho^2 
\widetilde{\omega}^4 - (m^2/r^2 + k^2) \widetilde{S} \,,\non\\[1mm]
&& \widetilde{A} \,\equiv\, \rho (\widetilde{\omega}^2 - \omega_{\rm A}^2) \,=\, 
\rho(\omega - \Omega^+_{\rm A}) (\omega - \Omega^-_{\rm A})  \,,\non\\[1mm]
&& \widetilde{S}\, \equiv\, \rho(\gamma p + B^2) (\widetilde{\omega}^2 - \omega_{\rm 
S}^2) \,=\, \rho(\gamma p + B^2)(\omega - \Omega^+_{\rm S}) (\omega - \Omega^-
_{\rm S}) \,, \En{16} \EEQAR
whereas the definitions of the remaining coefficients, as well as the fourth order ODEs 
exploited in the numerical analysis, may be found in Appendix~\S{A1}. 

The expressions for the tangential components $\xi_\theta$ and $\xi_z$ in terms of 
$\chi'$ and $\chi$, explicitly given by Eqs.~(13.93) and (13.94) of \Ron{GKP2019}, 
enter the crucial relation of the real frequency part $\sigma$ of the modes and \IT{the 
solution-averaged Doppler--Coriolis shift},
\BEQ \sigma = \overline{V} = {\int \!\rho \hs\big[\hs r\Omega_0\hs|\bfxi|^2 + {\rm 
i}\hs\Omega \hs(\chi^*\xi_\theta - \chi \hs\xi_\theta^*) \hs\big] {\hs}dr}\Big/{\!\int \rho 
\hs|\bfxi|^2 r{\hs}dr} \,,\En{17}\EEQ
which is real! The first part of the integral is the solution-averaged Doppler shift, to be 
distinguished from the local Doppler shift~$\Omega_0(r)$, and the second part is the 
Coriolis shift of the real part of the frequencies of the modes. The physical significance of 
the equality~\E{17} is that the solution-averaged Doppler--Coriolis shifted real part of 
the frequency has to vanish, $\sigma - \overline{V} = 0\hs$, for every instability.

\subsection{The Spectral Web method}{\Sn{IID}}

In \Ron{Goed2009a, Goed2009b} and \Ron{Goed2018a, Goed2018b}, a new method of 
solving the quadratic eigenvalue problem \E{6} + BCs was developed based on the fact 
that the operator $U$ is actually self-adjoint irrespective of whether $\bfxi$ satisfies the 
BCs or not. Hence, the expression $\overline{V}$ is real throughout the complex 
$\omega$-plane for {\em any} solution of Eq.~\E{6}. In particular, one may compute the 
value of $\overline{V}$ for arbitrary $\omega$ by `shooting' from one of the boundaries, 
say at $r = r_1$ in the present case, with solutions of Eq.~\E{6} that only satisfy the BCs 
there. For a given value $\nu = \nu_0$, the resulting nonlinear algebraic equation $\sigma 
- \overline{V}[\hs\bfxi(\bfr;\sigma + {\rm i}\nu_0)\hs] = 0$ may then be solved by 
straightforward iteration on the zeros, which gives one or more solutions $\sigma = 
\sigma_0(\nu_0)$. The collection of all these points $\{\sigma_0(\nu_0) + {\rm i} 
\hs\nu_0\}$ of the $\omega$-plane, where the real part of the Doppler--Coriolis shifted 
frequency vanishes, has been called {\em the solution path} since all eigenvalues have to 
lie on that path. The actual eigenvalues may then be obtained by a second iteration, along 
the solution path, to also satisfy the remaining BCs, i.e.\ at $r = r_2$ in the present case. 

The original solution path method~\R{Goed2009a, Goed2009b} has been modified 
substantially~\R{Goed2018a, Goed2018b} by combining the mentioned two steps into a 
single scheme involving {\em the complementary energy $W_{\rm com}$.} This is a 
complex surface integral representing the energy to be injected into or extracted from the 
plasma such  that energy conservation applies for the solution $\bfxi$ of the spectral 
equation for any arbitrary value of $\omega\hs$. The zeros of the imaginary part of the 
complementary energy provide the solution path (where $W_{\rm com}$ is real), 
whereas the zeros of the real part provide a second path, called {\em the conjugate path} 
(where $W_{\rm com}$ is imaginary): 
\BEQ W_{\rm com}[\bfxi(\bfr;\omega)] = 0 \quad\Rightarrow\; \left\{ \begin{array}{l}
{\rm Im}(W_{\rm com}) = 0 \quad\hbox\IT{(solution path)} \\[4mm] 
{\rm Re}(W_{\rm com}) = 0 \quad\hbox\IT{(conjugate path)} 
\end{array} \right. . \En{18}\EEQ
It is shown in \Ron{Goed2018a} that Eq.~\E{18}(a) for the solution path is equivalent to 
$\sigma - \overline{V}= 0$, whereas Eq.~\E{18}(b) for the conjugate path is equivalent 
to satisfying the remaining BC, i.e.\ at $r = r_2$ if one `shoots' from the left. The ideal 
MHD spectrum of unstable modes of rotating equilibria is then found by constructing the 
{\em Spectral Web}, consisting of a superposition of all the curves of the solution path 
and the conjugate path in the complex $\omega$-plane. The eigenvalues then emerge by 
computing the intersections of those curves, i.e.\ \IT{where both real and imaginary part 
of the complementary energy $ W_{\rm com}$ vanish.}

For the present investigation of the instabilities of the cylindrical slice model of an 
accretion disk, it is most expedient to exploit a \IT{mixed} integration scheme of the 
ODEs~\E{12} by `shooting' both from the left and from the right and joining the solution 
$\chi$ in some mixing point $r_{\rm mix}$ in the middle of the interval $(r_1, r_2)$, 
notably where $\chi$ has a large amplitude. One can then adapt the amplitude of the left 
solution $\chi^\ell$, satisfying $\chi^\ell(r_1) = 0$, to that of the right solution $\chi^{\rm 
r}$, satisfying $\chi^{\rm r}(r_2) = 0$, by renormalizing such that $\chi^\ell(r_{\rm 
mix}) = \chi^{\rm r}(r_{\rm mix})$. For an arbitrary value of $\omega$, the derivative 
of $\chi$ will not be  continuous at $r_{\rm mix}$, i.e.\ the total pressure perturbation 
$\Pi$ will exhibit a jump there, $\doublel \Pi \doubler \equiv \Pi(r_{\rm mix}^+) - 
\Pi(r_{\rm mix}^-)$. This jump determines the complementary energy:
\BEQ W_{\rm com}^{\rm mix}  = - \half \int \xi^* \,\doublel \Pi \doubler \,dS_{\rm mix} 
=  -\pi \Delta z \,\big(\chi^* \doublel \Pi \doubler\big) _{r_{\rm mix}}  \,. \En{19} \EEQ
The spectral web for the mixed solutions then follows by constructing the two paths from 
the two expressions~\E{18} for `all' values of $\omega$. Note that the numerical 
procedure involved is trivially parallel. Moreover, it may be restricted to the particular 
strip of the $\omega$-plane where physical solutions are expected.

\section{Accretion disk instabilities}{\Sn{III}}

\subsection{The accretion disk differential equation }{\Sn{IIIA}}

For the purpose of explicit analysis of the accretion disk instabilities, here we will 
simplify the basic ODE~\E{10} by exploiting the smallness of the parameters of the 
equilibrium~\E{4}. The only place where the pressure, i.e.\ the parameter $\beta$, enters 
the coefficients of the spectral differential equation~\E{10} is in the definition of the 
slow continuum frequency~\E{15}, i.e.\ in the coefficient $\gamma p/(\gamma p + 
B^2)^{1/2}$. We have already seen in Sec.~\S{IIC} that this coefficient is approximately 
unity so that the modes are essentially  incompressible. The slow and Alfv\'en 
singularities then coalesce, $\Omega_{\rm S}^+ \rightarrow \Omega_{\rm A}^+$, and 
similarly for the backward singularities, so that the singularity quotient from the 
expressions~\E{16} becomes: 
\BEQ \frac{N}{D} \approx - \frac{r \hs\rho \hs(\widetilde{\omega}^2 - \omega_{\rm 
A}^2)}{m^2 + k^2 r^2} \equiv - \frac{r \hs\rho \hs(\omega - \Omega_{\rm 
A}^+)(\omega - \Omega_{\rm A}^-)}{m^2 + k^2 r^2} \,. \En{20} \EEQ
We will indicate the coalesced continua by $\{\Omega_{\rm A}^+(r)\}$ and 
$\{\Omega_{\rm A}^-(r)\}$, i.e.\ we will not distinguish between slow and Alfv\'en 
anymore. This way, the more essential distinction between forward and backward 
continua is highlighted. We will show that the radial dependence of these continua 
determines the kind of instabilities that occur, with fundamental differences between the 
usual axisymmetric MRIs (where the tilde on $\widetilde{\omega}$ may be dropped 
since $\Omega_0 = 0$) and the present non-axisymmetric modes (where $\Omega_0 \ne 
0$ is crucial).

 In the incompressible approximation, the differential equation~\E{10} with the 
expressions \E{A1}--\E{A6} for the coefficients simplifies to:
\BEQ r \frac{d}{dr} \bigg[\hs \frac{\rho (\widetilde{\omega}^2 - \omega_{\rm 
A}^2)}{m^2 + k^2 r^2} \hs r \hs\frac{d\chi}{dr} \hs\bigg] - \bigg[\,\rho 
(\widetilde{\omega}^2 - \omega_{\rm A}^2) + \Delta - \frac{4 k^2 (B_\theta F + \rho 
r\Omega \widetilde{\omega})^2}{( m^2 + k^2 r^2)\rho (\widetilde{\omega}^2 - 
\omega_{\rm A}^2)} - 2 m r \bigg(\frac{B_\theta F + \rho r\Omega 
\hs\widetilde{\omega}}{r(m^2 + k^2 r^2)}\bigg)' \,\bigg] \, \chi = 0 \,, \En{21}\EEQ
where the function $\Delta(r)$ is defined in Eq.~\E{A5}. Introducing the squared 
epicyclic frequency $\kappa_{\rm e}^2 \equiv r (\Omega^2)' + 4 \Omega^2$ [{\hs}which 
may be negative, in general, but it is positive for the equilibrium profiles~\E{4}, where 
$\kappa_{\rm e}^2 = \Omega_1^2 {\hs}r^{-3}\hs$] and exploiting the smallness of the 
parameters $\epsilon$ and $k^{-1}$, the orders of magnitude of the different terms of 
Eq.~\E{21} are given by
\BEQ m^2 \sim 1 \ll k^2 r^2\,,\quad \Delta \approx - \rho r (\Omega^2)' \equiv - 
\rho(\kappa_{\rm e}^2 - 4 \Omega^2) \sim 1\,,\quad B_\theta F \sim \epsilon^2 \ll \rho r 
\Omega \hs\widetilde{\omega} \sim 1\,,\En{22}\EEQ
whereas the last term of Eq.~\E{21} (with the derivative) may be neglected. We then get
\BEQ \frac{r}{\rho}\hs \frac{d}{dr} \bigg[\hs \frac{\rho}{r} \hs(\widetilde{\omega}^2 - 
\omega_{\rm A}^2) \hs\frac{d\chi}{dr} \hs\bigg] - k^2 \hs\bigg[\, \widetilde{\omega}^2 
- \omega_{\rm A}^2 - \kappa_{\rm e}^2 
- \frac{4 \Omega^2 \omega_{\rm A}^2}{\widetilde{\omega}^2 - \omega_{\rm A}^2} 
\,\bigg] \, \chi = 0 \,. \En{23}\EEQ
Since the approximations \E{22} are extremely well satisfied for the accretion disk 
equilibria, Eq.~\E{23} can be considered as the basic ordinary differential equation for 
all instabilities of these configurations in the cylindrical representation. For that reason, 
we will call it the {\em accretion disk ODE} in the following. Note that the special 
equilibrium distributions~\E{4} have not been exploited yet in this equation.

Finally, that the pressure, i.e.\ compressibility, hardly affects the accretion disk 
instabilities does not imply that it may also be neglected in the equilibrium 
equation~\E{2}. Accordingly, all results presented will involve the exact solutions of that 
equation. Also, whereas the accretion disk ODE~\E{23} will be exploited for all 
analytical investigations, for the numerical results the solutions of the exact ODEs \E{10}  
or \E{12} will be exploited.

\begin{figure}[ht]
\FIG{\begin{center} 
\includegraphics*[height=12.6cm]{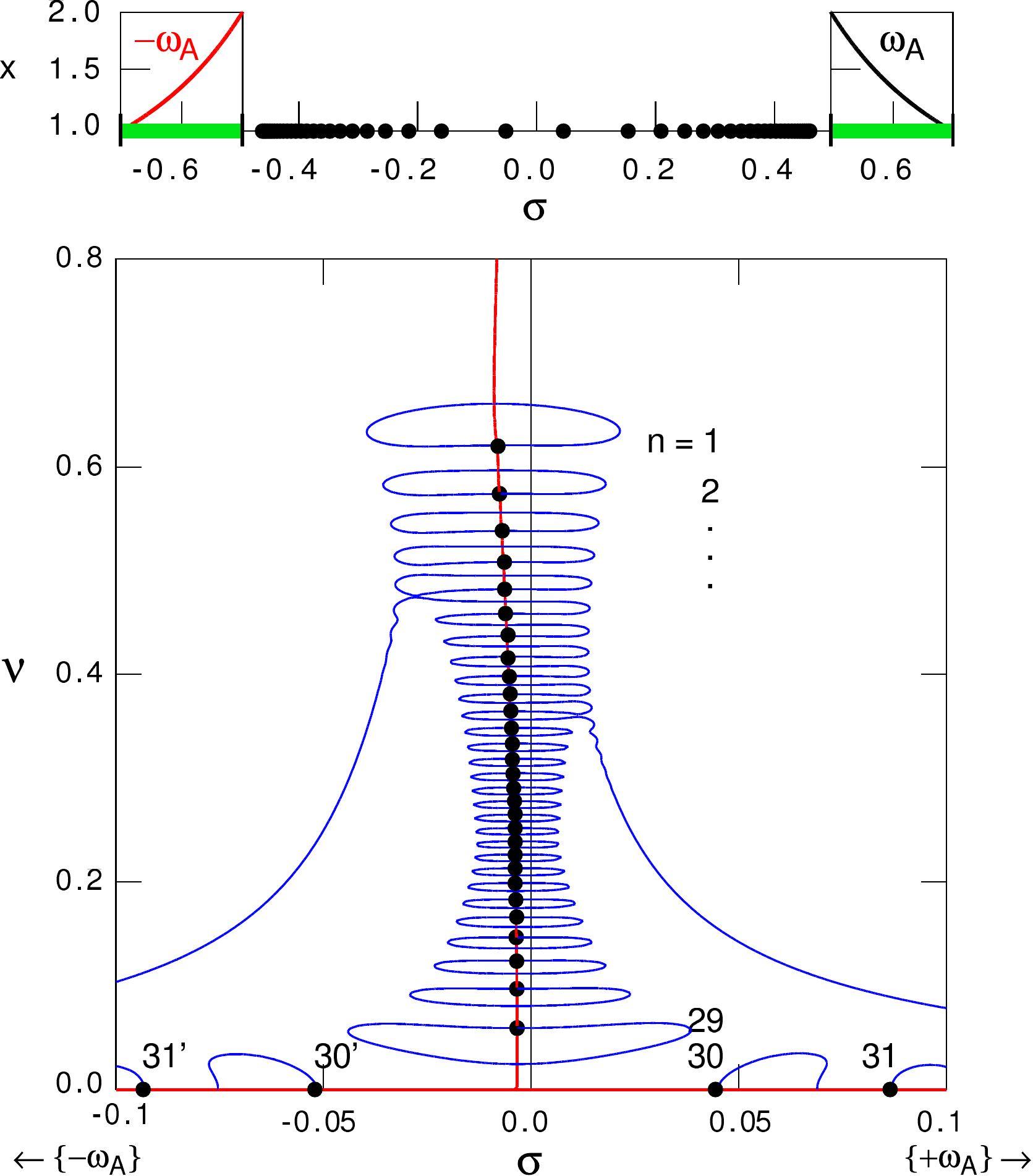}
\end{center}}
\caption{Spectral Web of the incompressible Magneto-Rotational Instabilities for an 
accretion disk with equilibrium parameters $\epsilon \equiv B_1 = 0.0141$, $\mu_1 
\equiv B_{\theta 1}/B_{z1}  =1$, $\beta \equiv 2p_1/B_1^2 = 100$, $\delta \equiv 
r_2/r_1 - 1 = 1$, and mode numbers $m = 0$ (axisymmetric!), $k = 70$. The top frame 
shows the radial profiles of the continuum frequencies and the continua themselves (in 
green), as well as the stable modes $n = 30 \ldots 126$ and $30' \ldots 126'$ (skipping 
with $\Delta n = 4$). Note the different horizontal scale of the top frame: the continua are 
located far away from the 29 MRIs of the bottom frame!}{\Fn{3}}
\end{figure}

\subsection{Axisymmetric Magneto-Rotational Instability}{\Sn{IIIB}}

To set the stage for the analysis of the non-axisymmetric instabilities to be discussed in 
Section~\S{IV}, it is instructive to recall the major characteristics of the axisymmetric 
MRIs following from Eq.~\E{23}. For the MRIs, since $m = 0$ so that $\Omega_0 = 0$, 
the function $\widetilde{\omega}(r)$ reduces to the eigenvalue $\omega$ and 
$\omega^2$ becomes real, as in static MHD, and waves and instabilities may be 
distinguished by the simple criterion $\omega^2 > 0$, or $\omega^2 < 0$. In that case, 
the ODE~\E{23} reduces to a second order system to determine the \IT{real} functions 
$\chi(r)$ and $\Pi(r)$. (These enormous simplifications may explain the popularity of 
MRIs to model the anomalous dissipation required for accretion.) A rather general 
dispersion equation may then be derived by means of a local WKB approximation of the 
form $\chi(r) = p(r) \hs\exp \big[\pm{\rm i} \int q(r) \,dr \hs\big]$, where radial variations 
of the equilibrium variables over distances of the order $q^{-1}$ are neglected. For $m = 
0$, where $\omega_{\rm A} = k B_z/\sqrt{\rho}$, this yields the local dispersion 
equation,
\BEQ (k^2 + q^2) (\omega^2 - \omega_{\rm A}^2)^2 - k^2 \kappa_{\rm e}^2 (\omega^2 
-\omega_{\rm A}^2) - 4 k^2 \Omega^2 \omega_{\rm A}^2 = 0 \,,\En{24}\EEQ
having the two solutions
\BEQ \omega^2 = \omega_{\rm A}^2 + \half \kappa_{\rm e}^2 E^{-1} \Big[{\hs} 1 \pm 
\big(1 + 16 E \hs\Omega^2 \omega_{\rm A}^2/\kappa_{\rm e}^4 \big)^{1/2} 
\,\Big]\,,\qquad E \equiv 1 +  q^2/k^2  \,, \En{25}\EEQ
where the plus sign corresponds to the frequencies of the stable modes of the epicyclic 
sub-spectrum and the minus sign corresponds to the stable and unstable modes of the 
MRI sub-spectrum. The expression~\E{25} provides reasonable estimates for the range 
of $q$ (or the number of zeros of the eigenfunctions, $n$) from the largest growth rate 
(smallest $n$) to the transition to stability. The `exact' unstable eigenvalues obtained 
from the numerical solutions presented below are in general complex though. 

\begin{figure}[ht]
\FIG{\begin{center} 
\includegraphics*[height=14.8cm]{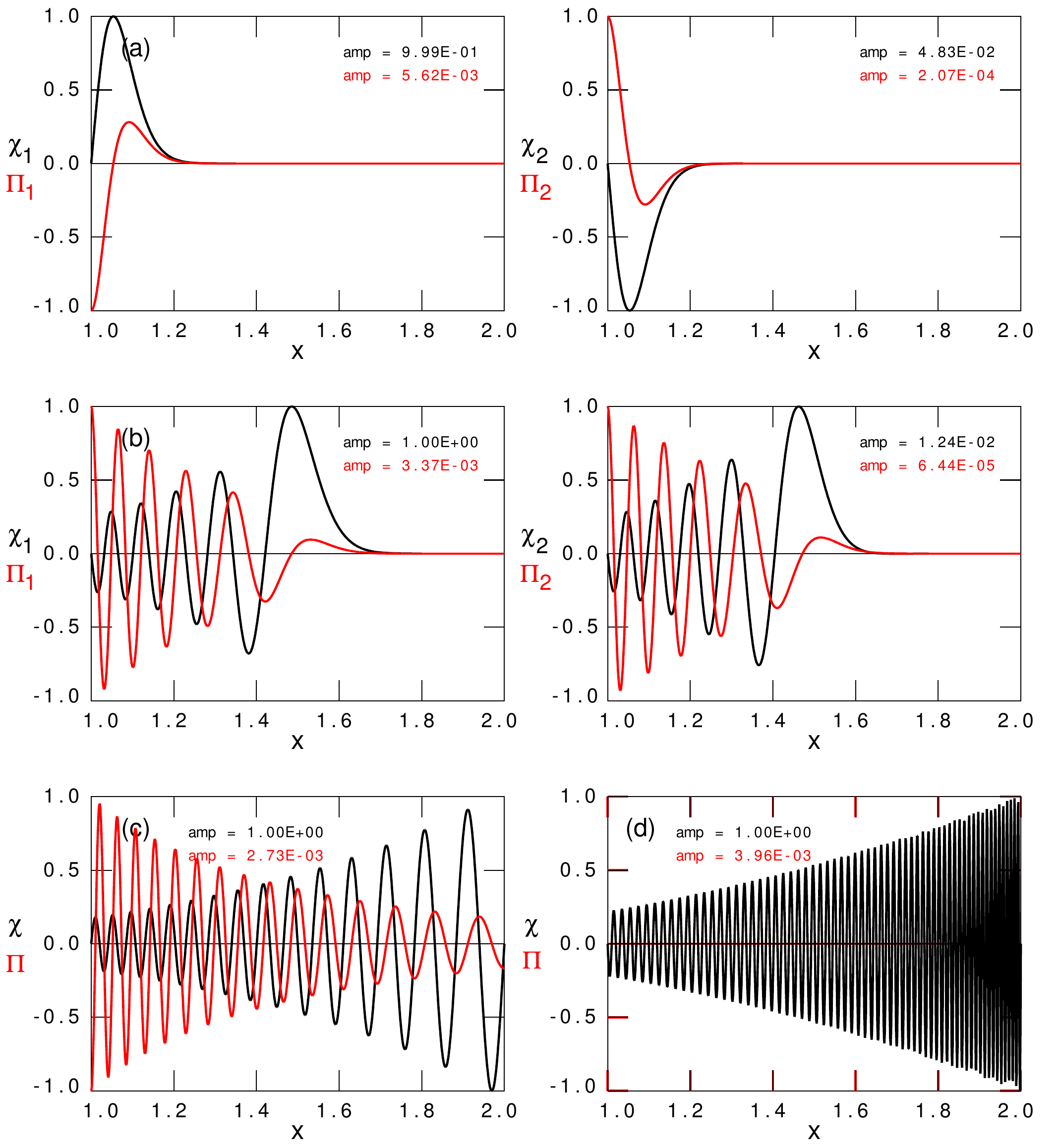}
\end{center}}
\caption{Eigenfunctions corresponding to the Spectral Web depicted in Fig.~\F{3}. 
Magneto-Rotational Instabilities: (a)~$n = 1\;$ ($\sigma = -8.288\times  10^{-3}\hs$, 
$\nu = 0.6208$); (b)~$n = 10\;$ ($\sigma = -5.249 \times 10^{-3}\hs$, $\nu = 0.3810$). 
Stable modes: (c)~$n = 30\;$ ($\sigma = 4.502 \times 10^{-2}\hs$), (d)~$n = 130\;$ 
($\sigma = 0.4609\hs$). In the latter frame, the function $\Pi$ has been omitted for 
clarity.}{\Fn{4}}
\end{figure}

The numerically computed MRI part of the Spectral Web is illustrated in Fig.~\F{3} for a 
representative configuration, whereas the corresponding eigenfunctions are shown in 
Fig.~\F{4} (these figures are for incompressible MRIs, the compressible counterparts are 
presented in \Ron{GKP2019}). This Spectral Web, consisting of the solution path (in 
red) and the conjugate path (in blue), clearly demonstrates the merits of this method: The 
29 unstable eigenvalues are clearly aligned along the solution path that deviates from the 
imaginary axis (which would be the solution path of the analytical solutions \E{25}), 
whereas each of them is located on a kind of pancake of the conjugate path. The epicyclic 
modes would be situated to the left of the $\{-\omega_{\rm A}\}$ continuum and to the 
right of the $\{+\omega_{\rm A}\}$ continuum in the top frame of Fig.~\F{3}. They are 
of no further interest here. On the other hand, only the most global modes ($n \le 29$) of 
the MRI sub-spectrum are unstable whereas the two sequences clustering towards the 
continua are stable (for simplicity, we will permit the misnomer `MRIs' also for these 
modes). 

The deviations shown in Fig.~\F{3} of the numerical solutions of the complex 
ODEs~\E{12}, or rather Eq.~\E{A7}, from the real analytic solutions \E{25} have also 
been found from an extended WKB method with complex frequencies, with proper 
matching at the turning points~\R{BSKG05}. This actually provides excellent 
approximations of the growth rates and eigenfunctions of the MRIs, but it fails to 
properly describe the approach to the continuum singularities. In general, for frequencies 
close to the continua, the two singularity coefficients $\widetilde{\omega}^2 - 
\omega_{\rm A}^2$ dominate the behavior of the solutions of Eq.~\E{23}. For the $m = 
0$ modes, only the static parts $\{\pm\omega_{\rm A}(r)\}$ of the continua enter. For 
the accretion disk equilibria of Sec.~\S{IIA}, the function $\omega_{\rm A}(r)$ is 
decreasing so that the stable modes cluster towards the edges $\pm\omega_{\rm A2}$ of 
the continua at $r_2$. Expansion for the modes localized about $r_2$, i.e.\ $\omega_{\rm 
A}(r) \approx \omega_{\rm A2} + \omega_{\rm A}' (r - r_2)$, transforms Eq.~\E{23} 
for $\widetilde{\omega} = \omega$ into the dominant differential equation
\BEQ \frac{d}{d\tilde{x}} \Big[\hs(\tilde{x} + \lambda) \hs\frac{d\chi}{d\tilde{x}} 
\hs\Big] 
+ \frac{p^2}{\tilde{x} +\lambda} \hs \chi = 0 \,, \qquad \tilde{x} \equiv \frac{r_2 - 
r}{\delta} \,,\quad \lambda \equiv \frac{\omega -\omega_{\rm A2}}{\delta 
\hs\omega_{\rm A}'} \,,\quad p \equiv\Big|\frac{k \hs\Omega}{\omega_{\rm 
A}'}\Big|_{r_2} \,,\En{26}\EEQ
having the solutions
\BEQ \chi = C_1 \sin[p \ln(x + \lambda)]  + C_2 \cos[p \ln(x + \lambda)]\,, \En{27}\EEQ
Satisfaction of the BCs at $r_1$ and $r_2$ yields the desired cluster spectra:
\BEQ \lambda \approx \exp(-n\pi/p) \quad\Rightarrow\quad \omega \approx \pm 
[\hs\omega_{\rm A2} + \delta \hs\omega_{\rm A}' \exp(-n\pi/p)\hs] \,.\En{28}\EEQ
Of course, these solutions only provide the asymptotic behavior for $n \gg1$ and small 
$\delta$, i.e.\ for the stable modes. For the global unstable modes (lower $n$), in 
particular the MRIs, the approximate WKB expressions~\E{25} are appropriate. Since 
the pertinent unstable part of the MRI spectrum is located far away from the continua, 
these instabilities are not affected by the details of the clustering at the continua. \IT{This 
is no longer the case for the non-axisymmetric modes!} For those, we will have to delve 
much deeper into the subtleties of the continua. This will turn out to have major physical 
consequences.

\section{Non-axisymmetric Super-Alfv\'enic Rotational Instability}{\Sn{IV}}

\subsection{Spectral Webs of the Super-Alfv\'enic Rotational 
Instabilities}{\Sn{IVA}}

Turning now to the non-axisymmetric instabilities, it would appear logical to present 
those in the same manner as the axisymmetric MRIs of the previous section, starting from 
the accretion disk ODE~\E{23} and then to introduce some further approximations to 
derive an approximate dispersion equation. {\em That approach is bound to fail!} A local 
WKB analysis cannot be applied to these modes because the continuous spectra 
$\{\Omega_{\rm A}^+(r)\}$ and $\{\Omega_{\rm A}^-(r)\}$ now are never far away 
from the actual unstable eigenvalues, but they enter in a crucial way. Hence, these 
singularities need to be incorporated in any analytical theory of the non-axisymmetric 
instabilities of accretion disks.  This is done in the next sub-section, Sec.~\S{IVB}, on 
the approximate asymptotic analysis of these instabilities.

\subsubsection{Central role of the Doppler frequency range}{\Sn{IVA1}}

In the present sub-section, we derive the Spectral Webs of the non-axisymmetric 
instabilities by numerical means, as described in Sec.~\S{IID}. Here, the search can be 
restricted to a strip in the complex $\omega$-plane centered about the real Doppler 
frequencies. As shown in \Ron{GBHK2004b}, the real Doppler frequency range
\BEQ \Omega_0 = m \Omega(r) \equiv m v_\theta(r)/r \En{29}\EEQ
is not a continuous spectrum in MHD, but it does play a very central role in the 
description of the non-axisymmetric instabilities. That range is necessarily located in 
between the forward and the backward Alfv\'en continua, since
\BEQ \Omega_{\rm A}^\pm \equiv \Omega_0(r) \pm \omega_{\rm A}(r) \,,\quad 
\omega_{\rm A} \equiv m B_\theta(r)/r + k B_z(r) \,.\En{30}\EEQ
We will show that, for the non-axisymmetric instabilities, the real frequency parts 
$\sigma$ of the eigenvalues necessarily lie inside the Doppler range $\{\Omega_0(r)\}$, 
so that $\sigma - \Omega_0(r)$ is much smaller than any of the frequencies 
$\Omega_0(r)$, whereas their growth rates $\nu$ are also of this smaller order of 
magnitude. Furthermore, the static Alfv\'en frequencies $\omega_{\rm A}$ are generally 
much smaller than the Doppler frequencies $\Omega_0$ so that the three frequency 
ranges $\{\Omega_{\rm A}^+(r)\}$, $\{\Omega_0(r)\}$, $\{\Omega_{\rm A}^-(r)\}$ 
overlap, at least partly. The closeness of the complex eigenvalues to all three of these real 
frequency ranges implies that the analysis of the non-axisymmetric instabilities becomes 
a very intricate problem with \IT{three near-singularities.} This particular terminology 
indicates that the two continuum singularities are not actual (which would require $\nu = 
0$) but just near because $\nu$ is small, whereas singularity of the Doppler frequency 
would require in addition that $\omega_{\rm A} = 0$, whereas $\omega_{\rm A}$ is just 
small compared to $\Omega_0$. Nevertheless, these three near-singularities determine 
the behavior of the solutions completely, as we will see. 

$[\,$In this discussion, we do not separately mention the slow magneto-sonic continua 
$\{\Omega_{\rm S}^\pm(r)\}$ anymore since they virtually coincide with the Alfv\'en 
continua, and exactly coalesce in the limit $\beta \rightarrow \infty$. That approximation 
is extremely well satisfied for the present modes, as well as for the MRIs, since $\beta \gg 
1$ for the equilibria. Of course, in the numerical analysis that approximation is not 
necessary, but it will be made anyway to simplify the presentation. The changes of the 
eigenvalues and the Spectral Web by compressibility are small enough to be neglected, 
unless explicitly mentioned.$\,]$

To prove that the real frequencies of the non-axisymmetric instabilities lie in the Doppler 
frequency range, we recall Eq.~\E{17} which demands that the solution path of these 
modes is given by the condition that the solution-averaged Doppler--Coriolis shifted real 
part of the frequency vanishes, $\sigma - \overline{V} = 0$. This condition coincides 
with the solution path condition~\E{18}(a), which is proved in \Ron{Goed2018a}. 
Clearly, without the contribution of the Coriolis term (the second term of the 
expression~\E{17}  involving $\xi_\theta$), this demands that the eigenfunction should 
have at least two parts with a different sign of the expression $\sigma - \Omega_0(r)$, 
i.e.\ $\sigma$ should be inside the range $\{\Omega_0(r)\}$. The Coriolis term would 
spoil this argument but, for the non-axisymmetric instabilities, it is small enough to be 
negligible compared to the Doppler term. This follows by applying the ordering for large 
mode numbers,
\BEQ \Omega_0 \sim m \gg \omega_{\rm A} \sim k \epsilon \,.\En{31}\EEQ
In this ordering, the order of magnitude of the Doppler term is ${\cal O}(m)$, whereas an 
estimate from the explicit expression for $\xi_\theta$ shows that the order of magnitude 
of the Coriolis term is only~${\cal O}(k^{-1})$, i.e.\ a factor $(m k)^{-1} \ll 1$, QED. 

For the MRIs, in contrast to the present non-axisymmetric modes, the above argument 
does not apply because the Doppler frequencies vanish identically ($m = 0$ by 
definition). With that, the rotation profile $\Omega(r)$ does not influence the eigenvalues 
directly but only indirectly through the near-Keplerian equilibrium. $[\,$This is actually 
one of the few cases where compressibility makes a difference. The solution path of the 
Spectral Web shown here in Fig.~\F{3}, for incompressible MRIs, deviates from the 
vertical imaginary axis by a factor of~$4$ compared to the much smaller deviation for 
the compressible MRIs that was shown in Fig.~13.17 of \Ron{GKP2019}.$\,]$

\begin{figure}[ht]
\FIG{\begin{center} 
\includegraphics*[height=14.5cm]{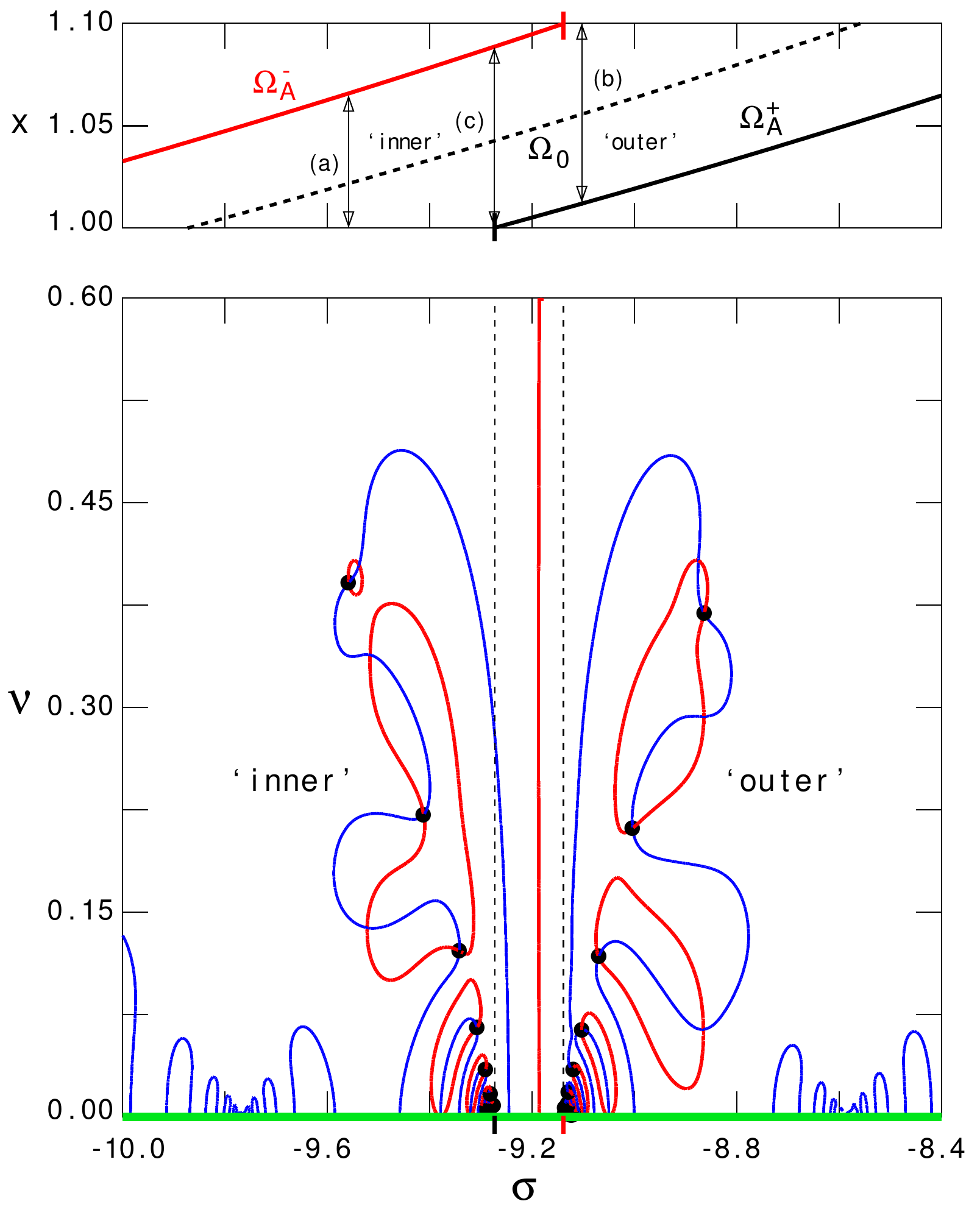}
\end{center}}
\caption{Spectral Web of the counter-rotating incompressible Super Alfv\'enic Rotational 
Instabilities for an accretion disk with  parameters $\epsilon \equiv B_1 = 0.0141$, 
$\mu_1 \equiv B_{\theta 1}/B_{z1}  =1$, $\beta \equiv 2p_1/B_1^2 = 100$, $\delta 
\equiv r_2/r_1 - 1 = 0.1$, and mode numbers $m = -10$, $k = 70$. The radial profiles of 
the overlapping continua are shown above the main figure. Their real frequencies are 
indicated in green in the main frame. The `inner' modes cluster towards the tip of the 
forward Alfv\'en continuum~$\{\Omega_{\rm A}^+\}$ at $r = r_1$, the `outer' modes 
cluster towards the tip of the backward Alfv\'en continuum~$\{\Omega_{\rm A}^-\}$ at 
$r = r_2$.}{\Fn{5}}
\end{figure}

Consequently, the non-axisymmetric modes are clearly distinguished from the MRIs by 
the dominance of the Doppler frequency range, with close `sidebands' of the backward 
and forward Alfv\'en frequency ranges, as expressed by Eqs.~\E{30} and \E{31}. This 
triple of near-singularities is the determining feature of the non-axisymmetric 
instabilities. The dominance of the Doppler frequency over the static Alfv\'en frequency, 
$|\Omega_0| \gg |\omega_{\rm A}|$, makes it appropriate to call these modes 
\IT{Super-Alfv\'enic Rotational Instabilities (SARIs).} 

Since the Doppler frequency is related to the rotation frequency through $\Omega_0 = m 
\hs\Omega$ these instabilities come in two flavors, viz.\ {\em counter-rotating SARIs} 
(for $m < 0$) with negative frequencies ($\sigma < 0$), and {\em co-rotating SARIs} 
(for $m > 0$) with positive frequencies ($\sigma > 0$). Furthermore, since 
$\Omega_0^\prime = m \hs\Omega^\prime$, the radial frequency profiles of the forward 
and backward Alfv\'en singularities are different for the two flavors, viz.\ increasing for 
the first kind so that $\Omega_{\rm A}^+(r_2) < \Omega_{\rm A}^-(r_1)$, and 
decreasing for the second kind so that $\Omega_{\rm A}^+(r_2) > \Omega_{\rm A}^-
(r_1)$. This is illustrated in the upper frame of Fig.~\F{5}, and in the upper frame of 
Fig.~\F{8} below.

Note that, whereas the unstable wave packages move at super-Alfv\'enic speeds, either 
counter- or co-rotating with the disk, the disk itself also rotates at super-Alfv\'enic speeds. 
The two speeds are well to be distinguished. In nonlinear context, the super-Alfv\'enic 
rotation of the disk will admit shocks, which demand symmetry breaking of the 
dynamics. This could be initiated by the non-axisymmetric SARIs, but not by the 
axisymmetric MRIs. The complex nonlinear interplay of SARIs with rotation of the disk 
is a subject that deserves separate investigation, clearly far beyond the scope of the 
present paper.

\subsubsection{Counter-rotating SARIs}{\Sn{IVA2}}

A representative Spectral Web of counter-rotating SARIs is shown in Fig.~\F{5}, with 
the radial distributions of the three near-singularities in the top frame. As demanded by 
theory, the eigenvalues (black dots) are located on the intersections of the solution path 
(in red) and the conjugate path (in blue). Along the real axis the forward Alfv\'en 
continuum $\{\Omega_{\rm A}^+(r)\}$ and the backward Alfv\'en continuum 
$\{\Omega_{\rm A}^-(r)\}$ are depicted in green. They overlap in a range bounded by 
the two extrema $\Omega_{\rm A}^+(r_1)$ and $\Omega_{\rm A}^-(r_2)$, which turn 
out to play an essential role in the spectral structure. These extrema are indicated by the 
black and red dashes, corresponding to the two thin dashed lines in the bottom frame. 
There are two infinite sequences of unstable eigenvalues, where one sequence (labeled 
`inner') clusters towards $\sigma = \Omega_{\rm A}^+(r_1)$, $\nu = 0$,  and the other 
one (labeled `outer') clusters towards $\sigma = \Omega_{\rm A}^-(r_2)$, $\nu = 0$. The 
equilibrium is the same as exploited for the MRIs of Figs.~\F{3} and \F{4}, except for 
the smaller radial range parametrized by~$\delta$. Nevertheless, the growth rates are 
comparable, whereas only the four most global MRIs would be unstable for $\delta = 
0.1$. In contrast, \IT{all SARIs of the two infinite sequences are unstable, down to 
approaching the cluster points at $\nu = 0$.} 

Whereas the positions of the eigenvalues are independent of the method of solution, the 
convoluted structure of the Spectral Web itself sensitively depends on the way the 
accretion disk ODE~\E{23} is solved. For the MRIs of Fig.~\F{3}, the integration was 
taken from the right since the left scheme runs into numerical instability caused by the 
low density on the outside. For the SARIs of Fig.~\F{5}, the mixed scheme~\E{19} of 
left and right solutions had to be used, with matching in the middle of the interval. This 
was done since the left integration scheme runs into numerical instability due to the 
proximity of the singularity $\sigma = \Omega_{\rm A}^-$ so that only the right branch 
labeled `outer' would be obtained. Vice versa for the right integration scheme, with 
problems at $\sigma = \Omega_{\rm A}^+$ so that only the left branch labeled `inner' 
would be obtained. With the mixed integration scheme, the matching point $r_{\rm 
mix}$ can be chosen anywhere, as long as it is not too close to the boundaries. For the 
particular case $r_{\rm mix} = 1.05$, used for this Spectral Web, the solution and 
conjugate paths are split into closed loops and loops ending on the real axis. It was 
proven in \Ron{Goed2018a} that, along each of these constituent pieces of the Spectral 
Web, the alternator $R \equiv (\doublel\Pi\doubler/\chi)_{r_{\rm mix}}$ is a monotonic 
function of arc length, with tangent-like branches. This permits labeling of the different 
modes along these pieces, reminiscent of the labeling by means of the number of radial 
nodes of the eigenfunctions for static equilibria, which still may be used to label the 
eigenfunctions of the MRIs. It is clear though that the mentioned split of the constituent 
paths of the Spectral Web inhibits a similar straightforward classification of the different 
SARIs: there are simply too many branching possibilities depending on the details of the 
radial profile of the Doppler frequency. However, an asymptotic numbering scheme of 
the different \{eigenfunction--eigenvalue\} pairs is introduced in the approximate 
asymptotic analysis of the next section, Sec.~\S{IVB}. We will return to this issue at the 
end of that section.

\subsubsection{`Virtual walls'}{\Sn{IVA3}}

Three eigenfunctions of counter-rotating SARIs, with radial intervals corresponding to 
the labels (a), (b) and (c) in the top frame of Fig.~\F{5}, are shown in Fig.~\F{6}.  
Anticipating the asymptotic numbering scheme, the orders of the functions in a particular 
sequence are given in the caption of this figure.  The eigenfunctions of Figs.~\F{6}(a) 
and (c) are taken from the `inner' sequence, whereas the eigenfunction of Fig.~\F{6}(b) is 
taken from the `outer' sequence. The nomenclature `inner' versus `outer' now also 
becomes clear. The radial eigenfunctions of the `inner' modes, clustering (for $|\nu| 
\downarrow 0$) to the extremum $\Omega_{\rm A}^+(r_1)$ of the forward Alfv\'en 
continuum, are intersected by the backward continuum $\{\Omega_{\rm A}^-(r)\}$ at 
some radius $r_2^* \le r_2$. In the range $r_2^* \le r \le r_2$, the amplitudes of the 
`inner' modes are effectively `cut-off' by the singularity so that the radius $r = r_2^*$ 
functions as a kind of `virtual wall'. Similarly, the eigenfunctions of the  `outer' modes are 
intersected by the forward continuum $\{\Omega_{\rm A}^+(r)\}$ at some radius $r_1^* 
\ge r_1$ and they cluster to the extremum $\Omega_{\rm A}^-(r_2)$ of the backward 
Alfv\'en continuum. In the range $r_1 \le r \le r_1^*$, the amplitudes of the `outer' modes 
are again `cut-off' by the singularity so that the radius $r = r_1^*$ also functions as a 
kind of `virtual wall'. The `cut-off' is not absolute, as illustrated in the blown up parts of 
the eigenfunctions shown in Figs.~\F{7}(a),(b1),(c1), but their amplitudes are extremely 
small in the `cut-off' region and, of course, vanish exactly at the prescribed boundary as 
demanded by the boundary condition. Hence, it is clear that, asymptotically (for small 
$\nu$), the eigenfunctions and eigenvalues would hardly change if the outer boundary at 
$r = r_2$ were moved inward to the point $r = r_2^*$ for the `inner' modes, or if the 
inner boundary were moved outward to the point $r = r_1^*$ for the `outer' modes. We 
will exploit this fact in the approximate analysis of Sec.~\S{IVB}.

\begin{figure}[ht]
\FIG{\begin{center} 
\includegraphics*[height=14.6cm]{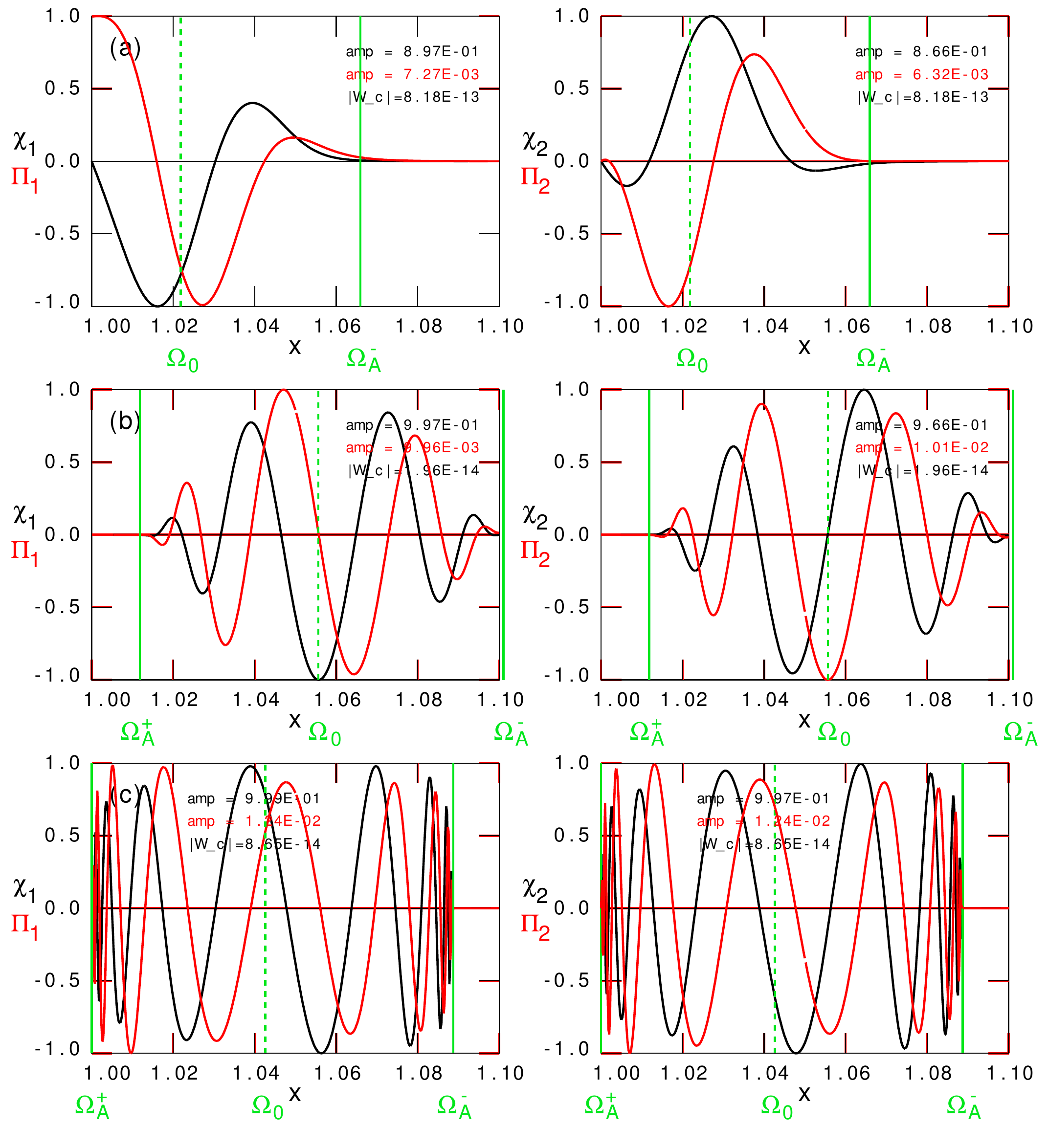}
\end{center}}
\caption{Three eigenfunctions of the Super-Alfv\'enic Rotational Instability 
corresponding to Fig.~\F{5}: (a)~$1^{\rm st}$ `inner' mode, $\sigma = -9.5587\hs$, $\nu 
= 0.39118$; (b)~$4^{\rm th}$ `outer' mode, $\sigma = -9.1031\hs$, $\nu = 0.06368$; and 
(c)~$10^{\rm th}$ `inner' mode, $\sigma = -9.2741\hs$, $\nu = 0.001316$. Note that the 
near-singularities $\Omega_{\rm A}^+$ for the `inner' modes (a) and (c), or 
$\Omega_{\rm A}^-$ for the `outer' mode (b), are located outside the physical interval. 
}{\Fn{6}}
\end{figure}

\begin{figure}[ht]
\FIG{\begin{center} 
\includegraphics*[height=15.5cm]{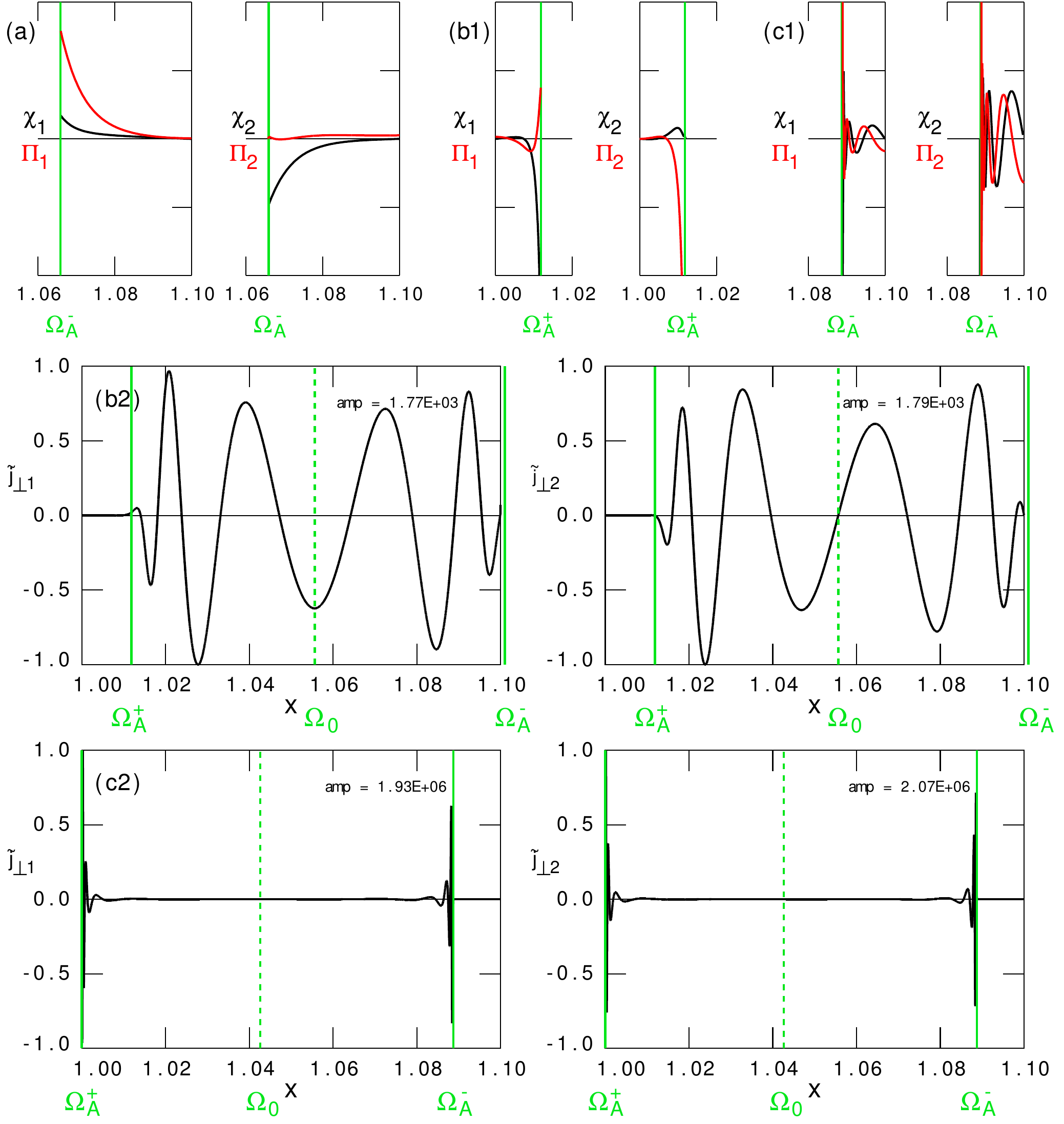}
\end{center}}
\caption{Blown up parts of the three eigenfunctions of Fig.~\F{6} in the cut-off regions 
with the approach beyond the near-singularity towards the actual outer boundary (for the 
`inner' modes), or the inner boundary (for the `outer' modes); to visualize this, the 
amplitudes were multiplied with the factors (a)~$30$, (b1)~$2 \times 10^3$, 
(c1)~$10^6$. The perpendicular current density distribution corresponding to the 
eigenfunctions of Fig.~\F{6}(b) and (c) is shown in the frames (b2) and (c2).}{\Fn{7}}
\end{figure}

Actually, the `cut-off' is only absolute in the limit $|\nu| \downarrow 0$, when the 
singularities produce a split in \IT{independent subintervals}, very much like the ones 
introduced in the seminal paper by~\Ron{New60} on the stability of the diffuse linear 
pinch. The physics of this phenomenon is that the singularity facilitates the induction of 
skin currents at the position of the singularity, which effectively counteract any motion 
across that radius. For finite $\nu$, the skin currents are spread out in the `cut-off' region, 
but they are still quite effective, as shown by the eigenfunction details of 
Figs.~\F{7}(a),(b1),(c1). The same holds for the influence of finite conductivity of the 
plasma when the induced currents become diffuse. Nevertheless, these `virtual wall' 
effects quite well explain, e.g., the localization of internal kink instabilities in tokamaks, 
replacing the actual wall by the virtual one at the singularity. For accretion disks, this 
effect is more than welcome since the introduction of rigid walls in the numerical 
calculation of MHD instabilities is anyway a rather questionable procedure which always 
needs some form of justification. On the other hand, the concept of a `virtual wall' can be 
justified easily since it is produced by a genuine physical effect, viz.\ the large 
conductivity of magnetized plasmas. We will see in Section~\S{V} how two such `virtual 
walls' (at $r = r_1^*$ and $r = r_2^*$) effectively produce the localization of  
`quasi-continuum SARIs' at rather arbitrary locations in the disk.  

In Section~\S{VC} on the Alfv\'en wave dynamics of the SARIs, we will discuss the 
relationship between the induction of the perpendicular current density perturbation 
$\tilde{j}_\perp$ and the appearance of `virtual walls' more extensively by means of the 
explicit expression for $\tilde{j}_\perp$ derived in Appendix~\S{A3}. Anticipating that 
analysis, the distribution of this current component for the two eigenfunctions of 
Figs.~\F{6}(b) and (c) is presented in Figs.~\F{7}(b2) and (c2). In fact, a sharp skin 
current-like distribution is evident in Fig.~\F{7}(c2) for the 10th `inner' mode with the 
very small value of $\nu$, very much like that following from the stability analysis 
by~\Ron{New60}, which is still at the basis of much present tokamak stability theory. 
However, further away from marginal stability, the broad oscillatory current distribution 
of Fig.~\F{7}(c2) for the 4th `outer' mode shows no sign any more of a skin current. This 
points to an important difference between tokamaks and accretion disks that is further 
discussed in  Section~\S{VC}.

In summary, {\em the SARIs are unstable waves rotating with super-Alfv\'enic velocities 
at the Doppler frequency, evaluated at approximately the middle of the 
\underline{reduced} radial interval.} The contributions of the forward and backward 
Alfv\'en waves, though localized at  the end points of the reduced interval, are averaged 
such that the wave packages as a whole move in phase with that particular value of the 
Doppler frequency. 

\subsubsection{Co-rotating SARIs}{\Sn{IVA4}}

The motion in phase with the Doppler frequency is either opposite to the direction of the 
rotation of the disk, for the counter-rotating SARIs just discussed, or in the same 
direction, for the co-rotating SARIs. A representative Spectral Web of the latter modes is 
shown in Fig.~\F{8}, again with the radial distributions of the three near-singularities in 
the top frame. The continua now decrease as a function of $\sigma$ so that the positions 
of the `inner' and `outer' modes in the Spectral Web are reversed with respect to that of 
Fig.~\F{5}. With that in mind, there is no need to show the eigenfunctions of the 
co-rotating SARIs since they are completely analogous to the counter-rotating ones 
shown in Fig.~\F{6}.

\begin{figure}[ht]
\FIG{\begin{center} 
\includegraphics*[height=13.7cm]{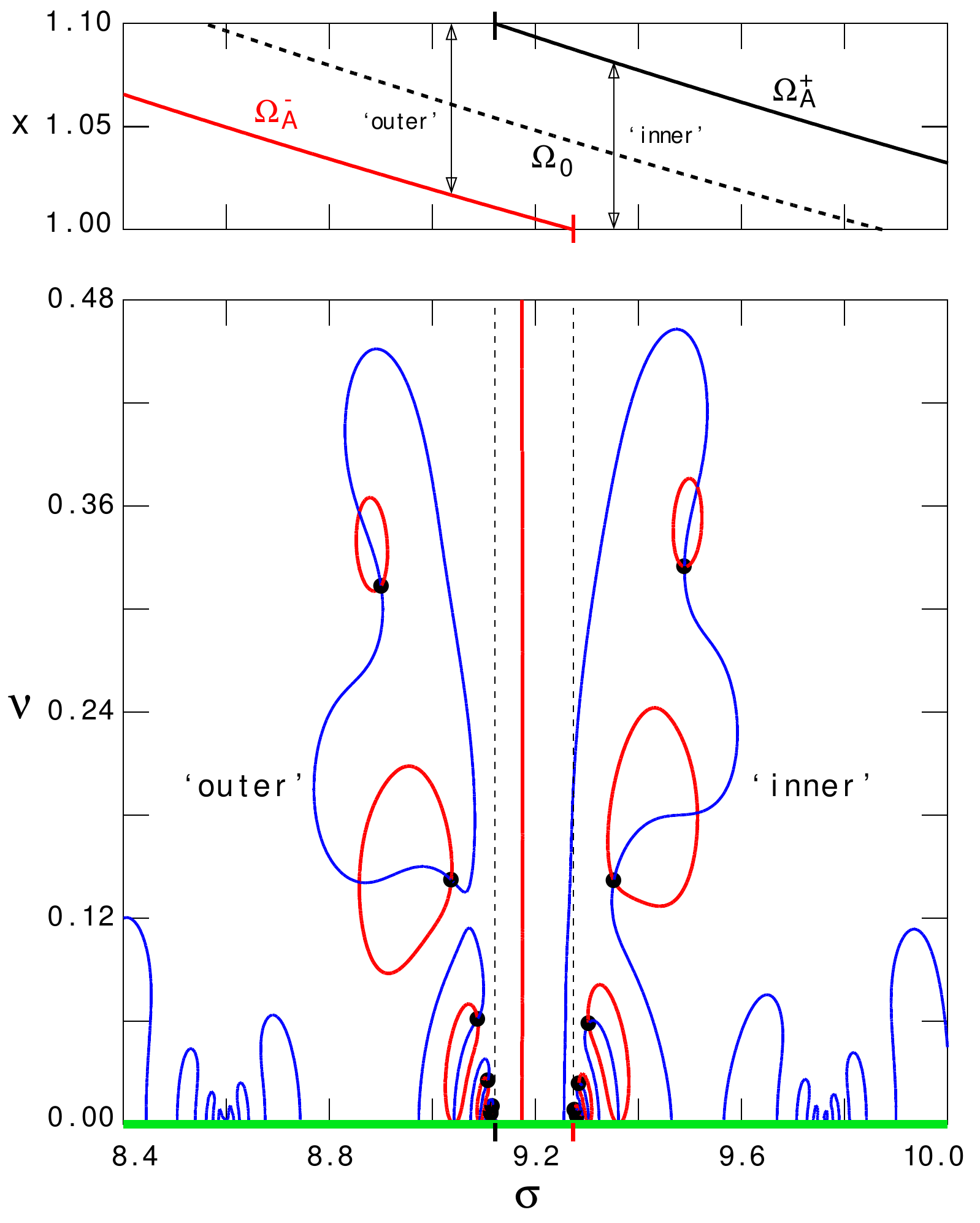}
\end{center}}
\caption{Spectral Web of the co-rotating Super-Alfv\'enic Rotational Instabilities for an 
accretion disk with parameters $\epsilon \equiv B_1 = 0.0141$, $\mu_1 \equiv B_{\theta 
1}/B_{z1}  = 1$, $\beta \equiv 2p_1/B_1^2 = 100$, $\delta \equiv r_2/r_1 - 1 = 0.1$, as 
in Fig.~\F{5}, but mode numbers $m = 10$, $k = 50$. The radial profiles of the 
overlapping continua are shown above the main figure. Their real frequencies are 
indicated by the green line in the main frame. The `outer' modes cluster towards the tip of 
the forward Alfv\'en continuum $\{\Omega_{\rm A}^+\}$ at $r = r_2$, the `inner' modes 
cluster towards the tip of the backward Alfv\'en continuum $\{\Omega_{\rm A}^-\}$ at 
$r = r_1$.}{\Fn{8}}
\end{figure}

The mode numbers $m$ and $k$ for the counter-rotating SARIs of Fig.~\F{5} and for 
the co-rotating SARIs of Fig.~\F{8} are chosen such that the values of the static Alfv\'en 
frequency~$\omega_{\rm A}$ are approximately the same for the two cases. 
Consequently, the maximum growth rates are of the same order of magnitude, in 
agreement with the over-estimated limit on the maximum growth rate given by the 
expression~\E{A12} of Appendix~\S{A2}. [$\hs$Incidentally, that expression is also 
valid for the MRIs of Fig.~\F{3}, where the maximum growth rate $0.6208$ is much 
closer to the estimated value $0.6988$ because the neglected first integral of 
Eq.~\E{A11} is quite small in that case, due to the small average amplitude of $|\chi|$ 
shown in Fig~\F{4}(a).$\hs$]

Except for the estimate of the maximum growth rate, given in Appendix~\S{A2}, not 
much explicit analysis can be presented on the SARIs. The four distinct cases of the 
asymptotic approach of the eigenvalues along straight lines to the two cluster points, viz.\ 
to $\Omega_{\rm A}^+(r_1)$ and $\Omega_{\rm A}^-(r_2)$ for the counter-rotating 
SARIs of Fig.~\F{5} and to $\Omega_{\rm A}^-(r_1)$ and $\Omega_{\rm A}^+(r_2)$ 
for the co-rotating SARIs of Fig.~\F{8}, is completely beyond the kind of local analysis 
usually exploited for the MRIs. This requires the systematic incorporation of the behavior 
at the cluster points, as given in the next section. We will take the example of the `inner' 
co-rotating SARIs to illustrate that analysis.

\smalltext{Shortcut \#1}{In summary, we now have showed numerically (using the 
Spectral Web technique) that non-axisymmetric SARIs are quite distinct from 
axisymmetric MRIs: a local WKB analysis does not apply since overlapping 
Doppler-shifted forward and backward continua interplay.  SARIs come in `outer' and 
`inner', co- and counter-rotating flavors. They self-create virtual walls of skin currents, 
becoming more evident for the higher modes of the infinite sequences of unstable modes. 
This makes them insensitive to one of our artificial boundaries, and they rotate 
super-Alfv\'enically at the Doppler frequency. The next section, which gives a complete 
analytic treatment of the SARIs, could be skipped on first reading, to continue with our 
search for quasi-modes that are truly local wave packages described in Section~\S{V}.}

\subsection{Approximate asymptotic analysis of the SARIs}{\Sn{IVB}}

For an analytic description of the SARIs, we again try to exploit a local approximation 
where radial variations of $\chi(r)$ occur over distances smaller than the scale length of 
equilibrium variations and consistency implies the assumption of a thin radial shell, 
$\delta \ll 1$. We also exploit the smallness of the deviations of the eigenvalues from the 
rotation frequencies, so that $|\widetilde{\omega}|^2 \sim \omega_{\rm A}^2 \sim 
\epsilon^2 \ll \kappa_{\rm e}^2 \sim 1$. Our basic differential equation~\E{23} then 
simplifies to
\BEQ \frac{d}{dr} \bigg[\hs(\widetilde{\omega}^2 - \omega_{\rm A}^2) 
\hs\frac{d\chi}{dr} \hs\bigg] + k^2 \bigg[\hs\kappa_{\rm e}^2 
- (\widetilde{\omega}^2 - \omega_{\rm A}^2) + \frac{4 \Omega^2 \omega_{\rm 
A}^2}{\widetilde{\omega}^2 - \omega_{\rm A}^2} \hs\bigg] \, \chi = 0 \,. \En{32}\EEQ
where the coefficients are assumed constant, except for the first and the last one, which 
produce the singularities, whereas the term $- (\widetilde{\omega}^2 - \omega_{\rm 
A}^2)$ will later be assumed constant with an average magnitude derived below. From 
the quadratic form corresponding to this ODE, estimates of the eigenvalues may be 
obtained (see Appendix~\S{A2}).

Equation~\E{32} still yields the correct MRI dispersion equation~\E{24} of the 
axisymmetric modes (where $\widetilde{\omega} = \omega$) in the WKB 
approximation, but we now focus on the expansion about the continuum frequencies 
$\Omega_{\rm A}^+(r)$ and $\Omega_{\rm A}^-(r)$ when $\widetilde{\omega} \ne 
\omega$. The basic differential equation~\E{32} is now essentially complex, hence of 
fourth order, where the companion system~\E{12}, or rather \E{A7}, determines the real 
and imaginary parts $\chi_1$, $\Pi_1$ and $\chi_2$, $\Pi_2$ of the eigenfunctions. In 
this case, the continua usually overlap and the unstable discrete modes have eigenvalues 
that are close to those continua, as shown by the Spectral Webs of Figs.~\F{5} and \F{8}, 
and  the eigenfunctions of Figs.~\F{6} and \F{7}. These pictures show that the radial 
variation of the singularity terms dictates the behavior of the SARIs, where {\em it is 
essential that both singularities, $\widetilde{\omega} = \omega_{\rm A}$ and 
$\widetilde{\omega} = - \omega_{\rm A}$, contribute.}

\subsubsection{Reduction to the Legendre equation}{\Sn{IVB1}}

For definiteness, we restrict the analysis to the sequence of `inner' co-rotating SARIs, 
where the various geometrical quantities are illustrated in Fig.~\F{9}. For a given real 
frequency component~$\sigma$ of the complex eigenvalue $\omega$, indicated by the 
vertical line, the intersection of the continuum function $\Omega_{\rm A}^-(r)$ at 
$r_1^*$ is outside the physical interval, whereas the continuum function $\Omega_{\rm 
A}^+(r)$ is intersected at $r_2^*$. Hence, the relevant radial interval is reduced to $(r_1, 
r_2^*)$. We consider the sequence of modes tending to the cluster point, $\sigma 
\rightarrow \Omega_{\rm A}^-(r_1)$, so that the solutions are effectively  determined by 
the two near-singularities $\Omega_{\rm A}^-(r_1)$ and $\Omega_{\rm A}^+(r_2^*)$. 
The left boundary condition is unchanged, $\chi(r_1) = 0$, whereas the right one should 
be replaced by $\chi(r_2^*) = 0$ to get the proper boundary value problem for the 
reduced interval. We introduce a unit radial variable for that interval:
\BEQ \tilde{x} \equiv (r - r_1)/\delta^* \,, \qquad \delta^* \equiv r_2^* - r_1 \,, 
\En{33}\EEQ
where $\delta^*$ is the dimensionless size  of the interval (recall that distances have been 
made dimensionless by dividing through $r_1$), so that $\tilde{x} = 0$ corresponds to $r 
= r_1$ and $\tilde{x} = 1$ corresponds to $r = r_2^*$. From Fig.~\F{9} it is clear that 
the eigenvalue problem would involve a variable interval size~$\delta^*$ since it would 
depend on the horizontal distance~$\sigma - \Omega_{\rm A}^-(r_1)$ to the cluster 
point. To avoid this complication, $r_2^*$ and $\delta^*$ are fixed at the asymptotic 
value corresponding to $\sigma = \Omega_{\rm A}^-(r_1) = \Omega_{\rm 
A}^+(r_2^*)$. [$\hs$Analogously for the `outer' modes, with interval $(r_1^*, r_2)$ and 
clustering towards $\Omega_{\rm A}^+(r_2)$.\hs] 

From the definition~\E{16} of the Alfv\'en factor $\tilde{A}$, the singular expression 
may be expanded about the end points $\tilde{x} = 0$ ($r = r_1$) and $\tilde{x} = 1$ ($r 
= r_2^*$) of the interval, i.e.
\BEQAR \hor{-10}\hbox{-- for $r \approx r_1$:} \qquad \widetilde{\omega}^2 - 
\omega_{\rm A}^2 &\approx& \big[\hs\omega - \Omega_{\rm A}^-(r)\hs\big] 
\big[\hs\Omega_{\rm A}^-(r_1) - \Omega_{\rm A}^+(r_1)\hs\big] \non\\
&\approx& 2 \delta^* \omega_{\rm A}(r_1) \hs\Omega_{\rm A}^{-\prime}(r_1) 
(\tilde{x} - \lambda) \,, \hor{3}\quad\hbox{where}\quad \lambda \equiv \frac{\omega - 
\Omega_{\rm A}^-(r_1)}{\delta^* \hs\Omega_{\rm A}^{-\prime}(r_1)} 
\,,\En{34}\\[2mm]
\hor{-10}\hbox{-- for $r \approx r_2^*$:} \qquad \widetilde{\omega}^2 - \omega_{\rm 
A}^2 &\approx& \big[\hs\omega - \Omega_{\rm A}^+( r)\hs\big] \big[\hs\Omega_{\rm 
A}^+(r_2^*) - \Omega_{\rm A}^-(r_2^*)\hs\big] \non\\
&\approx& - 2 \delta^* \omega_{\rm A}(r_2^*) \hs\Omega_{\rm A}^{+\prime}(r_2^*) 
(\tilde{x} - 1 - \hat{\lambda}) \,,\quad\hbox{where}\quad \hat{\lambda} \equiv 
\frac{\omega - \Omega_{\rm A}^+ (r_2^*)}{\delta^* \hs\Omega_{\rm 
A}^{+\prime}(r_2^*)} \,.\En{35}\EEQAR
The derivatives $\Omega_{\rm A}^{\pm\prime}$ of the continuum frequencies may be 
transformed into derivatives of the Doppler frequency and, hence, of the rotation 
frequency, $\Omega_{\rm A}^{\pm\prime} \approx \Omega_0^{\prime} = 
m\Omega^{\prime}$, since $|\omega_{\rm A}| \ll |\Omega_0|$. Also, we will assume 
$|\lambda|$ and $|\hat{\lambda}| \ll1$. Of course, the eigenvalue parameters $\lambda$ 
and $\hat{\lambda}$ are related since there is only one eigenvalue $\omega$:
\BEQ \lambda \approx \frac{\sigma - \Omega_{\rm A}^-( r_1) + {\rm i} \hs\nu}{\delta^* 
\hs\Omega_0^{\prime}(r_1)} \,,\qquad \hat{\lambda} \approx \frac{{\rm i} 
\hs\nu}{\delta^* \hs\Omega_0^{\prime}(r_2^*)} \quad\Rightarrow\quad 
\hat{\lambda}_2 = \lambda_2/h \,, \quad h \equiv 
\frac{\Omega^{\prime}(r_2^*)}{\Omega^{\prime}(r_1)} \approx 1 - \fivehalf \delta^* \,. 
\En{36}\EEQ
Defining the average of equilibrium quantities at the two end points, $\langle f \rangle 
\equiv \half \big[f(r_1) + f(r_2^*)\big]$, the two expressions \E{34} and \E{35} may be 
combined into a single approximate expression,
\BEQ \widetilde{\omega}^2 - \omega_{\rm A}^2 \approx 2 \delta^* \langle|\omega_{\rm 
A} \hs\Omega_0^{\hs\prime}|\rangle (\tilde{x} - \lambda) (\tilde{x} - 1 - \hat{\lambda}) 
\,, \En{37}\EEQ
which is considered to be valid over the whole interval. The absolute signs are introduced 
in this combined expression (not in the definitions of $\lambda$ and $\hat{\lambda}$!) to 
get positive multiplication factors in the basic parameters of the ODE presented below. 
This implies that that ODE is valid for the present co-rotating SARIs, where 
$\Omega_0^{\prime} < 0$, as well as for the counter-rotating SARIs, where 
$\Omega_0^{\prime} > 0$. However, the defining relations~\E{34} and \E{35} between 
the complex variables $\lambda$ and $\omega$ would differ in sign for the two cases. 
Note that $\langle{f}\rangle \langle{g}\rangle \approx \langle{f g}\rangle$ since the 
equilibrium variations are assumed small. With this approximation, the differential 
equation~\E{32} transforms into an explicit representation involving all basic 
parameters:
\BEQ \frac{d}{d\tilde{x}} \bigg[ (\tilde{x} - \lambda) (\tilde{x} - 1 - \hat{\lambda}) 
\hs\frac{d\chi}{d\tilde{x}} \bigg] + \bigg[ \frac{k^2 \delta^*}{2 |m|} \Big\langle 
\frac{\kappa_{\rm e}^2}{|\omega_{\rm A} \Omega^{\hs\prime}|} \Big\rangle - k^2 
\delta^{*2} (\tilde{x} - \lambda) (\tilde{x} - 1 - \hat{\lambda})
+ \frac{(k^2/m^2) \hs\langle \Omega/\Omega^{\hs\prime} \rangle^2}{(\tilde{x} - 
\lambda) (\tilde{x} - 1 - \hat{\lambda})} \bigg] \hs\chi = 0 \,, \En{38}\EEQ
where the mode number $m$ now appears since $\Omega_0^\prime = m 
\Omega^\prime$. Because of the approximation~\E{37}, the second term  of Eq.~\E{38} 
is negligible at the end points and has its maximum contribution in the middle of the 
interval. Hence, its line average,
\BEQ \overline{(\tilde{x} - \lambda) (\tilde{x} - 1 - \hat{\lambda})} \approx \int_0^1 
\tilde{x} (\tilde{x} - 1) \,d\tilde{x} = - \onesixth \,, \En{39}\EEQ
gives a contribution that can be included in the first term.

\begin{figure}[ht]
\FIG{\begin{center} 
\includegraphics*[height=3.6cm]{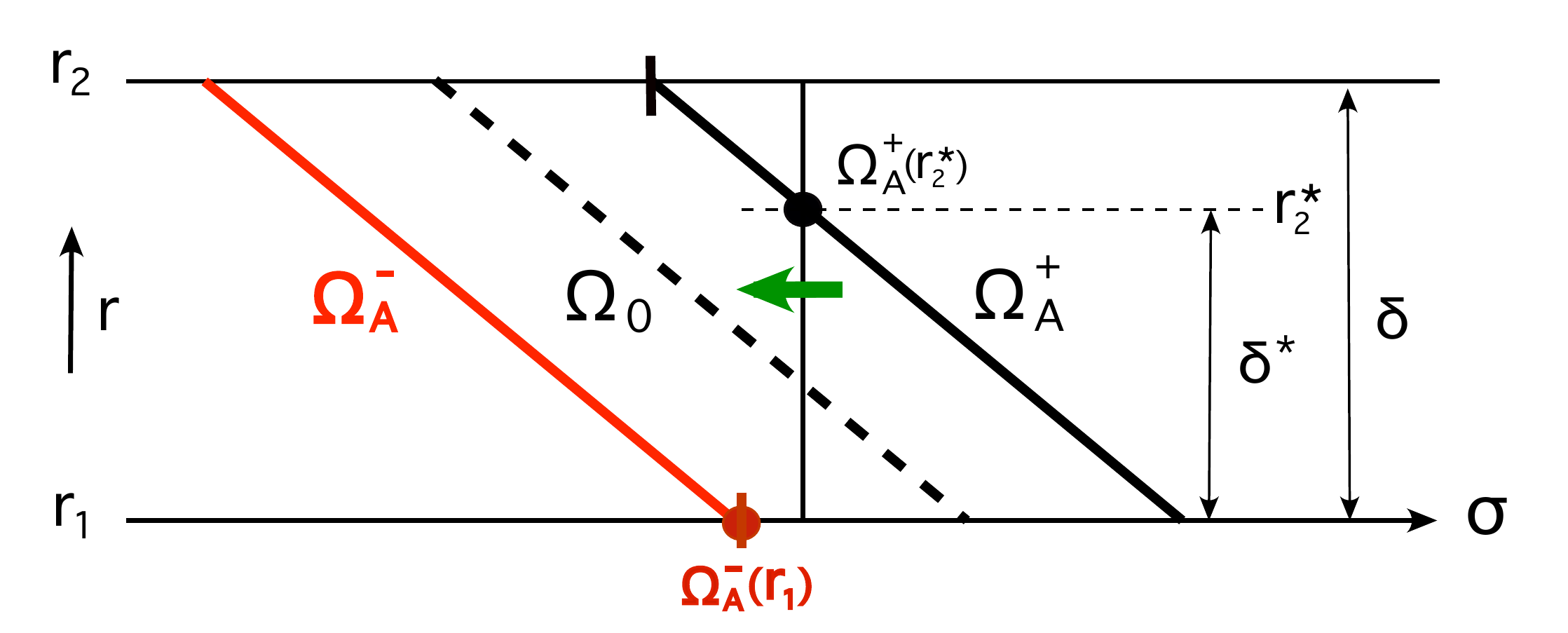}
\end{center}}
\caption{Schematic representation of the two continuum functions $\sigma = 
\Omega_{\rm A}^-(r) \equiv \Omega_0(r) - \omega_{\rm A}(r)$ and $\sigma = 
\Omega_{\rm A}^+(r) \equiv \Omega_0(r) + \omega_{\rm A}(r)$ with the two 
near-singularities $\Omega_{\rm A}^-(r_1)$ and $\Omega_{\rm A}^+(r_2^*)$ for an 
`inner' co-rotating Super-Alfv\'enic Rotational Instability. The vertical line indicates the 
real part $\sigma$ of the eigenvalue of the mode. The eigenvalues of the `inner' SARIs 
are dictated by the asymptotic approach $\sigma \rightarrow \Omega_{\rm A}^-(r_1)$ 
indicated by the green arrow.}{\Fn{9}}
\end{figure}

\begin{figure}[ht]
\FIG{\begin{center} 
\includegraphics*[height=5.4cm]{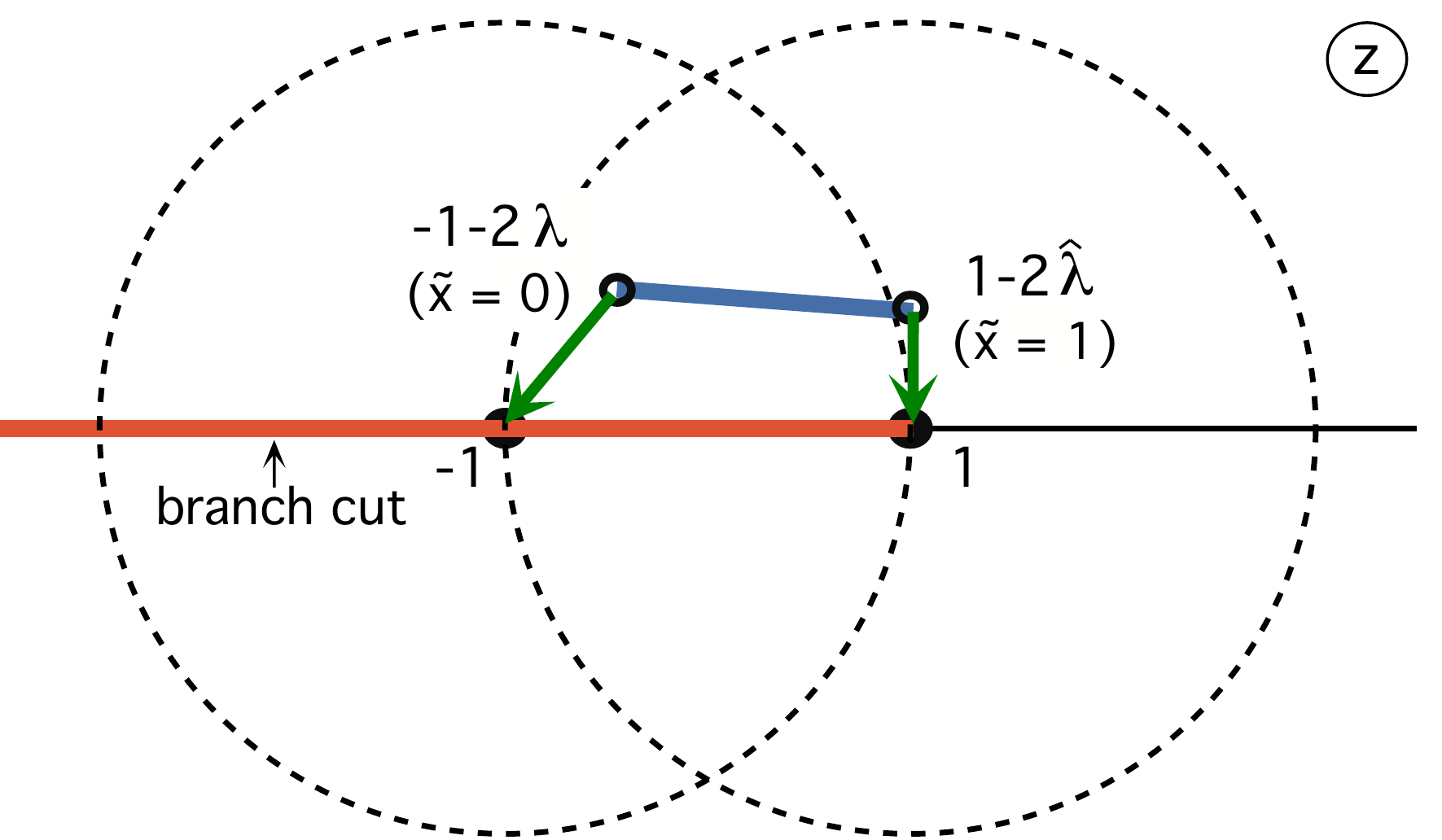}
\end{center}}
\caption{Asymptotic approach (green arrows) of the image in the $z$-plane of the end 
points $\tilde{x} = 0$ and $\tilde{x} = 1$ of the physical interval (in blue) to the two 
singularities $z = \pm 1$ of the associated Legendre functions. Their circles of 
convergence are dashed.}{\Fn{10}}
\end{figure}

Introducing the complex variable $z \equiv (2 \tilde{x} - 1 - \lambda - \hat{\lambda})/(1 
- \lambda + \hat{\lambda})$, where $\tilde{x} = 0$ corresponds to $z \approx - 1 - 2 
\lambda$ and $\tilde{x} = 1$ corresponds to $z \approx 1 - 2 \hat{\lambda}$, the 
singular expression becomes $(\tilde{x} - \lambda) (\tilde{x} - 1 - \hat{\lambda}) \equiv 
\onefourth (1 - \lambda + \hat{\lambda})^2 (z^2 - 1) \approx \onefourth (z^2 - 1)$. 
Equation~\E{38} then transforms into a differential equation in terms of the complex 
variable $z$:
\BEQ \frac{d}{dz} \Big[\hs(z^2 - 1) \hs\frac{d\chi}{dz} \hs\Big] + \Big[\hs u^2 + 
\onefourth - w^2(z^2 - \onethird) + \frac {v^2}{ z^2 - 1} \hs\Big] \, \chi = 0 \,, \En{40} 
\EEQ
where the physical parameters $u$, $v$ and $w$ are defined by
\BEQ u \equiv \Big[\hs\frac{k^2 \delta^*}{2 |m|} \Big\langle\frac{\kappa_{\rm 
e}^2}{|\omega_{\rm A}\hs\Omega^{\hs\prime}|}\hs\Big\rangle + \onesixth k^2 
\delta^{*2} - \onefourth \hs\hs\Big]^{1/2} \,,\qquad v \equiv \Big| \frac{2 k}{m} 
\Big\langle \frac{\Omega}{\Omega^{\hs\prime}} \Big\rangle \Big| \,,\qquad w \equiv 
\half k \delta^* \,. \En{41}\EEQ
Since we have assumed $\delta \sim \delta^* \ll 1$, the term $w^2(z^2 - \onethird)$, 
representing the deviation from the average given by Eq.~\E{39}, is {\em formally} 
negligible with respect to the term $u^2 + \onefourth$. (We will return to this at the end 
of this section.) Hence, Eq.~\E{40} transforms into Legendre's equation solely involving 
the coefficients $u$ and $v$, which are related to the canonical Legendre exponents 
$\nu_{1,2}$ and $\mu_{1,2}$ through 
\BEQ \nu(\nu+1) = - u^2 - \onefourth \;\;\Rightarrow\;\; \nu_{1,2} \equiv -\half \pm {\rm 
i}\hs u \,,\quad\hbox{and}\quad \mu^2 = - v^2 \;\;\Rightarrow\;\; \mu_{1,2} \equiv\pm 
{\rm i}\hs v \,. \En{42}\EEQ
The solutions are the associated Legendre functions $P_{-(1/2)\pm {\rm i}\hs u}^{\pm 
{\rm i}\hs v}(z)$ and $Q_{-(1/2)\pm {\rm i}\hs u}^{\pm {\rm i}\hs v}(z)$, where the 
different combinations may be chosen to suit the requirements of a particular boundary 
value problem. For our case, the pair $P_{-(1/2)+{\rm i}\hs u}^{\hs{\rm i}\hs v}(z)$ and 
$P_{-(1/2)+{\rm i}\hs u}^{-{\rm i}\hs v}(z)$ turns out to be the most expedient choice, 
so that a general solution of Eq.~\E{40} can be written as
\BEQAR
\chi &=& F_1(z) + C F_2(z) \,,\En{43}\\[2mm]
&& F_1 \equiv \Gamma(1 - {\rm i}\hs v) P_{-\frac{1}{2}+{\rm i}\hs u}^{{\rm i}\hs 
v}(z) \,,\quad  F_2 \equiv \Gamma(1 + {\rm i}\hs v) P_{-\frac{1}{2}+{\rm i}\hs u}^{-
{\rm i}\hs v}(z) \,, \non\EEQAR
where the $\Gamma$-factors are inserted for (later) algebraic convenience. The behavior 
of this solution, subjected to the BCs~\E{11} with $r_2$ replaced by $r_2^*$, is to be 
investigated for values of $\omega$ approaching the two singularities $z = -1$ and $z = 
1$, as illustrated in Fig.~\F{10}.

The convenience of the Legendre solution~\E{43} is that it provides the full connection 
between the boundaries of the physical interval at $\tilde{x} = 0$ and $\tilde{x} = 1$, so 
that we just need to impose the BCs there to determine the free parameters of the 
problem. Those are the complex eigenvalue parameter~$\lambda$ and the complex 
eigenfunction constant~$C$,
\BEQ \lambda = |\lambda|\hs e^{\hs{\rm i} \alpha} \,,\qquad C = |C|\hs e^{\hs{\rm i} 
\varphi}\,, \En{44}\EEQ
involving four so far unknown constants: $|\lambda|$, $\alpha$, $|C|$ and $\varphi$.

\subsubsection{Right boundary condition}{\Sn{IVB2}}

It is expedient to start with the BC at $\tilde{x} = 1$ ($r = r_2^*$) since a ready-to-use 
representation of the associated Legendre functions is given in \Ron{NIST2010} 
[Eq.~14.3.6] in terms of an algebraic function multiplied by a hypergeometric function 
that converges for  $|z-1| < 2$ (i.e., in the right circle of Fig.~\F{10}):
\BEQ [F_1(z)]_{\rm R} \equiv \Big(\frac{z+1}{z-1}\Big)^{\frac{1}{2}{\rm i}\hs v} 
F(\half + {\rm i}\hs u, \half - {\rm i}\hs u; 1-{\rm i}\hs v; \half - \half z) \,, \En{45}\EEQ
whereas $[F_2(z) ]_{\rm R}$ is obtained from $[F_1(z) ]_{\rm R}$ by replacing $v$ by 
$-v$. The dominant behavior approaching the singularity, $z \rightarrow 1$ for $\tilde{x} 
= 1$, follows from these expressions by evaluating the algebraic factor for $z \approx 1 - 
2 \hat{\lambda}$, with $|\hat{\lambda|} \ll 1$, and eliminating the hypergeometric 
function accordingly using the property $F(a,b;c;\hat{\lambda}) \rightarrow F(a,b;c;0) 
\equiv 1$. For the algebraic factors, it is to be noted that the parameters $\lambda_1$, 
$\lambda_2$,  and $\hat{\lambda}_2$ are negative, as follows from Eq.~\E{36} for the 
`inner' co-rotating SARIs, so that the approach of the singularities is from above the 
branch cut, as illustrated in Fig.~\F{10}. [$\hs$Recall that they are positive for the `inner' 
counter-rotating SARIs, so that the singularities would be approached from below the 
branch cut in that case.$\hs$] This yields
\BEQ [F_1(1 - 2 \hat{\lambda})]_{\rm R} \approx ({\rm i}|\hat{\lambda}_2|)^{-
\frac{1}{2}{\rm i}\hs v} \approx {\rm e}^{\frac{1}{4}\pi v} {\rm e}^{-\frac{1}{2}{\rm 
i\hs} v \ln |\hat{\lambda}_2|} \,,\quad [F_2(1 - 2 \hat{\lambda})]_{\rm R} \approx ({\rm 
i}|\hat{\lambda}_2|)^{\frac{1}{2}{\rm i}\hs v} \approx {\rm e}^{-\frac{1}{4}\pi v} 
{\rm e}^{\frac{1}{2}{\rm i\hs} v \ln |\hat{\lambda}_2|}\,,\En{46}\EEQ
so that the boundary condition at $\tilde{x} = 1$ determines two of the four constants:
\BEQAR && [F_1(1 - 2\hat{\lambda})]_{\rm R} + C \hs[F_2(1 - 2\hat{\lambda})]_{\rm 
R} \approx {\rm e}^{\frac{1}{2}{\rm i}\varphi} \big[{\rm e}^{\frac{1}{4}\pi v} {\rm 
e}^{-\frac{1}{2}{\rm i\hs} (v \ln |\hat{\lambda}_2| + \varphi)} + |C| \hs{\rm e}^{-
\frac{1}{4}\pi v} {\rm e}^{\frac{1}{2}{\rm i\hs} (v \ln |\hat{\lambda}_2| + \varphi)} 
\big]= 0 \nonumber\\[2mm]
&&\qquad\Rightarrow\quad |C| =  {\rm e}^{\frac{1}{2}\pi v}\,,\quad\hbox{and}\quad  
\cos\hs(v \ln |\hat{\lambda}_2| + \varphi) = -1 \,. \En{47}\EEQAR
The latter condition yields $v \ln |\hat{\lambda}_2| + \varphi = (2 n' + 1) \pi$, where $n'$ 
is any (positive or negative) integer. However, since $|\hat{\lambda}_2|$ is the scaled 
growth rate of the instabilities, we should demand that it monotonically decreases to zero 
for positive integers $n = 1, 2, \ldots$ referring to increasingly oscillatory behavior of the 
solutions. Writing $n' = n_0 - n$, where $n_0$ is to be determined yet, and noting that $-
\pi < \alpha < 0$, this gives 
\BEQ |\hat{\lambda}_2| \equiv - h |\lambda| \sin \alpha =  h |\lambda| \sin(\pi + \alpha) = 
{\rm e}^{-[(2 n - 2 n_0 - 1) \pi + \varphi]/v}   \quad\hbox{($n = 1, 2, \ldots$)} 
\,,\En{48}\EEQ
where the negative angle $\alpha$ in the complex $\lambda$-plane has been replaced by 
the positive angle $\pi + \alpha$ (later called $\alpha'$) of the locus of eigenvalues in the 
$\omega$-plane. There is some arbitrariness in where to start counting. For $n = 1$, the 
exponent should be negative to ensure monotonicity for the fastest growing mode, so that 
the integer $n_0$ should at least satisfy 
\BEQ n_0 \le \half(1 + \varphi/\pi) \,, \En{49}\EEQ
whereas a slightly sharper criterion is obtained by applying inequality \E{A12} of 
Appendix~\S{A2}. Equation~\E{48} implies that the logarithmic sequence of eigenvalue 
moduli decreases inversely proportional to the singularity exponent~$v$:
\BEQ \ln(|\lambda_{n+1}|/|\lambda_n|) = - 2 \pi/v \,,\En{50}\EEQ
i.e., assuming that the constants $\alpha$ and $\varphi$ (to be determined yet by the other 
BC) do not to depend on $n$, which turns out to be the case.

\subsubsection{Left boundary condition}{\Sn{IVB3}}

As stated above, the solution~\E{43} provides the full connection between the two 
boundaries. However, to apply the BC at $\tilde{x} = 0$ ($r = r_1$), the 
representation~\E{45} of the basic solution cannot be used since it is only valid in the 
right circle of convergence shown in Fig.~\F{10}, i.e.\ for $|z -1| < 2$. For the left BC we 
need an expression that converges in the left circle, i.e.\ for $|z+1| < 2$. Such a 
representation may be obtained from \Ron{NIST2010} [Eq.~15.8.4], or  \Ron{AS64} 
[Eq.~15.3.6], by converting the two hypergeometric functions presented there in terms of 
$z$ and $1-z$ to functions in terms of $\half - \half z$ and $\half + \half z$. This linear 
transformation from the representation for the right circle of convergence to the one for 
the left circle is valid for $|\arg(1-z)| < \pi$ and $|\arg(1+z)| < \pi$, i.e.\ above as well 
below but not on the branch cut. We then obtain the following expression for the basic 
solution for $|z+1| < 2$:
\BEQAR [F_1(z)]_{\rm L} &\equiv& \frac{{\rm i}\hs\pi}{\sinh(\pi v)} \hs\Big(\frac{z 
+1}{z - 1}\Big)^{\frac{1}{2}{\rm i}\hs v} \hs\bigg[ \hs\frac{\Gamma(1-{\rm i}\hs v) \hs 
F(\half + {\rm i}\hs u, \half - {\rm i}\hs u; 1+{\rm i}\hs v; \half + \half z)}{ 
\Gamma(1+{\rm i}\hs v) \hs\Gamma(\half + {\rm i}\hs u - {\rm i}\hs v) \hs\Gamma(\half 
- {\rm i}\hs u - {\rm i}\hs v)}
\non\\[2mm]
&& \qquad\qquad - \,(\half + \half z)^{- {\rm i}\hs v} \,\frac{F(\half + {\rm i}\hs u  - 
{\rm i}\hs v, \half - {\rm i}\hs u - {\rm i}\hs v; 1-{\rm i}\hs v; \half + \half 
z)}{\Gamma(\half + {\rm i}\hs u) \hs\Gamma(\half - {\rm i}\hs u)} \hs\bigg]
\,,\En{51}\EEQAR
whereas $[F_2(z) ]_{\rm L}$ is obtained from $[F_1(z) ]_{\rm L}$ by replacing $v$ by 
$-v$. 

The dominant behavior approaching the singularity, $z \rightarrow -1$ for $\tilde{x} = 
0$, follows from these expressions by evaluating the algebraic factors for $z \approx -1 - 
2 \lambda$, with $|\lambda| \ll 1$, and eliminating the hypergeometric functions as 
above. It is to be noted that the $-\lambda$ occurring inside the brackets of the second 
algebraic factor is to be evaluated {\em above the branch cut}, so that $(\half + \half z)^{-
{\rm i}v} \approx (-\lambda)^{-{\rm i}v} = {\rm e}^{\pi v} \cdot {\rm e}^{\alpha 
v}\hs{\rm e}^{-{\rm i}v \ln |\lambda|}$. The first solution at $\tilde{x} = 0$ then 
becomes
\BEQ [F_1(- 1 - 2 \lambda) ]_{\rm L} \approx {\rm i}\hs f^{-1} \hs\big[\hs \sqrt{1 + 
f^2} \hs\hs{\rm e}^{-\frac{1}{2}\alpha v} \hs{\rm e}^{\hs\frac{1}{2}{\rm 
i\hs}v\ln|\lambda|} \hs{\rm e}^{\hs{- \rm i\hs} \tau} - {\rm e}^{(\frac{1}{2}\alpha + \pi) 
v} \hs{\rm e}^{-\frac{1}{2}{\rm i\hs}v\ln|\lambda|} \hs\big] \,,\En{52}\EEQ
where $f$ is a function of $u$ and $v$, 
\BEQ f \equiv {\sinh(\pi v)}/{\cosh(\pi u)} \,,\En{53}\EEQ
and the angle $\tau$ involves the arguments $\gamma_v$ and $\delta_{u \pm v}$ of the 
$\Gamma$-functions,
\BEQAR &&\Gamma(1 + {\rm i\hs}v) = \sqrt{\pi v/\!\sinh(\pi v)} \,{\rm e}^{{\rm i\hs} 
\gamma_v} \,,\non\\[2mm]
&&\Gamma(\half + {\rm i\hs}u) = \sqrt{\pi/\!\cosh(\pi u)} \,{\rm e}^{{\rm i\hs} 
\delta_u} \,,\quad \delta_u = \gamma_{2u}- \gamma_u - 2 u \ln 2 \,, \non\\[2mm]
&&\tau \equiv 2 \gamma_v - \delta_{u+v} + \delta_{u-v} = 2 \gamma_v - 
\gamma_{2(u+v)} + \gamma_{u+v} + \gamma_{2(u-v)}  - \gamma_{u-v} + 4 v \ln 2 
\,.\En{54}\EEQAR
The second basic solution $[F_2(-1 - 2 \lambda) ]_{\rm L}$ follows again by replacing 
$v$ by $-v$. 

With the given approximations, the boundary condition at $\tilde{x} = 0$, 
\BEQ [F_1(-1 - 2\lambda) ]_{\rm L} + {\rm e}^{\frac{1}{2}\pi v} \hs{\rm e}^{{\rm 
i}\hs\varphi} \hs[F_2(-1 - 2\lambda) ]_{\rm L} = 0 \,, \En{55}\EEQ
may be symmetrized by judicious manipulation of the coefficients to yield:
\BEQAR 
&&\hor{-10}\sqrt{1 + f^2}\hs\big[\hs S\hs\hs{\rm e}^{-\frac{1}{2}{\rm 
i\hs}(v\ln|\lambda| + \varphi)}\hs{\rm e}^{{\rm i}(\varphi + \tau)} - S^{-1}\hs{\rm 
e}^{\frac{1}{2}{\rm i\hs}(v \ln|\lambda| + \varphi)}\hs{\rm e}^{-{\rm i}(\varphi + 
\tau)} \hs\big] \non\\[2mm]
&&\quad + \; T\hs\hs{\rm e}^{-\frac{1}{2}{\rm i\hs}(v\ln|\lambda| + \varphi)} 
- T^{-1}\hs{\rm e}^{\frac{1}{2}{\rm i\hs}(v\ln|\lambda| + \varphi)} = 0 \,, \qquad S 
\equiv {\rm e}^{\frac{1}{2}(\alpha + \frac{1}{2}\pi)v} \,,\quad T \equiv {\rm 
e}^{\frac{1}{2}(\alpha + \frac{3}{2}\pi)v} \,. \En{56}\EEQAR
Substituting the relationship~\E{48} from the boundary condition at $\tilde{x} = 1$, the 
exponential factors involving $|\lambda| $ may be transformed to eliminate the 
dependence on $n$,
\BEQ {\rm e}^{\frac{1}{2}{\rm i}(v\ln|\lambda| + \varphi)} = {\rm e}^{-{\rm i}(n - n_0 
- \frac{1}{2})\pi} {\rm e}^{-{\rm i} \eta} = {\rm i} \hs(-1)^{n-n_0}\hs{\rm e}^{-{\rm 
i}\eta} \,,\qquad \eta \equiv \half v\ln[h\sin(\pi + \alpha)] \,,\En{57}\EEQ
giving a complex equation for the determination of the angles $\alpha$ and $\varphi$:
\BEQ \sqrt{1 + f^2}\hs\hs\big[\hs S \hs{\rm e}^{{\rm i}(\varphi + \tau + \eta)} 
+ S^{-1} \hs{\rm e}^{-{\rm i}(\varphi + \tau + \eta)}\hs\big] 
+ T\hs\hs{\rm e}^{{\rm i\hs}\eta} + T^{-1} \hs{\rm e}^{-{\rm i\hs}\eta} = 0 \,. 
\En{58}\EEQ
Separating real and imaginary parts of this equation yields two expressions,
\BEQAR
\hor{-5}\sin(\varphi + \tau) &=& - \frac{\sinh(\half \pi v) \sin(2\eta)}{\sqrt{1 + 
f^2}\hs\sinh[(\alpha + \half\pi)v]} \,\hor{14.5}\equiv s(\alpha)\,,\non\\[2mm]
\hor{-5}\cos(\varphi + \tau) &=& - \frac{\sinh[(\pi + \alpha) v] - \sinh(\half \pi v) 
\cos(2\eta)}{\sqrt{1 + f^2}\hs\sinh[(\alpha + \half\pi)v]} \,\equiv c(\alpha)
\,, \En{59}\EEQAR
which, together, determine $\alpha$ and $\varphi$ in terms of the parameters $u$ and 
$v$.

The final equation determining $\alpha$ is obtained by adding the squares of these 
expressions:
\BEQAR 
P(\alpha) &\equiv& \sinh^2(\half\pi v) + \sinh^2[(\pi + \alpha) v] 
- 2 \sinh(\half\pi v) \sinh[(\pi + \alpha) v] \cos[v\ln(h\sin(\pi + \alpha))] \non\\[2mm]
&&\hor{51} - (1 + f^2) \sinh^2[(\alpha + \half\pi)v] = 0 \,. \En{60}\EEQAR
The angle $\alpha$ should have a value in the range $-\pi < \alpha < -\half \pi$ to 
correspond to the range $0 < \alpha' \equiv \pi + \alpha < \half \pi$ of the locus of 
eigenvalues of the `inner' co-rotating modes in the $\omega$-plane (see Fig.~\F{8}). A 
zero of $P(\alpha)$ in that range is guaranteed since $P(-\pi) < 0$ and $P(-\half\pi) > 0$ 
(for $h \ne 1$). Once this zero is found, the final equation determining the 
parameter~$\varphi$ follows directly from the relations~\E{59}: 
\BEQAR \varphi = - \tau + \arctan\big[{s(\alpha)}/{c(\alpha)}\big] \,. \En{61}\EEQAR
This completes the asymptotic analysis of the SARIs. The eigenvalue parameters 
$|\lambda|$ and $\alpha$ are determined by Eqs.~\E{48}, or \E{57}, and \E{60}, 
whereas the eigenfunction parameters $|C|$ and $\varphi$ are determined by 
Eqs.~\E{47} and \E{61}.

\subsubsection{Comparison with the Spectral Web results}{\Sn{IVB4}}

For the `inner' co-rotating SARIs, corresponding to the right branch of the Spectral Web 
depicted in Fig.~\F{8}, the numerical values of the input as well as the output parameters 
of the above analytic expressions are as follows,
\BEQAR
\hbox{input:}\hor{4} \quad && \epsilon = 0.0141 \,,\quad \beta = 100 \,,\quad \mu_1 = 
1\,,\quad \delta = 0.1 \,,\quad m = 10 \,,\quad k = 50 \non\\[1mm]
&& \Rightarrow\quad u = 3.8764\,,\quad v = 6.9584\,,\quad h = 0.81078 \,,\quad \delta^* 
= 0.087526\,; \non\\[1mm]
\hbox{output}: \quad && |C| = 5.5841\times 10^{4}\,,\quad \tau =  5.7643\hs\pi 
\;\;(\rightarrow - 0.2357 \hs\pi)\,,\quad \varphi = -1.7747\hs\pi \,,\non\\[1mm]
&& |\lambda_1| = 0.35152\,,\quad \alpha = -0.56194\hs\pi \,. \En{62}\EEQAR
To compare these numbers with the numerical Spectral Web results, the right branch of 
the Spectral Web of Fig.~\F{8} is replotted in double logarithmic coordinates 
$\log(|\Delta\sigma|), \log(\nu)$ in Fig~\F{11}, as appropriate for the exponential 
approach~\E{48} of the cluster point. In the logarithmic representation, $\Delta \sigma 
\equiv \sigma - \Omega_{\rm A}^+(r_1)$ is the real frequency shift with respect to the 
cluster point. A sequence of nine analytic eigenvalues is shown in green on a dashed line 
through those points. Note that the identical slope of $45^\circ$ for the locus of both the 
analytical and the numerical eigenvalues is accidental, just due to having chosen the same 
scale for the two logarithmic axes, it has nothing to do with the angle $\alpha'$. However, 
it is evident from the figure that the approximate eigenvalues (green dots) differ 
significantly from the `exact' (i.e.\ numerical) eigenvalues (black dots). In particular, the 
asymptotic value of the slope of the locus of the numerical `inner' eigenvalues shown in 
Fig.~\F{8} in regular coordinates is $\alpha' \equiv \pi + \alpha = 0.36063\hs\pi$, as 
opposed to the much larger analytical value $\alpha' = 0.43806\hs\pi$ following from the 
numbers~\E{62}. This is not surprising in view of the rather crude assumptions that had 
to be made in the approximations \E{37} and \E{39} to reduce the problem to the 
solution of the Legendre equation.

\begin{figure}[ht]
\FIG{\begin{center} 
\includegraphics*[height=10cm]{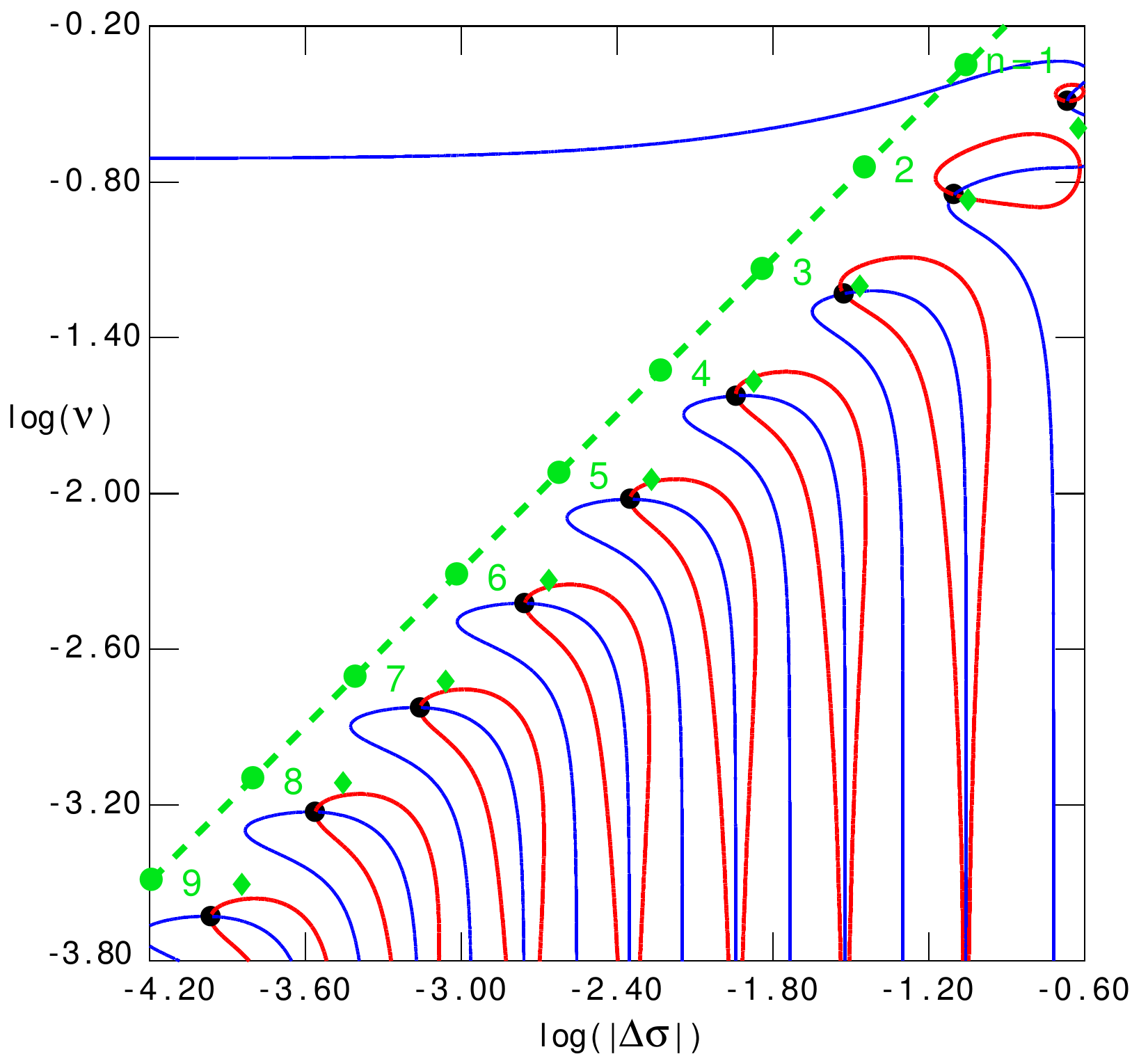}
\end{center}}
\caption{Logarithmic Spectral Web corresponding to the `inner' co-rotating SARIs of 
Fig.~\F{8} with nine analytic eigenvalues and their locus (green dots and dashed line). 
The eigenvalues obtained by numerical solution of the `extended' Legendre 
equation~\E{40} with $w = 2.1881$ are indicated by green diamonds. The horizontal 
scale refers to the frequency shift with respect to the cluster point, $\Delta \sigma \equiv 
\sigma - \Omega_{\rm A}^-(r_1)$.}{\Fn{11}}
\end{figure}

The influence of the approximation~\E{39}, that was necessary to neglect the 
contribution of the term multiplied by the parameter $w \equiv k \delta^*$ in the 
`extended' Legendre equation~\E{40}, can be estimated by solving that equation 
numerically for the pertinent value of $w$. That value is $2.1881$ for the present 
example, the corresponding eigenvalues (green diamonds) shown in Fig.~\F{11} are 
shifted significantly towards the exact eigenvalues (black dots). This suggests that the 
present approximate asymptotic analysis, based on the assumption of small variation of 
the equilibrium variables, is justified because the contribution of $w$ vanishes faster for 
$\delta^* \rightarrow 0$ than the other contributions according to the orders of the three 
parameters of the `extended' Legendre equation: $v^2 \sim 1$, $u^2 + \onefourth \sim 
\delta^*$, $w^2 \sim \delta^{*2}$.

Incidentally, note that the values of $u$ and $v$, when multiplied by $\pi$, produce large 
values of the hyperbolic functions and the arguments of the $\Gamma$~functions 
occurring in the auxiliary angle $\tau$. That angle then has to be reduced to the principal 
value by adding or subtracting the appropriate number of multiples of $2\pi$. Extreme 
care had to be taken to maintain the needed accuracy by proper cancelling of the 
oscillatory integrands of the $\Gamma$~function integrals.

Finally, two qualitative observations on the presented analytic properties of the SARIs 
are in order. First, the asymmetry of the equilibrium between the two end points of the 
interval (i.e.\ the two singularities) was partly accounted for by exploiting the ratio $h 
\equiv \Omega^\prime_{\rm top}/\Omega^\prime_{\rm bot} \ne 1$ of the rotation shears, 
defined in Eq.~\E{36}. However, the mentioned approximation~\E{37} implies that the 
index $v$ was forced to be equal for the two singularities, whereas for this particular 
equilibrium the two indices actually differ, viz.\ by a factor $v_{\rm top}/v_{\rm bot} = 
1.087$. The index $v$ is the only parameter occurring in the relation~\E{50}, which 
describes the logarithmic decay of the eigenvalues towards the cluster point. From 
Fig.~\F{11} it is clear that this decay is quite well described by the approximate analytic 
expression. There is a very surprising feature to this. Recall that Eq.~\E{50} was 
obtained by just applying the BC at $r = r_2^*$, so that, in a sense, the logarithmic 
sequence would be described by the {\em local} value of $v$ at that point. But the same 
expression would also have been obtained by applying the BC at $r = r_1$, resulting in a 
logarithmic sequence with {\em another local value} of $v$. From the `exact' (numerical) 
results shown in Fig.~\F{11} this is not the case! Apparently, the two singularities 
communicate in a {\em non-local} manner to establish just {\em one sequence with an 
averaged value of~$v$,} as in the approximate analysis. This is another manifestation of 
the intricate interaction of the three near-singularities in the dynamics of the SARIs 
mentioned at the end of Sec.~\S{IVA}.

Second, the alignment of the numbered sequence of approximate eigenvalues with the 
`exact' Spectral Web eigenvalues shown in Fig.~\F{11} has important consequences for 
the classification of the different solutions. Whereas numbering of the eigenvalues by 
counting the number of zeros of the associated eigenfunctions is a standard tool in the 
spectral analyses of static equilibria (which still applies for the MRIs, as illustrated in 
Fig.~\F{4}), this is impossible for the complex eigenfunctions of the SARIs. This is 
already evident from the fact that the number of zeros $(n_{\rm z1}, n_{\rm z2})$ of the 
real and the imaginary components $\chi_1$ and $\chi_2$ of $\chi$ are usually different, 
so that there appears to be no unique way of labelling the modes with one number of 
zeros. For example, for the modes shown in Fig.~\F{6} those numbers are $(2,2)$ for the 
1st `inner' mode, $(10,8)$ for the 4th `outer' mode, and $(23,20)$ for the 10th `inner' 
mode shown. Here, `number of zeros' is counted for the full interval, excluding the zero 
of the BC at $r_1$ but including the one at $r_2$, so that this number corresponds with 
the intuitive notion of `number of lobes' of the eigenfunction (whereas the `number of 
nodes' is just one less). However, the Legendre analysis provides the unique numbering 
of eigenvalues and associated eigenfunctions for the SARIs, that was already used in 
Fig~\F{6}. Note that, quite in contrast to the standard spectra with real functions and one 
singularity, {\em there is no zero-node (one zero) solution!} Both sequences start off with 
solutions with at least two zeros. Moreover, all successive eigenfunctions have about two 
additional zeros. Apparently, each of the two singularities simultaneously contributes one 
zero at a time, so that the number of zeros of the eigenfunctions of both sequences is 
incremented  with at least two as $n$ is incremented with one: quite a peculiar aspect of 
the SARIs! 
 
Of course, counting based on the Legendre analysis will only apply as long as the 
approximations made in this section are justified. In the following section, we will move 
far away from these assumptions so that this remnant of spectral counting also will break 
down.

\section{Quasi-Continuum Super-Alfv\'enic Rotational Instabilities}{\Sn{V}}

\subsection{Fragmentation of the Spectral Web}{\Sn{VA}}

The Super-Alfv\'enic Rotational Instabilities discussed in Section~\S{IV} appear to need 
an edge of one of the Alfv\'en continua so that they localize about the inner or outer 
boundary of the accretion disk (see the eigenfunctions of Fig.~\F{6}). Actually, 
localization at the inner boundary also happens, for different reasons, for the 
axisymmetric Magneto-Rotational Instabilities discussed in Section~\S{III} (see the 
eigenfunctions of Fig.~\F{4}). The analysis of Section~\S{IVB} was necessarily 
restricted to modes localized in narrow layers of the accretion disk ($\delta \ll 1$) so that 
the urgent question to be addressed is: Are the SARIs for larger values of~$\delta$ still 
localized at the boundaries of the region investigated? Since that region is a kind of 
necessary artefact of the numerical method used, it would be nice if we also found 
instabilities that are localized in the middle of the interval, far away from the boundaries 
of the disk. Hence, we now investigate modes for values of $\delta$ that are not small. 
This analysis has a number of surprises in store for us. It will surpass anything that has 
been investigated before in the study of the linear MRIs: the best is yet to come!

\begin{figure}[ht]
\FIG{\begin{center} 
\includegraphics*[height=14.6cm]{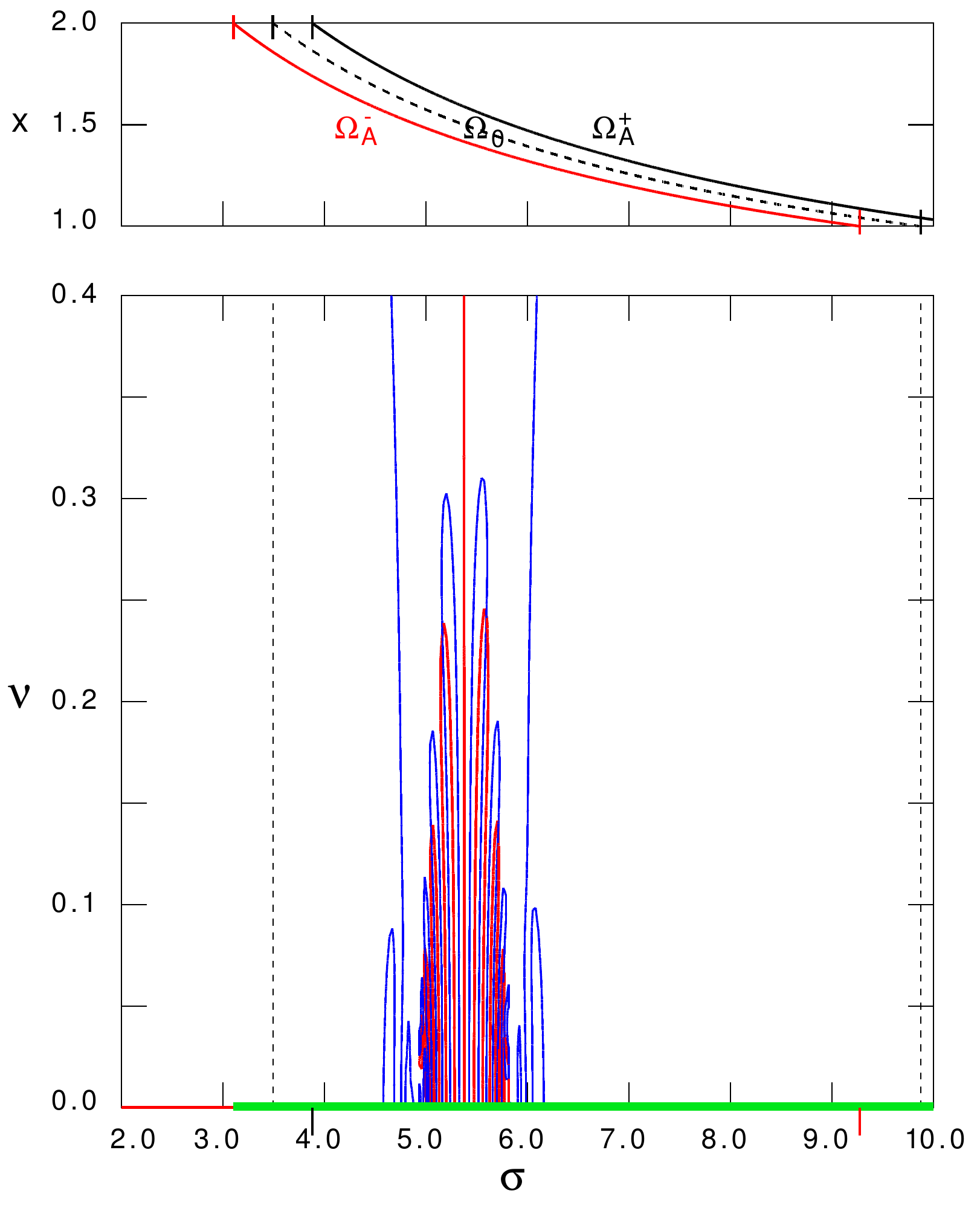}
\end{center}}
\caption{Spectral Web of the co-rotating Super-Alfv\'enic Rotational Instabilities for an 
accretion disk with  parameters $\epsilon \equiv B_1 = 0.1041$, $\mu_1 \equiv B_{\theta 
1}/B_{z1}  = 1$, $\beta \equiv 2p_1/B_1^2 = 100$, like in Fig.~\F{8}, but much larger 
thickness, $\delta \equiv r_2/r_1 - 1 = 1$, and mode numbers $m = 10$, $k = 50$, for 
fixed matching radius $r_0 = 0.5$. The radial profiles of the continua are shown in the 
top figure.}{\Fn{12}}
\end{figure}

\begin{figure}[ht]
\FIG{\begin{center} 
\includegraphics*[height=10.4cm]{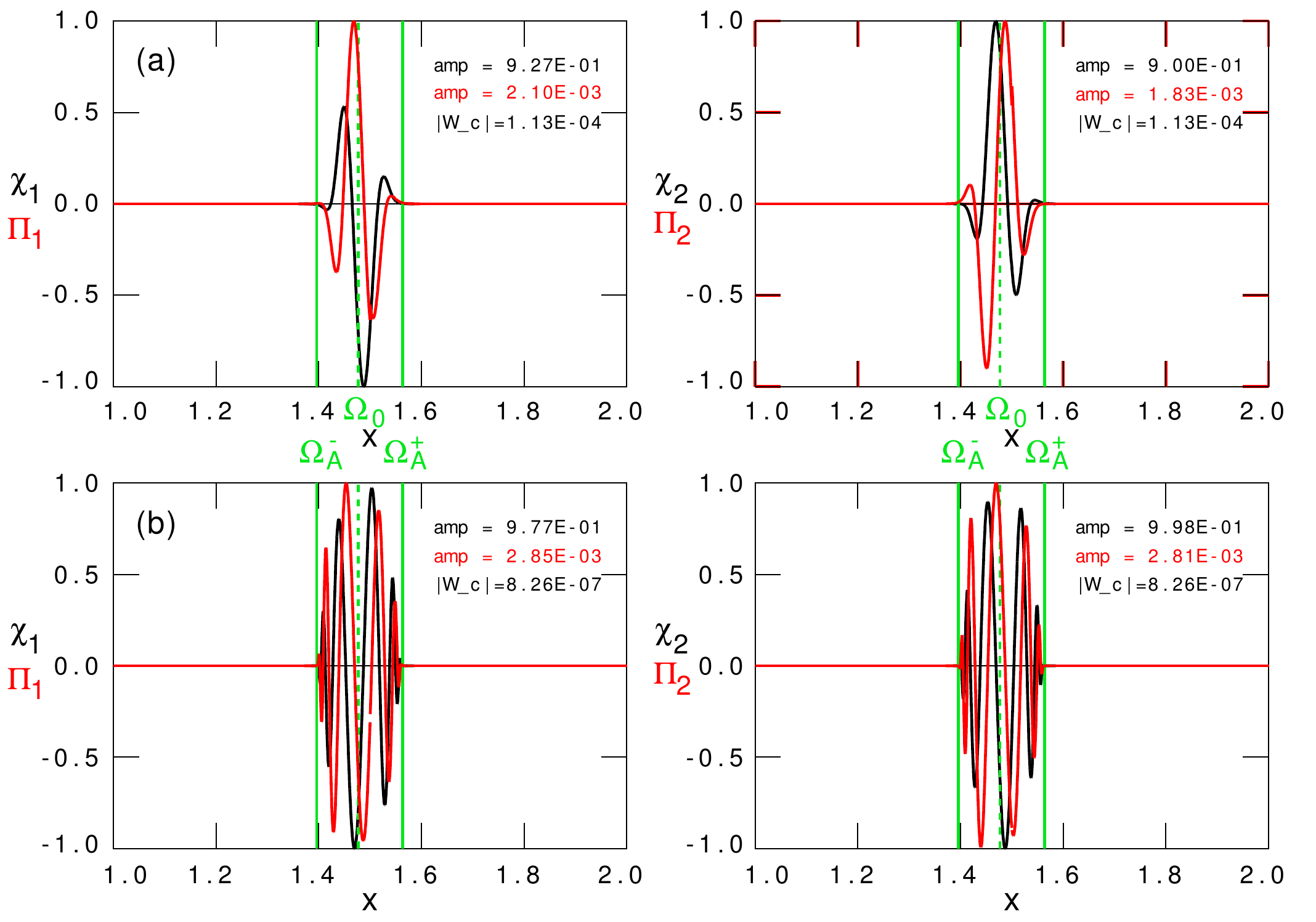}
\end{center}}
\caption{Quasi-discrete solutions corresponding to the Spectral Web of Fig.~\F{12}. The 
functions $\Pi_1$ and $\Pi_2$ display a tiny (invisible) jump such that $W_{\rm com} 
\ne 0$ but quite small, (a)~$\sigma = 5.5$, $\nu = 0.2$, (b)~$\sigma = 5.5$, $\nu = 
0.02$.}{\Fn{13}}
\end{figure}

In Fig.~\F{12}, the Spectral Web for the co-rotating SARIs is shown for the same 
equilibrium and mode numbers investigated in Sec.~\S{IVB}, except that the layer is 
widened to $\delta = 1$. In fact, concentration of the Spectral Web contours is found for a 
small range $\sigma \approx \Omega_0(r_0)$ about the Doppler frequency associated 
with the fixed matching point $r = r_{\rm mix} = r_0$ of the mixed integration scheme 
described in Sec.~\S{IID}. If that point is moved to the left where $\sigma \approx 
\Omega_0(r_2)$, the `outer modes' clustering towards the forward Alfv\'en singularity 
$\Omega_{\rm A}^+(r_2)$ appear. If that point is moved to the right where $\sigma 
\approx \Omega_0(r_1)$, the `inner modes' clustering towards the backward Alfv\'en 
singularity $\Omega_{\rm A}^-(r_1)$ appear. This checks with the above remark on the 
localization of the modes at the edges of the layer. Unfortunately, in the whole range in 
between those points, the Spectral Webs found just consist of {\em densely packed 
solution paths and conjugate paths, but they never cross.} This appears to suggest that 
there are simply no instabilities to be found in the middle of the layer! 

However, for each value of $\omega$ in the region of the `condensed' Spectral Web, 
solutions of the accretion disk eigenvalue problem~\E{10}, \E{11} are found (as e.g.\ 
illustrated in Fig.~\F{13}) that do not correspond to discrete eigenvalues but they do have 
quite small values of the complementary energy, typically $W_{\rm com} \sim 10^{-5}$ 
for this case. They also do have the required property of localization in the middle of the 
interval: they are `confined' by the two singularities, so that they may be characterized as 
being both `inner' and `outer'. Also note that the Spectral Web of Fig.~\F{12} is for fixed 
$r_0  = 1.5$. Varying that value of $r_0$, the whole range of the Doppler frequency, 
from $\sigma = 3.49$ to $\sigma = 9.87$, is covered with complex values of $\omega$ 
that all are close to genuine eigenvalues. We will now chose parameters such that this 
behavior is maximized so that modes appear for values of $\omega$ that cannot be 
distinguished from genuine eigenvalues since the measure of distinction, viz.\ the value 
of the complementary energy $W_{\rm com}$, will be extremely small (less than 
machine accuracy in most numerical results).  For that reason, they may be termed {\em 
quasi-discrete `eigenvalues'.} It is important to notice that, in the dynamical environment 
of an accretion disk, excitation energies of this order of magnitude are readily available 
so that the distinction between actual eigenvalues and quasi-discrete `eigenvalues' 
becomes irrelevant.

\begin{figure}[ht]
\FIG{\begin{center} 
\includegraphics*[height=14.8cm]{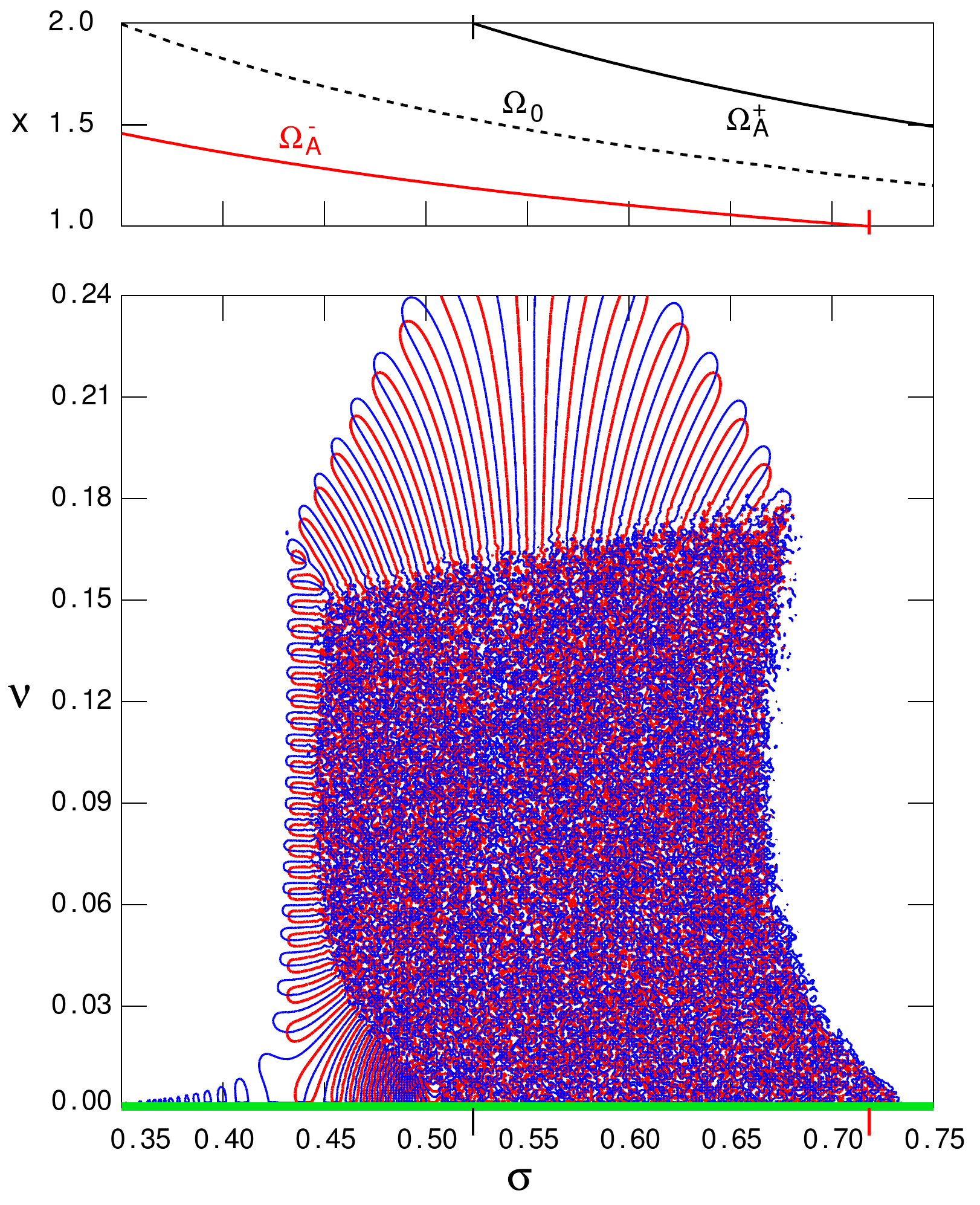}
\end{center}}
\caption{Fragmentation of the Spectral Web of the co-rotating Super-Alfv\'enic 
Rotational Instabilities for an accretion disk with  parameters $\epsilon \equiv B_1 = 
0.045$, $\mu_1 \equiv B_{\theta 1}/B_{z1}  = 10$, $\beta \equiv 2p_1/B_1^2 = 10$, 
$\delta \equiv r_2/r_1 - 1 = 1$, and mode numbers $m = 1$, $k = 50$, for fixed matching 
radius $r_0 = 0.5$. The radial profiles of the continua are shown above the main 
figure.}{\Fn{14}}
\end{figure}

\begin{figure}[ht]
\FIG{\begin{center} 
\includegraphics*[height=14.8cm]{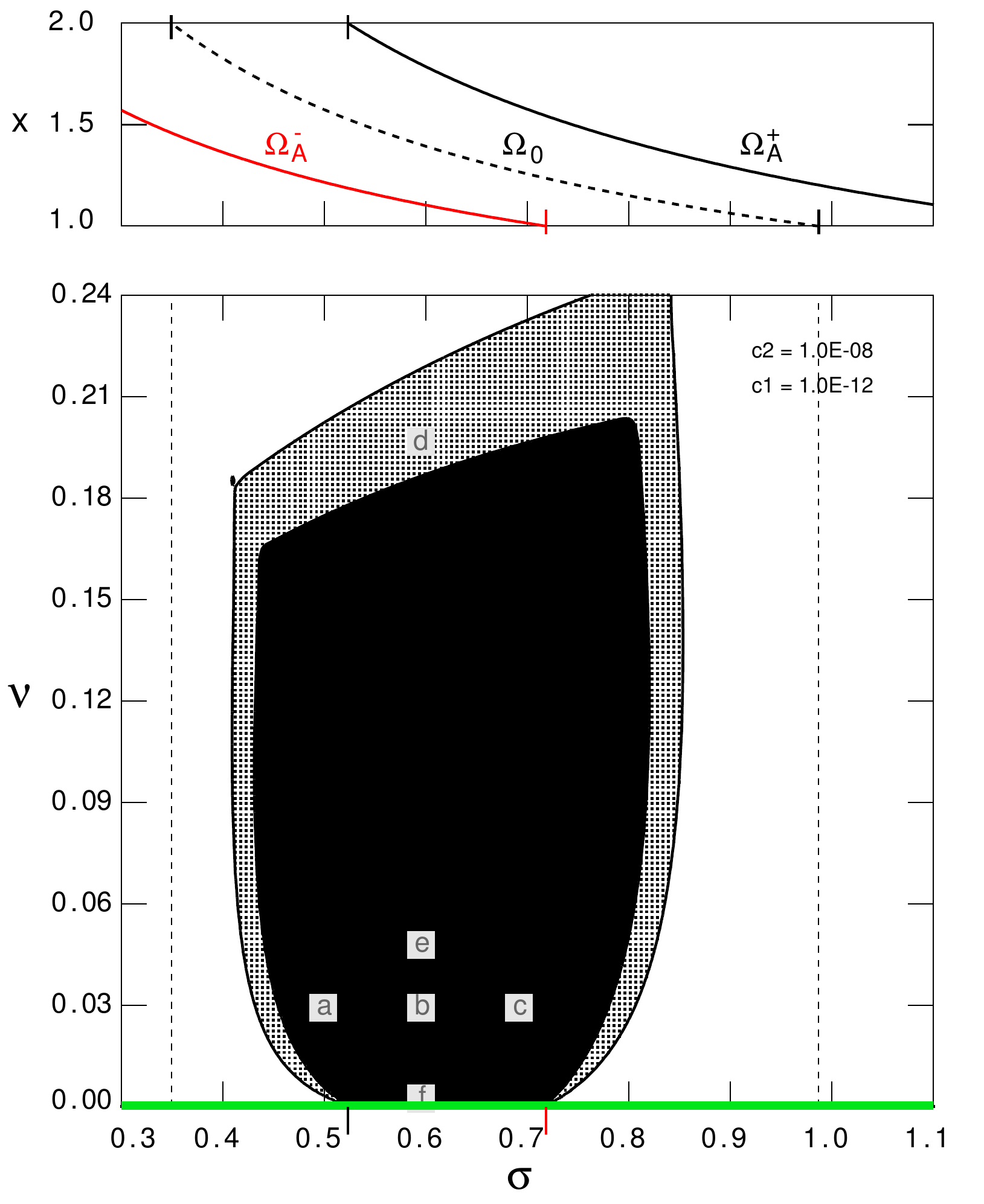}
\end{center}}
\caption{Quasi-continuum of the co-rotating Super-Alfv\'enic Rotational Instabilities for 
the same parameters as for Fig.~\F{14}, except that $r_0$ is varied to follow the radial 
variation of $\Omega_0$ (the dashed curve in the top frame).   Every point of the black 
area corresponds to a quasi-discrete `eigenmode' with $|W_{\rm com}| < c_1 \equiv 
10^{-12}$, every point of the grey area corresponds to a quasi-discrete `eigenmode' with 
$c_1 < |W_{\rm com}| \le c_2 \equiv 10^{-8}$. The labels (a)--(f) refer to the solutions 
presented in Fig.~\F{16}.} {\Fn{15}}
\end{figure}

In the next example, the equilibrium and mode numbers are chosen such that the Spectral 
Web degenerates into a completely fragmented configuration with myriads of small-scale 
structures of the solution and conjugate paths that never appear to cross though 
(Fig.~\F{14}). The values of $\epsilon$ and $p_1 \equiv \half \beta \epsilon^2$ are 
chosen about a factor of $2$ smaller than in the equilibrium exploited for Fig.~\F{12}. 
The ratio $\mu_1$ of the toroidal over the longitudinal magnetic field  is chosen a factor 
10 larger so that a completely different magnetic structure of the accretion disk 
equilibrium is obtained. Most important, the ratio $k/m$ of the longitudinal and toroidal 
mode numbers is chosen a factor $10$ larger by keeping the value of $k$ the same, but 
decreasing the value of $m$ to $1$. The rationale behind this choice will become clear in 
the analysis of exponential factors in Section~\S{VB}.

The fragmentation of the Spectral Web shown in Fig.~\F{14} appears to represent a 
disastrous collapse of the concepts we have presented in this paper. However, quite the 
opposite is true. The very reason of the fragmentation is not the disappearance of 
crossings of the two paths, but the fact that whether or not they cross becomes irrelevant 
because the absolute value $|W_{\rm com}|$ of the complementary energy becomes 
extremely small there. Since the complementary energy distinguishes between 
quasi-modes (small $|W_{\rm com}|$) and actual eigenvalues (vanishing $|W_{\rm 
com}|$),  that distinction becomes irrelevant when the value of $|W_{\rm com}|$ 
becomes of the order of the machine accuracy of the computer exploited to derive these 
results. This is the case for the quasi-modes that are found anywhere in the fragmented 
region.

At this point, a significant simplification of the method of the Spectral Web is in order. 
Since the separate solution and conjugate paths, and their crossings, make no sense 
anymore for this type of Super-Alfv\'enic Rotational Instabilities, it is expedient to 
replace their plots with the contours of $|W_{\rm com}|$ instead. Those contours 
delineate regions of the complex $\omega$-plane where quasi-modes are to be found that 
cannot be distinguished from genuine eigenmodes when reasonable measures of 
accuracy, or better, reasonable measures for a closed system are exploited. Remember 
that $W_{\rm com}$ is the amount of energy that needs to be provided or extracted to 
bring the system in resonance. That energy vanishes for eigenmodes (when  the system is 
closed), and it is ridiculously small for quasi-modes (when the system is nearly closed). 
As already mentioned, such small amounts of the complementary energy are readily 
available from fluctuations in any dynamical system. In Fig.~\F{15}, the resulting 
diagram is shown of contours of $|W_{\rm com}| = 10^{-8}$ and $|W_{\rm com}| = 
10^{-12}$ for the same case that produced the Spectral Web of Fig.~\F{14}. One 
extension was exploited though: the matching radius $r_0$ joining left and right solutions 
of the ODEs (which was kept fixed for Fig.~\F{14}) was varied to correspond to the 
position of the Doppler `resonance', $\sigma - \Omega_0(r_0) = 0$. This way, a 
significantly larger area of the $\omega$-plane with quasi-modes is found. That extension 
could have been exploited for Fig.~\F{14} as well, but the corresponding figure is 
omitted here because the resulting Spectral Web is even more convoluted than the one 
shown. Moreover, the quasi-continuum representation of the complementary energy 
contours is just as instructive as the Spectral Web contours.

\begin{figure}[ht]
\FIG{\begin{center} 
\includegraphics*[height=14.4cm]{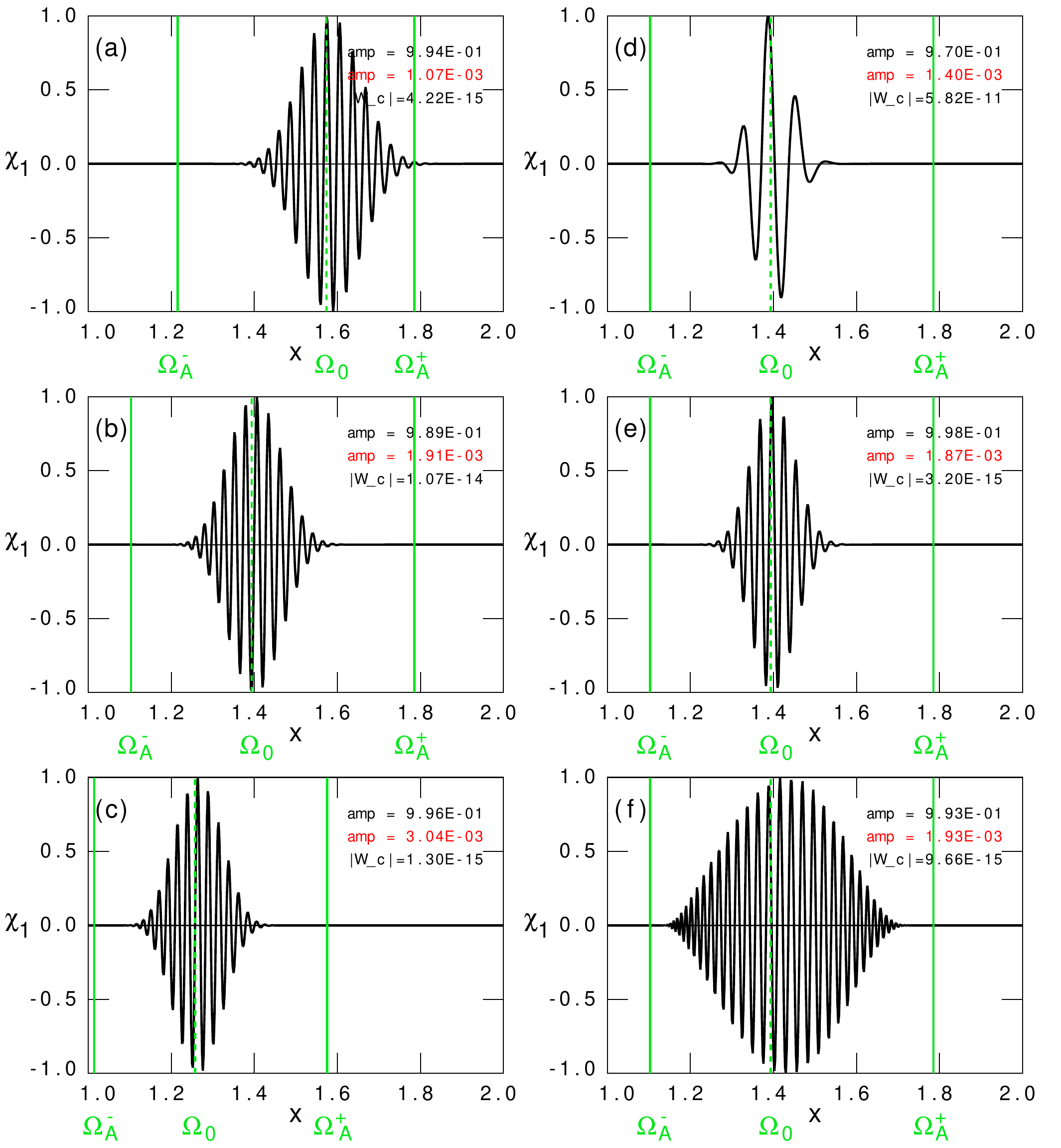}
\end{center}}
\caption{Quasi-continuum `eigenfunctions' corresponding to Fig.~\F{15}. Left column: 
fixed growth rate $\nu = 0.03$ but varying frequency, (a)~$\sigma = 0.5$, (b)~$\sigma = 
0.6$, (c)~$\sigma = 0.7$. Right column: fixed frequency $\sigma= 0.6$ but varying 
growth rate, (d)~$\nu = 0.2$, (e)~$\nu = 0.05$, (f)~$\nu = 0.005$. Similar pictures of 
$\chi_2$, $\Pi_1$  and $\Pi_2$ have been omitted, the amplitudes of $\Pi_1$ are 
indicated in red.}{\Fn{16}}
\end{figure}

In Fig.~\F{16} the quasi-continuum `eigenfunctions' of the SARIs are shown for values 
of~$\omega$ corresponding to the labels (a)--(f) of Fig.~\F{15}. The left column shows 
that, for fixed growth rate $\nu$, the localization of the modes shifts from the right to the 
left position in the disk following the value of $\sigma$ corresponding to the Doppler 
frequency in the top frame. The right column shows that, for fixed frequency $\sigma$, 
the `eigenfunctions' become ever more oscillatory as the continua on the real axis are 
approached. Note that, with the exception of the one shown in Fig.~\F{16}(d), all 
`eigenfunctions' of Fig.~\F{16} have values of $|W_{\rm com}| \sim 10^{-14}$: the 
omitted functions $\Pi_1$ and $\Pi_2$ would show a discontinuity of this order of 
magnitude at the matching radius $r_0$ that is completely invisible! \HG{}{For the same 
reason, the position and the amplitude of that discontinuity, which depend on the choice 
of $r_{\rm mix}$, do not influence the visible shape of those functions (as long as 
$r_{\rm mix}$ is not close to $r_1^*$ or $r_2^*$).} Consequently and quite 
conveniently, all these quasi-continuum `eigenfunctions' are obtained by a {\em single 
shot solution} of the accretion disk eigenvalue problem~\E{10}, \E{11} for arbitrary 
values of $\omega$ in the quasi-continuum range. This is in sharp contrast to all genuine 
eigenfunctions shown in the previous sections that are obtained by {\em time-consuming 
iteration} for isolated values of $\omega$ to get $|W_{\rm com}|$ to become small 
enough to qualify for an eigenvalue. How the single shot `miracle' comes about is 
described in the analysis of the next section.

In summary: We have found a {\em genuine} 2D continuum (without any holes!) of 
quasi-discrete modes in the lower unstable part of the complex $\omega$-plane. This 
continuum will be called {\em `quasi-continuum'} for short. The modes have small jumps 
in the total pressure perturbation such that the complementary energy is extremely small. 
This continuum is well to be distinguished from the classical continuous spectrum, where 
the modes have  non-square integrable singularities, \HG{}{like in the 2D continuous 
spectrum found by \Ron{Lifschitz1997} for hyperbolic flows}. As will be clarified in the 
next section, the only condition in addition to the general instability criteria for the SARIs 
formulated in the previous sections, is that the ratio of the mode numbers is large, $|k|/|m| 
\gg 1$ (recall that all variables have been made dimensionless, so that $k$ is actually 
$\bar{k} \equiv k r_1$). The superposition of {\em all} quasi-continua for all values of 
$k$ and $m$ satisfying this condition will occupy a vast area of the complex 
$\omega$-plane!


\smalltext{Shortcut \#2}{Again, the next section, which provides a full asymptotic 
analysis of the 2D quasi-continua, could be skipped on first reading. The physical 
implications of the quasi-continua are picked up in Section~\S{VC}, where we explain 
that these local instabilities, completely unaware of any artificial boundary conditions, set 
up intricate electric current distributions in between the forward and backward resonant 
locations, fully aware of the local magnetic field orientation. This makes them resemble 
truly  local, growing wave packages that surf super-Alfv\'enically along the disk.}

\subsection{Asymptotic analysis of the quasi-continuum SARIs}{\Sn{VB}}

For the analysis of the quasi-continuum SARIs a considerable revision of the analysis of 
Sec.~\S{IVB} is in order. The Legendre equation is no longer applicable since the 
variation of the coefficients of the accretion disk ODE~\E{23} over the interval cannot 
be approximated anymore as was done in that section. However, the dominant 
contribution of the Alfv\'en singularities remains and, with it, some striking similarities of 
the solutions. To demonstrate that, we will write the solutions of the general spectral 
equation in the same form as in the first part of Eq.~\E{43}:
\BEQ \chi = F_1(z) + C F_2(z) \,,\En{63}\EEQ
where the complex variable $z$ of Sec.~\S{IVB} is exploited again. Now, $F_1$ and 
$F_2$ are not Legendre functions, but the solutions of the accretion disk ODE~\E{23}, 
or rather the general spectral differential equation~\E{10}, subject to the boundary 
conditions~\E{11}. As we have seen, the latter effectively may be replaced by the 
boundary conditions
\BEQ \chi(r_1^*) = 0 \;\;\hbox{(left)}\,,\qquad \chi(r_2^*) = 0 \;\;\hbox{(right)} 
\En{64}\EEQ
at the `virtual walls' at $r = r_1^*$ and $r = r_2^\star$. The positions of these `walls' are 
determined by the backward and forward Alfv\'en continuum singularities $\sigma = 
\Omega_{\rm A}^-( r_1^*)$ and $\sigma = \Omega_{\rm A}^+( r_2^*)$ for co-rotating 
modes, similar to that shown in Fig.~\F{9} (but with two intersections with the continua 
whereas the overlap region $\Omega_{\rm A}^+( r_2) \le \sigma \le \Omega_{\rm A}^-( 
r_1)$ is now significantly larger than depicted in that figure). For the contra-rotating 
modes, the positions of $\Omega_{\rm A}^-$ and $\Omega_{\rm A}^+$ are 
interchanged.

\subsubsection{Large and small solutions}{\Sn{VB1}

We again expand the singular factor $\widetilde{\omega}^2 - \omega_{\rm A}^2 \equiv 
[\omega - \Omega_{\rm A}^+(r)] [\omega - \Omega_{\rm A}^-(r)]$, like in Eqs.~\E{34} 
and \E{35}, with $r_1 \rightarrow r_1^*$, but of course without the averaging procedure 
exploited in Sec.~\S{IVB} that gave the Legendre equation. For this case, we introduce 
the scaled radial variable $\tilde{x} \equiv (r - r_1^*)/\delta^*$, where $\delta^* \equiv 
r_2^* - r_1^*$, generalizing Eq.~\E{33}, and  again relate that variable to the complex 
variable $z \equiv (2 \tilde{x} - 1 - \lambda - \hat{\lambda})/(1 - \lambda + 
\hat{\lambda})$. Since now $\sigma = \Omega_{\rm A}^-(r_1^*) = \Omega_{\rm 
A}^+(r_2^*)$, the parameters $\lambda$ and $\hat{\lambda}$ are both imaginary (the 
parameter $\alpha$ of Eq.~\E{44} no longer appears):
\BEQ \lambda = -{\rm i}\hs|\lambda| \equiv \frac{{\rm i}\hs\nu}{\delta^* 
\hs\Omega_{\rm A}^{-\prime}(r_1^*)} \,, \qquad \hat{\lambda} = -{\rm 
i}\hs|\hat{\lambda}| \equiv \frac{{\rm i}\hs\nu}{\delta^* \hs\Omega_{\rm 
A}^{+\prime}(r_2^*)} = -{\rm i}{\hs}|\lambda|/h \,, \quad h \equiv \frac{\Omega_{\rm 
A}^{+\prime}(r_2^*)}{\Omega_{\rm A}^{-\prime}(r_1^*)} \approx 
\Big(\frac{r_2^*}{r_1^*}\Big)^{-5/2}\,.\En{65}\EEQ
Hence, the images of the points $\tilde{x} = 0$ and $\tilde{x} = 1$ will be situated 
vertically above the points $z = -1$ and $z = 1$ in the $z$-plane depicted in Fig.~\F{10}. 
As before, the approximations are for Alfv\'en frequency small compared to the Doppler 
frequency, $|\omega_{\rm A}| \ll |\Omega_0|$, but that assumption is not really 
necessary for the analysis.
  
We exploit Frobenius expansions about the singular points $z = -1$ and $z = +1$. Close 
to the backward singularity on the left, where $z \approx - 1 + 2 (\tilde{x} - \lambda$) 
and $\tilde{x} \ll 1$, the dominant part of the accretion disk ODE then yields a left 
solution $\chi_{\rm L}$ built of two independent solutions $[F_1]_{\rm L}$ and 
$[F_2]_{\rm L}$ with imaginary indices:
\BEQAR
&& \frac{d}{dz} \Big[ (z + 1) \hs\frac{d\chi}{dz} \Big] + \Big[ \,\ldots\, +  
\frac{v_1^2}{4\hs(z + 1)}\hs\Big] \hs\chi = 0 \,, \qquad v_1 \equiv \Big|\frac{2k 
\Omega(r_1^*)}{\Omega_{\rm A}^{-\prime}(r_1^*)}\Big| \approx \Big|\frac{4k 
r_1^*}{3m}\Big| \non\\[1mm]
&&\qquad\Rightarrow\quad \chi_{\rm L}: \quad[F_{1,2}]_{\rm L} \equiv \big[-\half (z 
+ 1)\big]^{\hs\pm\frac{1}{2}{\rm i\hs} v_1} f_{1,2}(z)\,. \En{66}\EEQAR
Close to the forward singularity on the right, where $z \approx 1 - 2 (1 - \tilde{x} + 
\hat{\lambda})$ and $1 - \tilde{x} \ll 1$, the accretion disk ODE then yields a right 
solution $\chi_{\rm R}$ built of two other independent solutions $[F_1]_{\rm R}$ and 
$[F_2]_{\rm R}$ with imaginary indices:
\BEQAR
 &&\frac{d}{dz} \Big[(z - 1)\hs\frac{d\chi}{dz} \Big] + \Big[ \,\ldots\, + 
\frac{v_2^2}{4\hs(z - 1)}\hs\Big] \hs\chi = 0 \,, \qquad v_2 \equiv \Big|\frac{2k 
\Omega(r_2^*)}{\Omega_{\rm A}^{+\prime}(r_2^*)}\Big|  \approx \Big|\frac{4k 
r_2^*}{3m}\Big| \non\\[1mm]
&&\qquad\Rightarrow\quad \chi_{\rm R}: \quad[F_{1,2}]_{\rm R} \equiv \big[\half (z - 
1)\big]^{\hs\mp\frac{1}{2}{\rm i\hs} v_2} g_{1,2}(z) \,. \En{67}\EEQAR
Here, $f_{1,2}(z)$ and $g_{1,2}(z)$ are regular analytic functions with power series 
expansions that converge in overlapping regions of the $z$-plane, similar to the 
hypergeometric functions exploited in Sec.~\S{IVB} as illustrated in Fig.~\F{10}. We do 
not need any information about these functions other than that they are regular and 
normalized by imposing the conditions $f_{1,2}(-1) = 1$ and $g_{1,2}(1) = 1$. In 
principle, they could be obtained from the numerical solution of the ODE that is 
exploited to construct the Spectral Web. 

To avoid confusion, it is important to spend a few words on the notation used to 
distinguish the different quantities. In the parameters $r_{1,2}^*$ and $v_{1,2}$, the 
subscripts $1$ and $2$ refer to the left and right end points of the interval $0 \le \tilde{x} 
\le 1$. For the functions $F_{1,2}$, $f_{1,2}$ and $g_{1,2}$, they refer to the two 
independent solutions of the ODE, where the upper sign of the powers in Eqs.~\E{66} 
and \E{67} corresponds to the subscript $1$ and the lower sign corresponds to the 
subscript $2$. The use of the notation $F_{1,2}$ for the left as well as for the right 
solutions will be justified shortly.

At this point, it should be observed that the distinguishing feature of the present analysis, 
compared to that of Sec.~\S{IVB} exploiting the Legendre functions, is the magnitude of 
the parameters $v_1$ and $v_2$, which is now much larger: $v_1 \gg 1$ and $v_2 \gg 
1$. In terms of the physical variable $\tilde{x}$, this implies that the above solutions 
may be termed `large' (indicated by the upper sign) and `small' (indicated by the lower 
sign), exploiting the terminology introduced by~\Ron{New60} for similar singular 
functions. For example, from Eq.~\E{66}:
\BEQ [F_{1,2}]_{\rm L} \approx \big[\!-{\rm i}\hs(|\lambda| - {\rm i} 
\tilde{x})\hs\big]^{\hs\pm\frac{1}{2}{\rm i\hs} v_1} f_{1,2}(z) 
= {\rm e}^{\pm\frac{1}{4}\pi v_1}
\hs{\rm e}^{\pm\frac{1}{2}v_1\arctan(\tilde{x}/|\lambda|)}
\hs{\rm e}^{\pm\frac{1}{2}{\rm i\hs}v_1\ln\sqrt{|\lambda|^2 + \tilde{x}^2}} 
f_{1,2}(z)\,,\En{68} \EEQ 
where, away from $\tilde{x}  =  0$, the second exponential factor gives another boost (up 
or down) of the first factor:
\BEQ {\rm e}^{\pm\frac{1}{2}v_1\arctan(\tilde{x}/|\lambda|)} = \big\{\, 1 \;\hbox{(for 
$\tilde{x} = 0$)} \;\;\rightarrow\;\; {\rm e}^{\pm\frac{1}{4}\pi v_1} \;\hbox{(for 
$|\lambda| \ll \tilde{x} \ll 1$)} \,\big\}\,.\En{69} \EEQ
Analogously, from Eq.~\E{67} for the right solutions at $\tilde{x} = 1$:
\BEQ [F_{1,2}]_{\rm R} \approx \big[{\rm i}\hs(|\hat{\lambda}| + {\rm i} (1 - 
\tilde{x}))\hs\big]^{\hs\pm\frac{1}{2}{\rm i\hs} v_2} g_{1,2}(z) = {\rm 
e}^{\pm\frac{1}{4}\pi v_2}
\hs{\rm e}^{\pm\frac{1}{2}v_2\arctan[(1 - \tilde{x})/|\hat{\lambda}|]}
\hs{\rm e}^{\mp\frac{1}{2}{\rm i\hs}v_2\ln\sqrt{|\hat{\lambda}|^2 + (1 - \tilde{x})^2}} 
g_{1,2}(z) \,,\En{70} \EEQ 
where, away from $\tilde{x}  =  1$, the second exponential factor again gives a boost of 
the first factor:
\BEQ {\rm e}^{\pm\frac{1}{2}v_2\arctan[(1 - \tilde{x})/|\hat{\lambda}|]} = \big\{\, 1 
\;\hbox{(for $\tilde{x} = 1$)} \;\;\rightarrow\;\; {\rm e}^{\pm\frac{1}{4}\pi v_2} 
\;\hbox{(for $|\hat{\lambda}| \ll 1 - \tilde{x} \ll 1$)} \,\big\}\,.\En{71} \EEQ
Note that, with the parameters used to construct the Spectral Web and quasi-continuum of 
Figs.~\F{14} and \F{15}, the indices themselves are large, $v_1 = 84.42$ and $v_2 = 
100.8$, but {\em the exponential factors are huge:} ${\rm e}^{\frac{1}{4}\pi v_1} = 3.9 
\times10^{57}$ and ${\rm e}^{\frac{1}{4}\pi v_2} = 6.2\times 10^{68}$, i.e.\ perfect 
for expansions! This implies that the solution~$[F_1]_{\rm L}$ is large at $\tilde{x} = 
0$ and huge away from it, whereas  $[F_2]_{\rm L}$ is small at $\tilde{x} = 0$ and tiny 
away from it. Similarly, the solution~$[F_1]_{\rm R}$ is large at $\tilde{x} = 1$ and 
huge away from it, whereas  $[F_2]_{\rm R}$ is small at $\tilde{x} = 1$ and tiny away 
from it. The crucial conclusion is that {\em the solution that is large at the left end point 
is also large at the right end point, and similarly for the small solution.} Hence, apart 
from a constant multiplicator, the functions $[F_1]_{\rm L}$ and $[F_1]_{\rm R}$ refer 
to the same independent (large) solution of the ODE, and the functions $[F_2]_{\rm L}$ 
and $[F_2]_{\rm R}$ refer to the other independent (small) solution. This is the 
justification for choosing the same symbol for the solutions left and right, inspired by the 
nomenclature of Sec.~\S{IVB} where the different representations~\E{45} and \E{51} of 
the same function $F_1(z)$ already illustrate the crucial property just discussed.

\subsubsection{No discrete modes}{\Sn{VB2}

With $F_1(z)$ and $F_2(z)$ two independent solutions of the accretion disk 
ODE~\E{23}, we can now construct the solutions that satisfy the left and the right BC, 
respectively. For the left solution,
\BEQ \chi_{\rm L}(\tilde{x}) = A \big\{[F_1(z)]_{\rm L} + C_1 [F_2(z)]_{\rm 
L}\big\}\,, \quad C_1 = |C_1| {\rm e}^{{\rm i}\varphi_1} \,, \En{72} \EEQ 
satisfaction of the BC at $\tilde{x} = 0$ yields
\BEQAR
&&\chi_{\rm L}(0) = A \big\{ {\rm e}^{\frac{1}{4}\pi v_1}
\hs{\rm e}^{\frac{1}{2}{\rm i\hs}v_1\ln|\lambda|} + |C_1| \hs{\rm e}^{{\rm 
i\hs}\varphi_1} \hs{\rm e}^{-\frac{1}{4}\pi v_1}
\hs{\rm e}^{-\frac{1}{2}{\rm i\hs}v_1\ln|\lambda|}\big\} = 0 \,,\non\\[1mm] 
&& \qquad\Rightarrow\quad |C_1| = {\rm e}^{\frac{1}{2}\pi v_1} \,,\qquad {\rm 
e}^{{\rm i\hs}\varphi_1} = - {\rm e}^{{\rm i\hs}v_1\ln|\lambda|} \,,
\En{73} \EEQAR
i.e.\ the large contribution is balanced by the small contribution at $\tilde{x} = 0$ by 
means of the huge factor $|C_1|$. However, away from that point, for $\tilde{x} \gg 
|\lambda|$, the `large' contribution completely dominates over the `small' one because of 
the second exponential factor:
\BEQ \chi_{\rm L}(\tilde{x}) = A \hs{\rm e}^{\frac{1}{4}\pi v_1} \hs{\rm 
e}^{\frac{1}{2}{\rm i\hs}v_1\ln|\lambda|} \hs\big\{ {\rm e}^{\frac{1}{4}\pi v_1}
\hs{\rm e}^{\frac{1}{2}{\rm i\hs}v_1\ln(\tilde{x}/|\lambda|)} f_1(z) - \hs{\rm e}^{-
\frac{1}{4}\pi v_1}
\hs{\rm e}^{-\frac{1}{2}{\rm i\hs}v_1\ln(\tilde{x}/|\lambda|)} f_2(z) \big\} 
\,, \En{74} \EEQ 
Hence, in the middle of the interval, in particular about the matching point $\tilde{x} = 
\tilde{x}_{\rm mix}$, {\em the left solution is dominated by the independent solution 
that is `large' at $\tilde{x} = 0$.} Completely analogously for the right solution,
\BEQ \chi_{\rm R}(\tilde{x}) = B \big\{[F_1(z)]_{\rm R} + C_2 [F_2(z)]_{\rm 
R}\big\}\,, \quad C_2 = |C_2| {\rm e}^{{\rm i}\varphi_2} \,, \En{75} \EEQ 
satisfaction of the BC at $\tilde{x} = 1$ yields
\BEQ |C_2| = {\rm e}^{\frac{1}{2}\pi v_2} \,,\qquad {\rm e}^{{\rm i\hs}\varphi_2} = - 
{\rm e}^{-{\rm i\hs}v_2\ln|\hat{\lambda|}} \,,
\En{76}\EEQ
and the solution away from that point, for $1 - \tilde{x} \gg |\hat{\lambda}|$, becomes
\BEQ \chi_{\rm R}(\tilde{x}) = B \hs{\rm e}^{\frac{1}{4}\pi v_2} \hs{\rm 
e}^{\frac{1}{2}{\rm i\hs}v_2\ln|\hat{\lambda|}} \hs\big\{ {\rm e}^{\frac{1}{4}\pi v_2}
\hs{\rm e}^{\frac{1}{2}{\rm i\hs}v_2\ln[(1-\tilde{x})/{|\hat{\lambda}|}]} g_1(z) - 
\hs{\rm e}^{-\frac{1}{4}\pi v_2}
\hs{\rm e}^{-\frac{1}{2}{\rm i\hs}v_2\ln[(1-\tilde{x})/|\hat{\lambda}|]} g_2(z) \big\} 
\,, \En{77} \EEQ 
so that in the middle of the interval, {\em the right solution is dominated by the 
independent solution that is `large' at $\tilde{x} = 1$.}

It remains to match the two solutions \E{74} and \E{77} at the matching point $\tilde{x} 
= \tilde{x}_{\rm mix}$, which we will choose to be the Doppler point $r = r_0$ so that 
$\sigma = \Omega_0(r_0) = \Omega_{\rm A}^-(r_1^*) = \Omega_{\rm A}^+(r_2^*)$. 
For an eigenvalue, this implies that we have to impose the conditions 
$\doublel\chi\doubler \equiv \chi_{\rm R}(\tilde{x}_{\rm mix}) - \chi_{\rm 
L}(\tilde{x}_{\rm mix}) = 0$ and $\doublel\chi'\doubler \equiv \chi^\prime_{\rm 
R}(\tilde{x}_{\rm mix}) - \chi^\prime_{\rm L}(\tilde{x}_{\rm mix}) = 0$. Separating 
the large (exponential) contributions from the small (order unity) contributions, these 
jumps are determined by
\BEQAR
&&\doublel\chi\doubler = B \big(b_1{\rm e}^{\frac{1}{2}\pi v_2} + b_2 \big) - A 
\big(a_1{\rm e}^{\frac{1}{2}\pi v_1} + a_2 \big) \,,\En{78}\\[2mm]
&&\doublel\chi'\doubler = B \big(p_1 b_1{\rm e}^{\frac{1}{2}\pi v_2} + p_2 b_2 \big) 
- A \big(p_1 a_1{\rm e}^{\frac{1}{2}\pi v_1} + p_2 a_2 \big) \,,\quad p_1 \equiv 
\frac{F^\prime_1}{F_1} \,,\quad p_2 \equiv \frac{F^\prime_2}{F_2} \,.
\En{79} \EEQAR
where we do not need any information about the constants $a_{1,2}$, $b_{1,2}$ and 
$p_{1,2}$ other than that they are complex and of order unity. Formally, they are 
supposed to be known through the solutions of the ODE. It is important that, in contrast 
to the pairs $\{a_1, a_2\}$ and $\{b_1, b_2\}$ which are different due to the different left 
and right representations, the logarithmic derivatives $p_1$ and $p_2$ are the same left 
and right since they refer to the same function (apart from the constant amplitude factor). 
It is trivial now to impose continuity of $\chi$ at $\tilde{x} = \tilde{x}_{\rm mix}$. 
From the expression~\E{78}, this just determines the right amplitude~$B$ in terms of the 
left amplitude~$A$. Inserting this into the jump expression~\E{79} for $\chi'$, and 
exploiting the smallness of ${\rm e}^{-\frac{1}{2}\pi v_1}$ and ${\rm e}^{-
\frac{1}{2}\pi v_2}$, the leading order term cancels but a smaller part remains: 
\BEQ \doublel\chi'\doubler \approx A a_1 {\rm e}^{\frac{1}{2}\pi v_1} (p_1 - 
p_2)\big[(a_2/a_1)\hs {\rm e}^{-\frac{1}{2}\pi v_1} - (b_2/b_1) \hs{\rm e}^{-
\frac{1}{2}\pi v_2}\big]  \,. \En{80}\EEQ
This jump is small compared to the order of magnitude of the function~$\chi$ itself, $\chi 
\sim A a_1 {\rm e}^{\frac{1}{2}\pi v_1}$, because of the small term in square brackets. 
In general, this jump does not vanish, as would be required for a discrete eigenvalue!

Imposing continuity of the two jumps would be similar to the procedure applied in 
Sec.~\S{IVB}, except that it was hidden there because the Legendre functions 
automatically transport the solution continuously from the right to the left. A more 
fundamental difference is that the eigenvalue was complex there, $\lambda = 
|\lambda|\hs{\rm e}^{\hs{\rm i\hs}\alpha}$ with two free parameters, whereas here, 
$\lambda = - {\rm i} \hs|\lambda|$ with only one free parameter. Whereas continuity of 
$\doublel\chi\doubler$ can always be satisfied, in general (except maybe for coincidental 
values of the parameters), continuity of the expression~\E{79} for 
$\doublel\chi'\doubler$ cannot be satisfied because there are not enough free parameters 
in this problem. Stated differently, the original eigenvalue problem~\E{63}--\E{64} will 
not have a solution, as is also evident from the fact that it contains only one complex 
constant $C$, whereas the present analysis contains two essentially different complex 
constants $C_1$ and $C_2$. We wind up with a very negative conclusion: {\em In 
general, there are no eigenvalues in the region we are looking for, viz.\ away from the 
edges of the continua.}

\subsubsection{Instead, a continuum of quasi-modes emerges}{\Sn{VB3}

As has been illustrated already by Figs.~\F{14}--\F{16}, the Spectral Web method 
provides a powerful way out of this conundrum. So far, eigenvalues were determined by 
the condition that the complementary energy vanishes, $W_{\rm com} = 0$. This is 
associated with the intersections of the solution paths with the conjugate paths, which 
conspicuously are absent now. For the mixed integration scheme, exploiting the left and 
right solutions of the accretion disk ODE, the complementary energy is given by the 
expression~\E{19}, involving the jump $\doublel\Pi\doubler$ of the total pressure 
perturbation, defined above Eq.~\E{12}. This expression assumes that the left and right 
solutions have been matched by imposing $\doublel\chi\doubler = 0$, so that 
\BEQAR
 && \doublel\Pi\doubler \equiv Q_0 \hs\doublel\chi'\doubler \,, \qquad Q_0 \equiv - 
(N/D)_{r_0} \approx \Big[\frac{\rho(\widetilde{\omega}^2 - \omega_{\rm 
A}^2)}{r(m^2/r^2 + k^2)} \Big]_{r_0} = -\frac{\rho_0(\omega_{\rm A, 0}^2 + 
\nu^2)}{r_0(m^2/r_0^2 + k^2)} \non\\[2mm]
&& \qquad\Rightarrow\quad W_{\rm com} = - \pi \Delta z \hs|\chi_0|^2  
\hs\frac{\doublel\Pi\doubler}{\chi_0} = \pi \Delta z \hs|\chi_0|^2 {\hs}Q_0 \hs\hs\frac{ 
\doublel\chi'\doubler}{\chi_0} \,.\En{81} \EEQAR
Inserting $\doublel\chi'\doubler$ from Eq.~\E{80} and $\chi_0$ from the second term of 
Eq.~\E{78}  then yields
\BEQAR
W_{\rm com} &\approx& \pi \Delta z \hs|\chi_0|^2{\hs}Q_0 \hs\hs(p_1 - 
p_2)\big[(a_2/a_1)\hs {\rm e}^{-\frac{1}{2}\pi v_1} - (b_2/b_1) \hs{\rm e}^{-
\frac{1}{2}\pi v_2}\big] \non\\[2mm]
&\approx& \pi \Delta z \hs|\chi_0|^2{\hs}Q_0 \hs\hs (p_1 - p_2)\hs(a_2/a_1) \hs{\rm 
e}^{-\frac{1}{2}\pi v_1} \quad\Rightarrow\quad  |W_{\rm com}| \sim {\rm e}^{-
\frac{1}{2}\pi v_1} \,,
\En{82}\EEQAR
where the largest (the one with $v_1$) of the two small exponentials survives. The very 
last expression exhibits the key result: {\em asymptotically, the complementary energy is 
exponentially small!} For example, $|W_{\rm com}| = 6.7 \times 10^{-50}$ when $\nu 
\ll 1$ in Figs.~\F{14} and \F{15}.

In conclusion, the asymptotic analysis confirms the numerical results found in the 
previous section~\S{VA}: The Spectral web `condenses' into a 2D continuum of 
quasi-discrete modes with eigenfunctions that cannot be distinguished from genuine 
eigenfunctions because $|W_{\rm com}|$ is extremely small, i.e.\ the discontinuity of 
$\chi'$ at the matching radius is completely determined by the solution that is 
exponentially small everywhere. 

It should be noted that the asymptotic analysis is for $\nu \rightarrow 0$, whereas the 
numerical results shown in Sec.~\S{VA} are for finite values of $\nu$, away from the 
continua on the real axis. For these values of $\nu$, the quasi-modes have their maximum 
amplitudes in the middle of the interval, as illustrated by all `eigenfunctions' shown in 
Figs.~\F{16}. Hence, their oscillatory properties are rather described by expanding the 
accretion disk ODE about the Doppler point $r = r_0$ in the middle than about the 
Alfv\'en frequencies at $r_1^*$ and $r_2^*$. This yields
\BEQAR 
&& \frac{d}{d\tilde{x}} \Big[ (\omega_{\rm A}^2 + \nu^2) \hs\frac{d\chi}{d\tilde{x}} 
\Big] - k^2 \delta^{*2} \Big[ \kappa_{\rm e}^2 + \omega_{\rm A}^2 + \nu^2  
- \frac{4 \Omega^2 \omega_{\rm A}^2}{\omega_{\rm A}^2 + \nu^2} \Big] \hs\chi = 0 
\non\\
&&\qquad\Rightarrow\quad \chi \sim {\rm e}^{{\rm i} q \tilde{x}} \,,\quad q =  k 
\delta^* \bigg[\frac{4 \Omega^2 \omega_{\rm A}^2}{(\omega_{\rm A}^2 + \nu^2)^2} - 
\frac{\kappa_{\rm e}^2}{\omega_{\rm A}^2 + \nu^2} - 1\bigg]^{1/2} \,. 
\En{83}\EEQAR
In the middle of the interval, the wave length $2\pi/(q \delta^*)$ corresponds to that of 
the quasi-mode eigenfunctions depicted in Fig.~\F{16}. From this expression, oscillatory 
behavior is limited to
\BEQ \nu \le \nu_{\rm max} = \Big[\hs\half \sqrt{\kappa_{\rm e}^4 + 16 \Omega^2 
\omega_{\rm A}^2} - \half \kappa_{\rm e}^2  - \omega_{\rm A}^2\Big]^{1/2} \approx 
|\omega_{\rm A}| \sqrt{{4\Omega^2}/{\kappa_{\rm e}^2} - 1} \approx |\omega_{\rm 
A}| \sqrt{3}\,, \En{84}\EEQ
which should be an upper limit to the quasi-mode continuum. Not surprisingly, this 
agrees with the expression~\E{A12} of Appendix~\S{A2}, except that the parameters are 
now evaluated at the Doppler point where $\sigma = \Omega_0(\tilde{x}_0)$. Formally, 
this yields the same approximate criterion for instability as for the MRIs:
\BEQ 0 <  \omega_{\rm A}^2 < 4 \Omega^2 - \kappa_{\rm e}^2 \approx 3 \Omega^2 \,, 
\En{85}\EEQ
where the left inequality guarantees non-vanishing growth rate and the approximation on 
the right is for the accretion disk equilibrium described in Sec.~\S{IIA}. The important 
difference with the MRIs is that, for the SARIs, the Alfv\'en frequency $\omega_{\rm A} 
\equiv (m B_\theta/r + k B_z)/\sqrt{\rho}$ contains both the mode number $m$ and the 
toroidal field component $B_\theta$.

The expression~\E{84} explains why the upper limit of the quasi-continuum shown in 
Fig.~\F{15} goes up with increasing frequency~$\sigma$. At $\nu = 0$, the 
quasi-continuum is limited by the singularity $\sigma = \Omega_{\rm A}^+(r_2)$ on the 
left (black dash) and by the singularity $\sigma = \Omega_{\rm A}^-(r_1)$ on the right 
(red dash). For larger values of $\nu$, it widens in the directions of the two vertical 
dashed lines at $\sigma = \Omega_0(r_2)$ and $\sigma = \Omega_0(r_1)$, but it is 
limited from above by the curve that is increasing because $\nu_{\rm max}$ from 
Eq.~\E{84} is smaller on the left than on the right: $\omega_{\rm A}(r_2) < 
\omega_{\rm A}(r_1)$ for the equilibria investigated.

\subsection{Alfv\'en wave dynamics of the quasi-continuum SARIs}{\Sn{VC}}
 
To our knowledge, the present 2D continuum of quasi-modes has not been described 
before. It should be considered as a sub-class of modes outside the spectrum itself. Quite 
fundamentally, the spectrum consists of the collection of proper (discrete) and improper 
(continuum) modes and any complex frequency outside is part of the so-called resolvent 
set (see, e.g., Sec.~{6.3.1} of \Ron{GKP2019}). In a physical context, the resolvent set 
may be considered as the collection of complex frequencies for which the initial value 
problem can be solved or, from another perspective, where the driven problem has a 
finite response. In terms of the differential equations involved, a frequency either belongs 
to the spectrum, corresponding with solutions to the homogeneous equation, or to the 
resolvent set, corresponding with solutions to the inhomogeneous equation. This 
dichotomy is called the Fredholm alternative. It is exactly expressed by the concept of the 
complementary energy $W_{\rm com}$ of the Spectral Web, which should be 
considered as the energy that is needed to bring the open system (described by the 
inhomogeneous version of the differential equation) into resonance. That energy vanishes 
for discrete modes, which are natural resonators. What we have found here is a sub-class 
of modes which, strictly speaking, are not discrete modes, but which require just a tiny 
amount of energy to bring them into resonance. Consequently, for all practical purposes, 
they cannot be distinguished from genuine discrete modes.

\begin{figure}[ht]
\FIG{\begin{center} 
\includegraphics*[height=14.2cm]{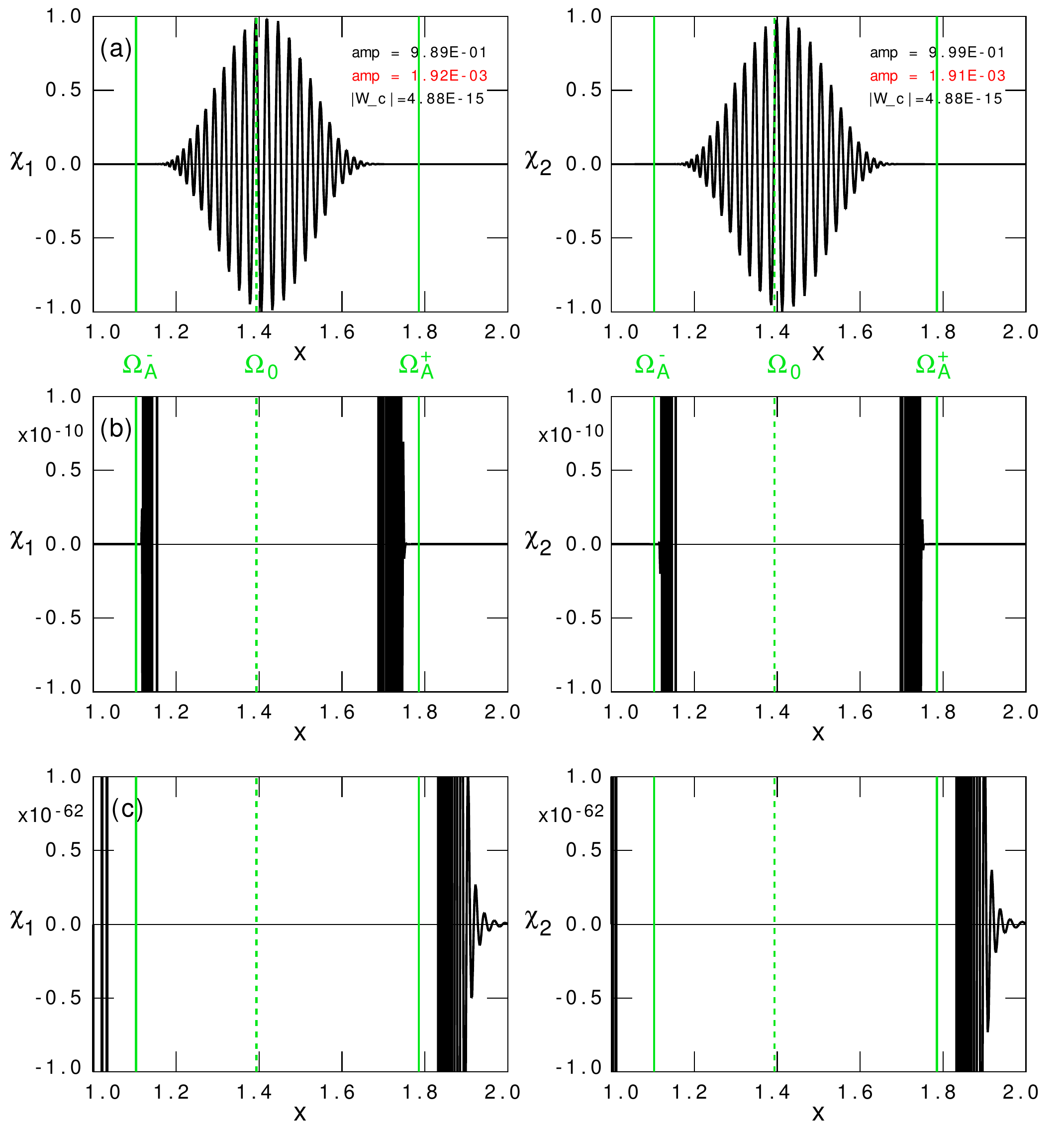}
\end{center}}
\caption{(a) Quasi-continuum `eigenfunction' of the co-rotating Super-Alfv\'enic 
Rotational Instability for the same parameters as in Fig.~\F{14}, $\sigma = 0.6$, $\nu = 
0.01$. For clarity of the oscillations, the plots of the associated variables $\Pi_{1,2}$ are 
suppressed, but their amplitudes are indicated in red. The eigenfunction is blown up 
(b)~with a factor of  $10^{10}$ to highlight the `explosion' at the singularities, (c)~with a 
factor of $10^{62}$ to demonstrate satisfaction of the BCs at the actual 
boundaries.}{\Fn{17}}
\end{figure}

An important physical question remains to be addressed: What causes the extreme 
localization of the quasi-continuum Super-Alfv\'enic Rotational Instabilities about the 
Doppler point brought about by what we have called `virtual walls' at the positions of the 
Alfv\'en/slow singularities? We have argued in Sec.~\S{IVA}, discussing the similar 
behavior of the eigenfunctions of the regular SARIs shown in Figs.~\F{7}, that induced 
currents at the positions of the singularities cause this, in analogy to the stabilization of 
internal kink modes in tokamaks. To analyze this behavior, it is necessary to compute the 
distribution of the perpendicular current density perturbation $\tilde{j}_\perp(r)$ since 
this yields radial confinement by the Lorentz force. This function may be constructed 
from the solution pair $\{\chi, \Pi\}$ of the accretion disk ODE. This requires rather 
involved algebra, presented in Appendix~\S{A3}. Exploiting the resulting 
expression~\E{A16} for $\tilde{j}_\perp$, the current density distribution of 
Fig.~\F{7}(c2) for a regular SARI with an eigenvalue $\omega$ close to the continuum 
singularities exhibited, in fact, skin currents at the positions of the singularities. However, 
in the broad current density distribution of the SARI with a larger growth rate shown in 
Fig.~\F{7}(b2) skin currents were absent. Nevertheless, that mode is still `confined' by a 
`virtual wall' at $r = r_1^*$. Evidently, this requires explanation beyond that given in 
Sec.~\S{IVA}.

Turning to the quasi-continuum SARIs, the extreme behavior of these modes, caused by 
the presence of the near-singularities $\sigma = \Omega_{\rm A}^-(r_1^*)$ and $\sigma 
= \Omega_{\rm A}^+(r_2^*)$ is exhibited, once more, in Fig.~\F{17}. These 
singularities are not even actual, or close, since the imaginary growth rate component of 
the `eigenfrequency' $\omega$ is finite, $\nu \ne 0$. Their importance is shown in the 
blow-up with a factor of $10^{10}$ in Fig.~\F{17}(b), showing the `explosions' of the 
amplitude of the solution at those positions. The additional blow-up, with a factor of 
$10^{62}$, in Fig.~\F{17}(c) shows that the boundary conditions at $r = r_1$ and $r = 
r_2$ are actually satisfied, but also that the amplitude of the solution is completely 
negligible in the two regions $(r_1, r_1^*)$ and $(r_2^*, r_2)$ outside the singularities. 
How could small amplitudes $\sim 10^{-10}$ of the solution at the near-singularities, 
compared to the large $\sim 1$ amplitude at the Doppler position, cause the 
`confinement' demonstrated by all eigenfunctions of the SARIs shown in this paper? 

\begin{figure}[ht]
\FIG{\begin{center} 
\includegraphics*[height=14.4cm]{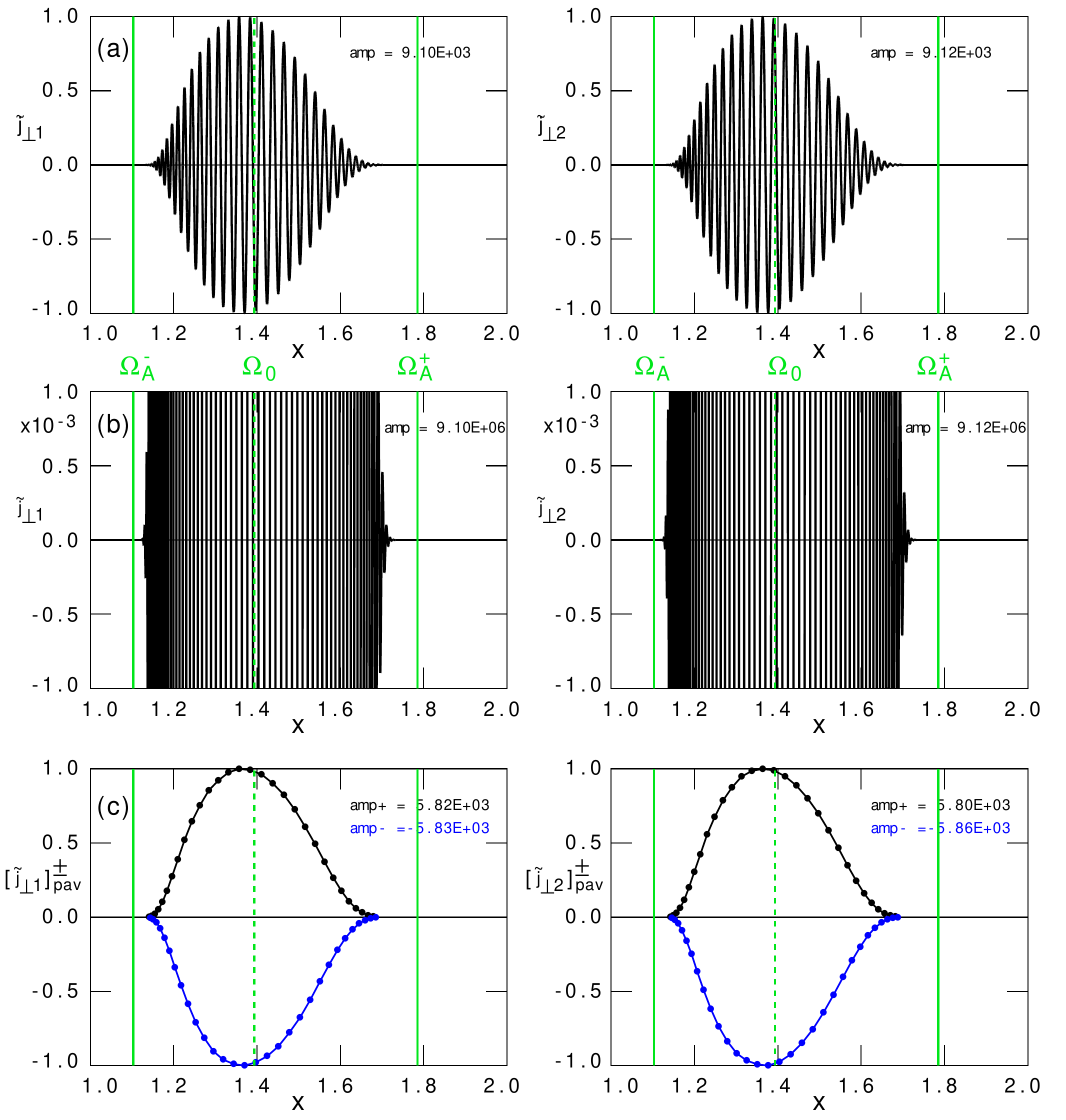}
\end{center}}
\caption{(a)~Perpendicular current density perturbation $\tilde{j}_\perp$ corresponding 
to the quasi-continuum `eigenfunction' shown in Fig.~\F{17}, $\sigma = 0.6$, $\nu = 
0.01$; (b)~The same plot blown up with a factor of~$10^3$ to focus on the currents 
flowing close to the singularities; (c)~Period-averaged perpendicular current density 
distribution.}{\Fn{18}}
\end{figure}

\begin{figure}[ht]
\FIG{\begin{center} 
\includegraphics*[height=14.4cm]{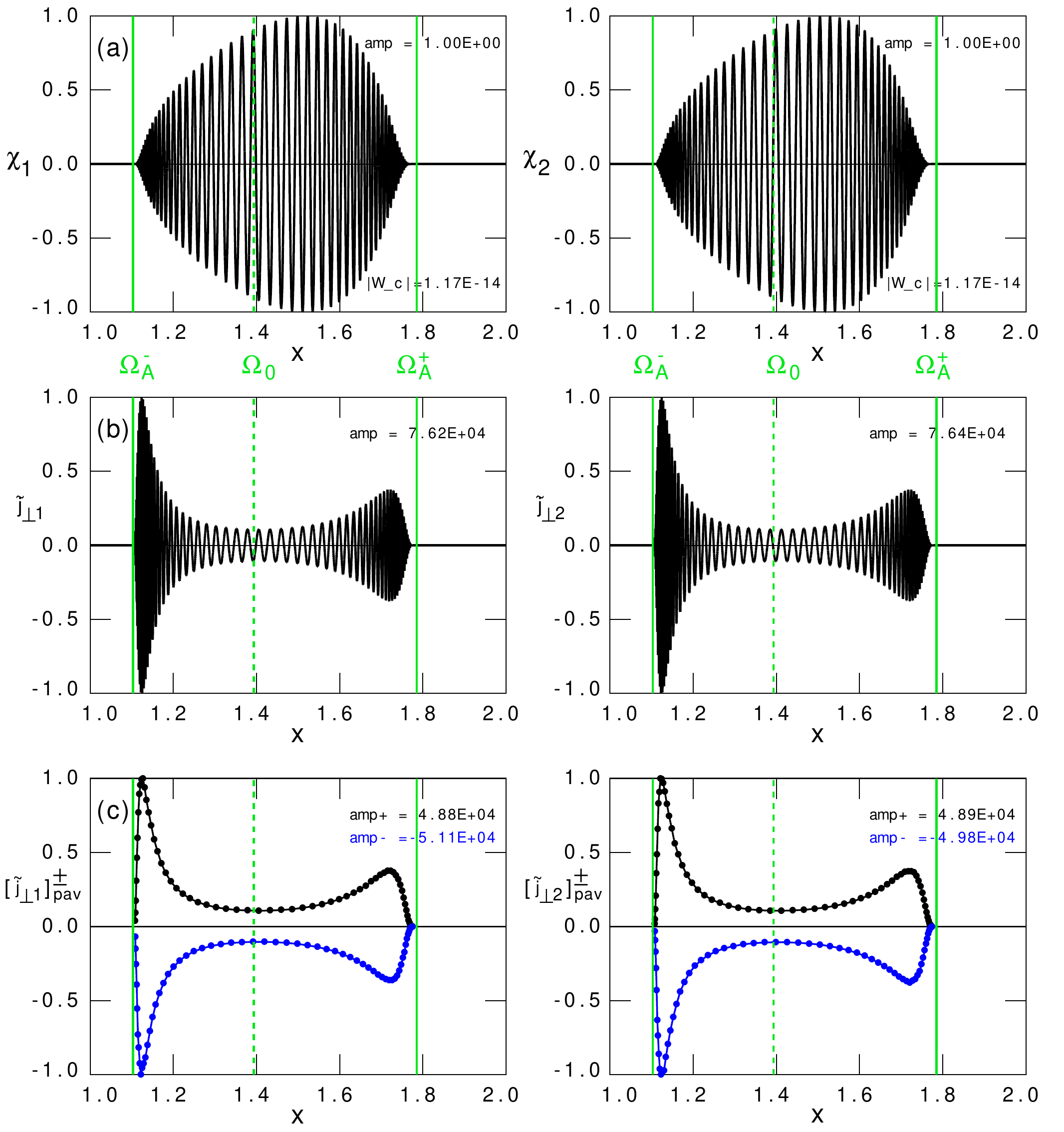}
\end{center}}
\caption{(a)~Quasi-continuum `eigenfunction' of the co-rotating Super-Alfv\'enic 
Rotational Instability for the same parameters as in Fig.~\F{14}, i.e.\ $\epsilon \equiv 
B_1 = 0.045$, $\mu_1 \equiv B_{\theta 1}/B_{z1}  = 10$, $\beta \equiv 2p_1/B_1^2 = 
10$, $\delta \equiv r_2/r_1 - 1 = 1$, $m = 1$, $k = 50$, $\sigma = 0.6$, but $\nu = 
0.001$; (b)~Perpendicular current density perturbation $\tilde{j}_\perp$; (c)~Period-
averaged perpendicular current density distribution.}{\Fn{19}}
\end{figure}

To answer this question, we study the behavior of the perpendicular current distributions 
presented in Figs.~\F{18} and \F{19}. These figures correspond to the same equilibrium 
and mode numbers as for the mode shown in Fig.~\F{17},  with $\sigma = 0.6$ and $\nu 
= 0.01$ for Fig.~\F{18}, but $\nu = 0.001$ for Fig.~\F{19}. They demonstrate, more 
clearly than the figures~\F{7}(b2) and \F{7}(c2) for the regular SARIs, the effective 
localization of both the `eigenfunction' $\chi$ and the perpendicular current density 
$\tilde{j}_\perp$ for the quasi-continuum SARIs. The confinement is now {\em well 
within the singularities} at the `virtual walls' at $r_1^*$ and $r_2^*$! 

The first frame of Fig.~\F{18} shows the perpendicular current density perturbation 
$\tilde{j}_\perp$, according to the expression~\E{A16}. It appears to follow the spatial 
oscillations of the `eigenfunction' of Fig.~\F{17}(a) rather closely, implying {\em the 
excitation of a large number of current filaments of opposite directions.} The blow-up 
with only a factor $10^3$ of the same current distribution, shown in Fig.~\F{18}(b), 
shows that the dependence on the second derivative of $\chi$ (represented by the first 
term of the expression~\E{A16} for $\tilde{j}_\perp$) causes a significant increase of the 
magnitude of the current density close to the singularities. However, skin currents at the 
singularities are absent. Clearly, an additional effect is needed to bridge the gap between 
the positions of the singularities (at $r_1^*$ and $r_2^*$) and the large amplitude 
perturbation within. The opposite directions of the perpendicular current in the filaments 
implies that each pair of currents will correspond to radial Lorentz forces $\bfe_r \cdot 
(\tilde{\bfj} \times \bfB) \equiv \tilde{j}_\perp B$ of opposite directions. Maybe, there is 
a systematic imbalance between the positive and the negative current channels which 
would produce a net inward Lorentz force? To decide on this, a careful analysis of the 
average current densities in the current channels was undertaken.

The period-averaged components $[\hs\tilde{j}_{\perp 1}]_{\rm pav}$ and 
$[\hs\tilde{j}_{\perp 2}]_{\rm pav}$ shown in Figs.~\F{18}(c) and \F{19}(c) are 
constructed as follows. Consider the $N$ {\em spatial} periods of oscillation of, e.g., the 
component~$\tilde{j}_{\perp1}$, numbered by $n$. Each period begins at $x = x_{{\rm 
b},n}$ where $\tilde{j}_{\perp1}$ becomes positive, it becomes negative in the middle 
of the interval at $x = x_{{\rm m},n}$, and it ends at $x = x_{{\rm e},n}$ where it 
becomes positive again. Evaluating the positive and negative contributions separately, in 
search of a systematic difference indicative of localized sheet currents, this yields
\BEQ [\hs\tilde{j}_{\perp1}]_{\rm pav}^{+,n} = (x_{{\rm m},n} - x_{{\rm b},n})^{-1} 
\!\!\int_{x_{{\rm b},n}}^{x_{{\rm m},n}} (\tilde{j}_{\perp1})^+ \, dx  \,,\qquad 
[\hs\tilde{j}_{\perp1}]_{\rm pav}^{-,n} = (x_{{\rm e},n} - x_{{\rm m},n})^{-1} 
\!\!\int_{x_{{\rm m},n}}^{x_{{\rm e},n}} (\tilde{j}_{\perp1})^- \, dx \,,\En{86}\EEQ
where the intervals are centered at, resp., $x_n^+ \equiv \half(x_{{\rm b},n} + x_{{\rm 
m},n})$ and $x_n^- \equiv \half(x_{{\rm m},n} + x_{{\rm e},n})$. Analogous 
expressions may be derived for $[\hs\tilde{j}_{\perp2}]_{\rm pav}^{\pm,n}$. The 
expressions~\E{86} are depicted in Figs.~\F{18}(c) and \F{19}(c), in black for the 
positive period-averaged components and in blue for the negative ones. Note that the 
black curves are defined only at the points $x = x_n^+$ and the blue curves only at the 
points $x = x_n^-$, whereas the points in between are obtained by simple interpolation. 
The curves exactly follow the envelopes of the oscillating current density components 
depicted in Figs.~\F{18}(a) and \F{19}(b). More important, reflecting the curve of the 
negative period-averaged components with respect to the $x$-axis, it precisely coincides 
with the curve of the positive period-averaged components: {\em there is no systematic 
difference indicative of localized sheet currents!} Moreover, the total positive and 
negative current perturbations flowing in the plasma exactly cancel (i.e.\ to machine 
accuracy): {\em there is no net perpendicular current $\tilde{j}_\perp$ flowing in the 
plasma either.} 

So far, we have only considered the {\em spatial} dependence of the different modes. 
The view point of spatial-period-averaging the perpendicular currents in the different 
channels yields perfect balance of the radial Lorentz forces. These forces are alternating 
between inward or outward, depending on the direction of the current in a particular 
channel, but they appear to yield a rather static picture. However, if we include the {\em 
temporal} behavior of the modes, dictated by their wave dependence ${\rm e}^{{\rm 
i}\sigma t}$, the picture becomes quite different. Since the directions of all currents in 
the different channels reverse sign in the second half-period of the temporal oscillation, 
automatically running waves are generated that propagate away from the singularities 
towards the Doppler point. For the co-rotating SARIs, these waves may be considered to 
be backward Alfv\'en waves travelling from $r = r_1^*$ to the right and forward Alfv\'en 
waves travelling from $r = r_2^*$ to the left, respectively. Pounding on the central part 
with the large current filaments, these Alfv\'en waves produce a wave pressure that 
effectively confines the perturbation within the `virtual walls'. This resembles the similar 
mechanism of ponderomotive stabilization of external kink modes in 
tokamaks~\R{DIpp1988}, where impinging ion cyclotron waves effectively produce the 
stabilization corresponding to a close-fitting conducting wall on the plasma, whereas the 
wall is actually far away. Similarly, the dynamics of the Alfv\'en waves associated with 
the SARIs produce `virtual walls' confining the perturbation within the actual boundaries.  

It might be objected that singular Alfv\'en waves only propagate along the magnetic field, 
not accross. For example, Alfv\'en wave heating occurs by means of excitation by fast 
magneto-sonic waves propagating towards the singularity. Here, the situation is reversed. 
Since the plasma is approximated to be incompressible, fast waves degenerate into 
`waves' that propagate the Lorentz forces instantaneously. In a sense, the fast contribution 
becomes part of the Alfv\'en waves, as evidenced by the sizeable radial component of all 
eigenfunctions shown in this paper.

Of course, separating the mode into a central part with the large current filaments and 
inward propagating Alfv\'en waves on the periphery is just a figure of speech since there 
is no separate plasma column and waves impinging on it, but the mode just keeps itself 
together by its wavy character represented by ${\rm e}^{{\rm i}\sigma t}$ while  
exponentially growing with the factor ${\rm e}^{\nu t}$. Nevertheless, the analogy with 
stabilization of internal kink modes at marginal stability ($\nu = 0$) and ponderomotive 
stabilization of external kink instabilities ($\nu \ne 0$)  by waves helps to clarify the 
different perpendicular current distributions shown in Figs.~\F{7}(b2),(c2) for the regular 
SARIs and in Figs.~\F{18}(a),(c) and Figs.~\F{19}(b),(c) for the quasi-continuum 
SARIs. In fact, as shown in Fig.~\F{7}(c2) for SARIs with a very small growth rate 
$\nu$, the current distribution is very close to skin currents at the singularities, producing 
the separation into independent sub-intervals~\R{New60} responsible for stability of 
internal kink modes. The wavy character of these currents is already visible in the 
small-scale oscillations of the peaked current distribution for that case, but it is more 
evident in the current distribution for the quasi-continuum modes illustrated in 
Fig.~\F{19}(b),(c). For the latter case, though rather close to marginal stability ($\nu = 
0.001$), the current distribution is still largest close to the singularities but it is already 
spreading out by the oscillations of the current channels. Further away from the 
singularities, for $\nu = 0.01$ illustrated in Fig.~\F{18}, the current distribution has 
become peaked at the Doppler point, i.e.\ furthest away from the `virtual walls'.

\begin{figure}[ht]
\FIG
{\begin{center} 
\hor{-50}\includegraphics*[height=8.4cm]{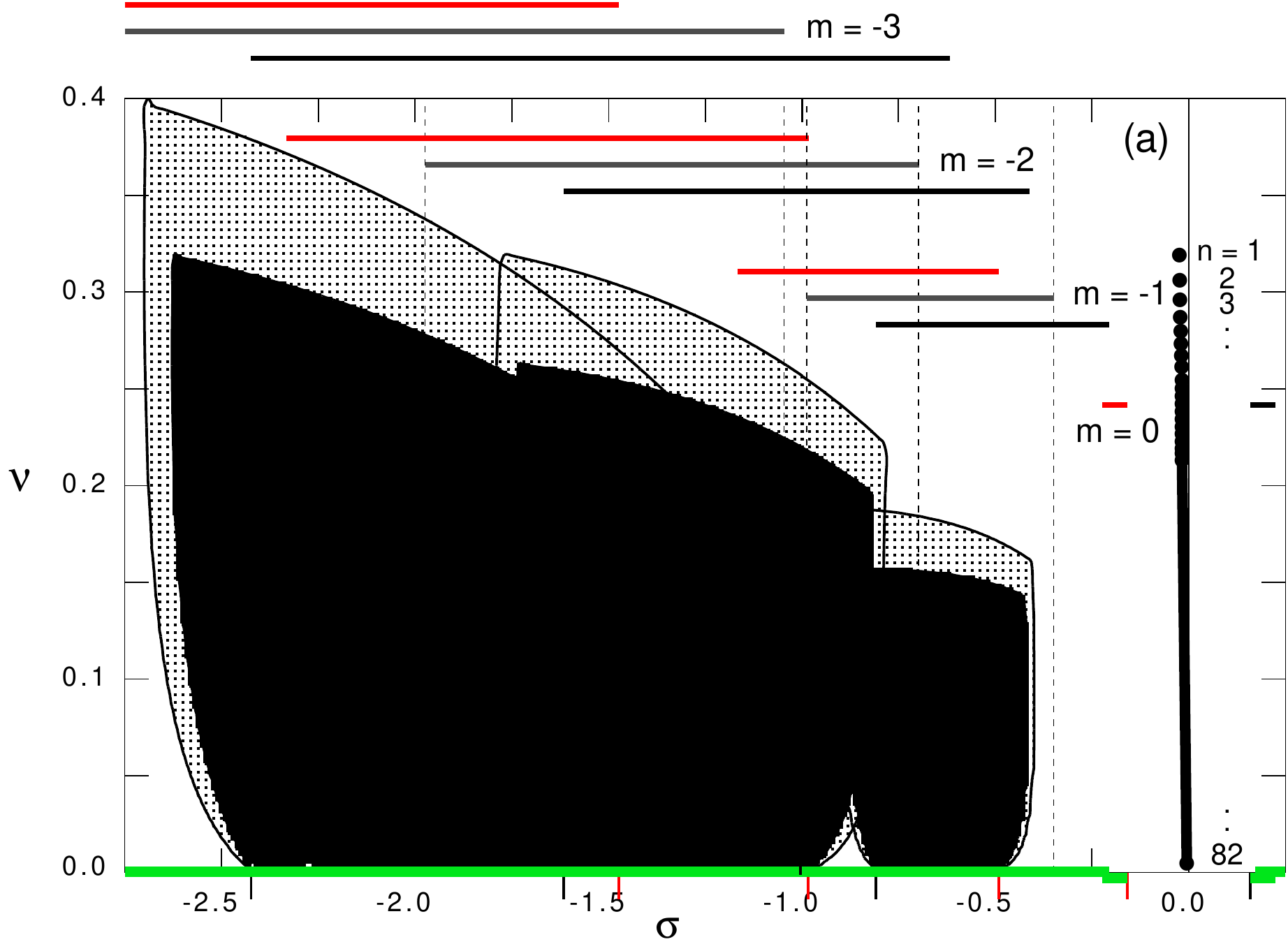}\\[4mm]
$\hor{35}$\includegraphics*[height=8.4cm]{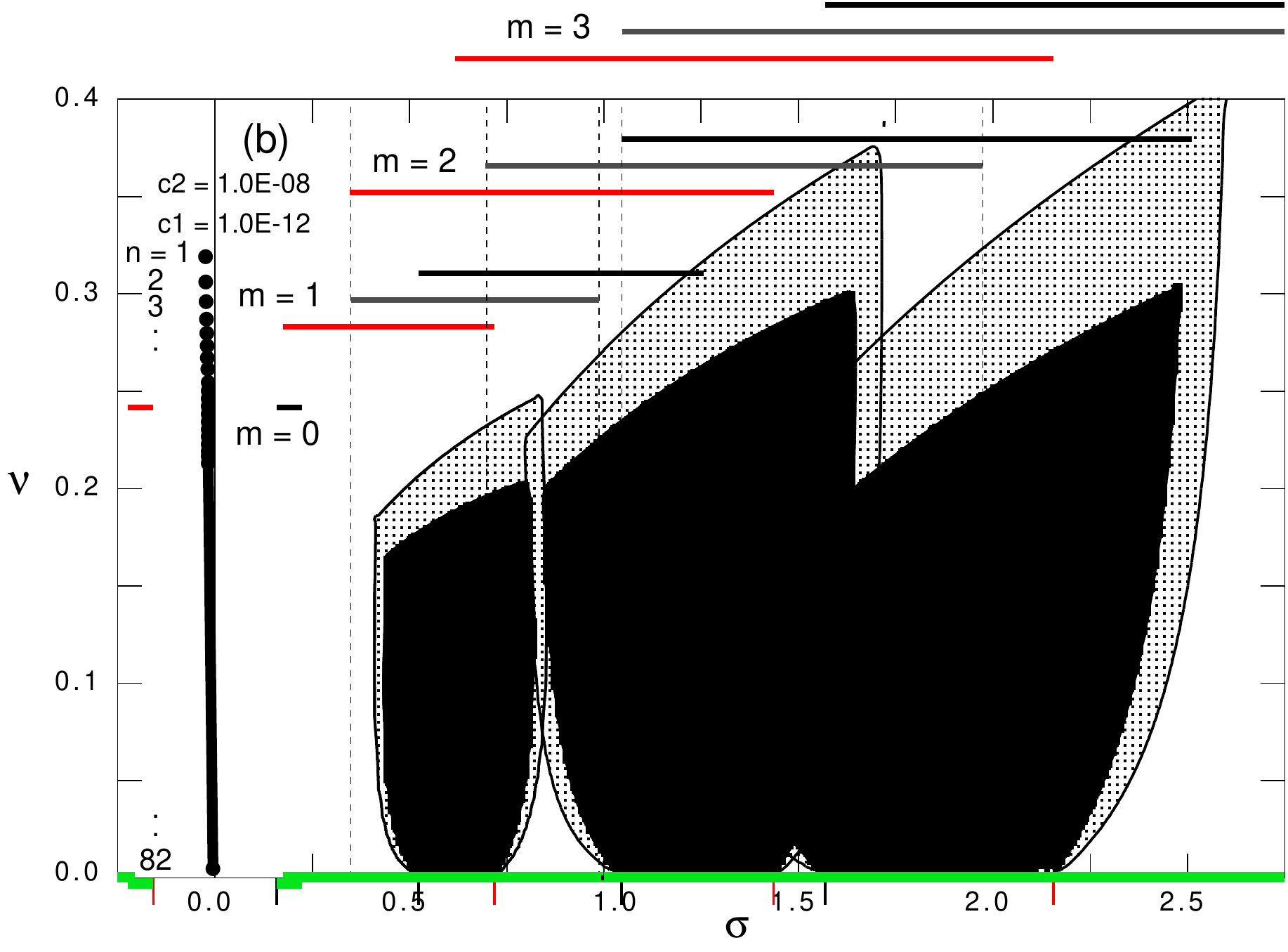}
\end{center}}
\caption{Quasi-continuum Super-Alfv\'enic Rotational Instabilities and discrete 
Magneto-Rotational Instabilities for the same equilibrium as in Figs.~\F{14}--\F{19}; 
$\epsilon = 0.045$, $\mu_1  = 10$, $\beta = 10$, $\delta = 1$. (a)~Contra-rotating $m = -
3, -2, -1$ SARIs for $k/|m| = 50$, and $m = 0$ MRIs for $k = 50$ with $n = 1, 2, \ldots 
82$. (b)~MRIs and co-rotating $m = 1, 2, 3$ SARIs. The overlapping continua are 
colored green, the ranges of $\Omega_{\rm A}^-$ (red), $\Omega_0^-$ (grey) and 
$\Omega_{\rm A}^+$ (black) are indicated above the figures.}{\Fn{20}}
\end{figure}

\HG{}{\subsection{Composite quasi-continua}{\Sn{VD}}}

For the same equilibrium parameters as exploited in the Spectral Web of Fig.~\F{14}, 
and the corresponding quasi-continuum of Fig.~\F{15}, a composition is shown in 
Fig.~\F{20} of three quasi-continua of counter-rotating SARIs ($m < 0$) and three 
quasi-continua of co-rotating SARIs ($m > 0$) together with the complete discrete 
spectrum of MRIs ($m = 0$) for a fixed value of $k$. For $|k| \gg |m|$, there are probably 
also infinite sequences of discrete SARIs clustering at the tips of the continua but they are 
swamped by the `sea' of quasi-continua: there is no way to tell them apart. This picture 
provides a glimpse of how the lower unstable part of the complex $\omega$-plane would 
become filled with quasi-continua above the real continua $\{\Omega_{\rm A}^\pm\}$ 
when all mode numbers would be incorporated. Note that the vertical scale of 
Fig.~\F{20} is of the same order of magnitude for the SARIs and the MRIs. This is in 
agreement with the expressions~\E{84} and \E{A12} for the maximum growth rate, 
which is similar for the two kinds of instability. Hence, the MRIs may be considered to 
be degenerate SARIs where the assumption of axisymmetry of the perturbations has 
eliminated the most important feature for the onset of turbulence, viz.\ the possible 
excitation of modes from the two-dimensional `sea' of non-axisymmetric 
quasi-continuum SARIs illustrated in Fig.~\F{20}. The latter modes are localized in all 
three directions: vertically (large $k$), toroidally ($m \ne 0$ and possibly large), and 
radially about the Doppler radius. Moreover, because of the occurrence of `virtual walls', 
they are radially detached from the actual boundaries of the disk so that any narrow 
annular band of the disk is unstable with respect to SARIs. Except for the vertical 
localization, none of these features is applicable to the MRIs.

\section{Summary and outlook}{\Sn{VI}}

\subsection{Summary}{\Sn{VIA}}

We found that the analytical solution of the accretion disk ODE~\E{23} for 
non-axisymmetric Super-Alfv\'enic Rotational Instabilities (SARIs) is much more 
intricate than the usual elementary analysis of the axisymmetric Magneto-Rotational 
Instabilities (MRIs). The reason is that the Doppler-shifted Alfv\'en/slow continua,
\BEQ \{\Omega_{\rm A}^\pm(r)\}  \equiv \{\Omega_0(r) \pm \omega_{\rm A}(r)\} 
\,,\En{87}\EEQ
completely control the analysis of the SARIs. Recall that the slow continua 
$\{\Omega_{\rm S}^\pm(r)\}$ are tacitly included since they coalesce with the Alfv\'en 
continua in the approximation of incompressibility, which is quite adequate for these 
modes. Whereas the equilibrium itself is super-Alfv\'enic for reasonable values of the 
mode numbers,
\HG{}{\BEQ \Omega(r)  > \omega_{\rm A}(r) \qquad \hbox{(for all 
$r$})\,,\En{88}\EEQ}
for MRIs as well as SARIs, the distinguishing feature of SARIs is that the Doppler 
frames are rotating at super-Alfv\'enic velocities:
\HG{}{\BEQ \Omega_0(r)  \equiv m \Omega(r) > \omega_{\rm A}(r) \qquad \hbox{(for 
all $r$})\,.\En{89}\EEQ}
Hence, on average, the modes are also propagating with the Doppler velocity, either 
co-rotating ($m > 0$) or counter-rotating ($m < 0$) with respect to the background 
rotation. This implies breaking of the axisymmetry of the configuration, of importance 
for the occurrence of shocks. This might be initiated by the SARIs, but that is clearly 
beyond the scope of the present paper. It also implies that {\em the forward and backward 
continua overlap,} which is instrumental for the occurrence of two infinite sequences of 
discrete SARIs emanating from the two tips of those continua.

The analysis of the discrete SARIs in Section~\S{IV}, even for minor inhomogeneity 
($\delta \ll 1$) to stay as closely as possible to the local analysis of the MRIs in 
Section~\S{IIIB}, immediately runs into the dominance of the two sets of continuum 
singularities $\{\sigma - \Omega_{\rm A}^+(r) = 0\}$ and $\{\sigma - \Omega_{\rm 
A}^-(r) = 0\}$ on the real axis of the $\omega$-plane. For finite values of the growth rate 
$\nu$, they are situated relatively far away in the complex $\omega$-plane, yet they turn 
out to have a major effect on the localization of the modes. At those values of $\sigma$, 
the Alfv\'en waves emitted from the tips of the continua affect the modes in such a way as 
to behave like confined by a {\em virtual wall} inside the actual boundaries. This could 
be documented by a detailed analysis of the accretion disk ODE, which was transformed 
into the Legendre equation by analytic continuation of the real radial variable $x$ 
($\equiv r$) into the complex $z$-plane. This produced completely explicit solutions in 
terms of the Legendre functions~\E{43}, propagating the solution from one boundary to 
the other. For the discrete SARIs, e.g.\ for a mode from the `inner' sequence emanating 
from $\sigma = \Omega_{\rm A}^+(r_1)$, see Fig.~\F{6}(a) or (c), one of those 
boundaries was the actual one, say $r = r_1$, whereas the other one corresponded to a 
virtual wall at $r = r_2^*$ ($ < r_2$). The crucial parameter describing the behavior at 
the singularities is the imaginary exponent ${\rm i} v$, where $v$ is defined in 
Eq.~\E{41}, which produces solutions that rapidly oscillate at the singularities and grow 
explosively away from them (all referring to the spatial domain). This behavior is not 
restricted to the Legendre equation, but it is characteristic for all solutions of the 
accretion disk equation~\E{23}, and also for the general ODEs~\E{12}, at the Alfv\'en 
singularities.

Quite unexpectedly, the analysis of the discrete SARIs for large inhomogeneity ($\delta 
\sim 1$) in Section~\S{V} runs into the problem of {\em fragmentation of the Spectral 
Web} (Fig.~\F{14}). Our method of computing eigenvalues by constructing the 
intersections of the solution path, from the condition ${\rm Im}(W_{\rm com}) = 0$, and 
the conjugate path, from the condition ${\rm Re}(W_{\rm com}) = 0$, appears to break 
down for certain values of the parameters. The two paths fragment into countless little 
islands, but there are no intersections anymore! The generalization in Section~\S{VB} of 
the Legendre analysis of Section~\S{IVB} provides the answer to why this happens. 
Again, the two independent solutions of the accretion disk ODE can be distinguished as 
`small' and `large' according to their behavior at the virtual wall end points $r_1^*$ ($> 
r_1$) and $r_2^*$ ($< r_2$), but they have the peculiar property that a solution that is 
`small' at the left end point is still `small' at the right end point. This implies that the 
boundary conditions at both end points can be satisfied for arbitrary values of $\omega$ 
in a wide neighborhood of the continuous spectra with just an exponentially small closing 
error. That error is the absolute magnitude $|W_{\rm com}|$ of the complementary 
energy, which is the energy that would have to vanish if the mode was to qualify as a 
discrete mode. Hence, these modes form {\em a genuine two-dimensional} (extending in 
frequency $\sigma$ as well as growth rate $\nu$) {\em continuum of quasi-discrete 
modes that just require a tiny amount of energy to be brought into resonance.} These 
extended regions in the complex eigenfrequency plane belong to the resolvent set of the 
operator, and their physical role in the actual solution of initial value calculations (as 
done by any numerical time integration of the MHD equations) is thus unquestionable. 
To our knowledge, these quasi-modes have not been described before in the literature, 
neither in the physics on instabilities, nor in the mathematics on spectral theory.

The parameters mentioned for which the quasi-continuum SARIs occur are the 
generalizations of the imaginary exponent ${\rm i} v$ of the singular expansions of the 
Legendre equation, viz.\ the two exponents ${\rm i} v_1$ and ${\rm i} v_2$ defined in 
Eqs.~\E{66} and \E{67}. For super-Alfv\'enic Doppler frequencies, these exponents 
become quite simple and transparent:
\BEQ v_i  \approx \Big|\frac{4k r_i^*}{3m}\Big| \qquad \hbox{($i = 
1,2$)}\,.\En{90}\EEQ
Consequently, for values of $|k| \gg |m|$, these numbers become large, and the 
exponential factors in which they occur even become exponentially large, which implies 
exponentially small complementary energies according to Eq.~\E{82}. All this according 
to asymptotic analysis close to the continua ($\nu \rightarrow 0$). Further away in the 
complex $\omega$-plane, numerical analysis provides contour plots of the boundaries of 
the quasi-continua for $|W_{\rm com}| < c$, where $c$ is typically a small number 
dictated by what one considers to be small enough to correspond to `eigenfunctions' of 
quasi-modes that cannot be distinguished from genuine eigenfunctions of discrete modes. 
This way, for the quasi-continuum SARIs, Spectral Web plotting, as in Fig.~\F{14}, is 
superseded by contour plotting of the absolute values of the complementary energy, as in 
Fig.~\F{15} for a single value and in Fig.~\F{20} for six values of $m$.

\HG{}{The quasi-continuum SARIs occur when the mode numbers satisfy the condition 
$|k| \gg |m|$ and the forward and backward Alfv\'en continuum frequency ranges 
$\Omega_{\rm A}^+(r)$ and $\Omega_{\rm A}^-(r)$ overlap. The latter requirement 
implies distinct conditions on the magnitudes of the mode numbers $m$ and $k$ and the 
equilibrium parameters $\epsilon$, $\mu_1$ and $\beta$,  viz.\ $\Omega_{\rm A}^+(r_1) 
< \Omega_{\rm A}^-(r_2)$ for the counter-rotating SARIs ($m < 0$) and $\Omega_{\rm 
A}^+(r_2) < \Omega_{\rm A}^-(r_1)$ for the co-rotating SARIs ($m > 0$). Inserting the 
explicit equilibria of Sec.~\S{IIA}, these can be combined into the following composite 
condition for the occurrence of quasi-continuum SARIs:
\BEQ 1 \ll \frac{|k|}{|m|} < \frac{\epsilon^{-1}\sqrt{1 + \mu_1^2}\,(1 - \frac{5}{8} 
\epsilon^2 \beta)\,[{\hs}1 - (r_1/r_2)^{3/2}] - {\rm sgn}(m) \hs\mu_1\,[{\hs}1 + 
(r_1/r_2)^{3/2}\hs]}{1 + (r_1/r_2)^{1/2}}\,.\En{91}\EEQ
When this condition is satisfied, 2D continua of quasi-discrete SARIs emanate from the 
overlapping continuous spectra $\{\Omega_{\rm A}^\pm\}$ on the real axis of the 
$\omega$-plane. How far the quasi-continua protrude into the complex $\omega$-plane 
depends on the value of $|W_{\rm com}|$ that one considers small enough to qualify for 
quasi-modes, and eventually (when $|W_{\rm com}|$ is no longer small) on the 
inequality \E{84}. The RHS of the inequality~\E{91} is dominated by the factor 
$\epsilon^{-1}$, which should be large for quasi-modes to be possible. This implies 
rather small values of the vertical magnetic field $B_z$, unless there is also a sizeable 
toroidal field component~$B_\theta$ (i.e.\ $\mu_1 \ne 0$ and large). The influence of the 
parameter $\beta$, measuring the magnitude of the kinetic pressure $p$ over the 
magnetic pressure $\half B^2$, is rather modest. Even at equipartition ($\beta = 1$), quite 
large regions of quasi-continuum SARIs occur, which has been verified by running the 
compressible version of ROC. Clearly, the explicit RHS inequality of the 
condition~\E{91} only applies for the particular class of equilibria defined in 
Sec.~\S{IIA}. Other equilibria will require a modification of this condition, but it is 
important to notice that all major effects of non-axisymmetric modes on an equilibrium 
with both vertical and toroidal components have been incorporated in the present 
analysis. Hence, the occurrence of quasi-continua will not be invalidated by the 
ramifications of all the possible flows in accretion disks~\R{Simon2009, 
Hollerbach2010}, if only they have overlapping continua. This might even apply to pure 
hydrodynamic disks~\R{Lyra2019,Umurhan2016} if the velocity profiles are non-monotonic enough to 
yield a flow continuum $\{\Omega_0\}$ folding over onto itself.
}

For the old question on which linear modes are responsible for the turbulent interaction 
producing the dissipation that is needed for accretion, it is evident that these 2D continua 
of non-axisymmetric quasi-modes, that just require a tiny amount of exciting energy, is 
extremely relevant. For ever larger ratio of $|k|/|m|$ that exciting energy becomes ever 
smaller so that the `sea' of quasi-continuum SARIs occupies ever larger portions of the 
complex $\omega$-plane (i.e.\ as long as the above mode number restrictions~\E{91} are 
respected). Clearly, the dynamical processes in the disk automatically will tune into the 
mode numbers and associated frequencies for which the response is maximum, thus 
initiating non-axisymmetric turbulence. We conclude with the conjecture that, {\em to all 
probability, the onset of 3D turbulence in accretion disks is not governed by the 
excitation of discrete axisymmetric Magneto-Rotational Instabilities but by the excitation 
of modes from the two-dimensional continua of quasi-discrete non-axisymmetric 
Super-Alfv\'enic Rotational Instabilities.} The final proof of this conjecture would 
require nonlinear or quasi-linear analysis beyond this paper.

\subsection{Outlook}{\Sn{VIB}}

We hope that this work helps to initiate a new research program of MHD spectroscopy 
for astrophysical plasma configurations. For example, it is well-known that merely 
including sheared flow in hydro or MHD configurations introduces Kelvin--Helmholtz 
pathways to instability, while (external or any effective) gravity field introduces 
Rayleigh--Taylor, Parker, quasi-Parker, and quasi-interchange modes \R{GKP2019}. 
Since the basic reference for accretion disk studies is the disk-height-averaged 
hydrodynamic viscous description from \Ron{ShakSun73}, a thorough MHD 
spectroscopic study of that configuration is called for. The full complexity of including 
magnetic fields in accretion disk configurations on the modes they support is thus far 
more complex than previously realized, and besides the well-known axisymmetric MRIs, 
we now identified both discrete SARI and quasi-continuum SARI modes as essential new 
ingredients. We look forward to future spectroscopic studies that should make a causal 
link to further nonlinear MHD behavior, by solving the full MHD equations with initial 
conditions targeted to excite a well-known single mode. For the particular context of 
accretion disks, these follow-up simulations must also quantify the role of the various 
modes in the angular momentum transport they might realize.

Perhaps the most important new aspect that came into focus through this study, is that not 
all relevant information is to be found in pure eigenmodes of the linearized MHD system. 
The concept of complementary energy, that can be defined for any complex frequency 
$\omega$, i.e.\ the energy required to bring the stationary equilibrium state into 
resonance with this particular frequency, is herewith established as more central to linear 
theory than previously realized. Exact eigenmodes, like the countable, discrete, unstable 
$m=0$ MRI, happen to be individual frequencies where this complementary energy 
vanishes: a perfect resonance. In the cylindrical disk limit studied here, the exact 
eigenmodes also show infinite sequences of discrete unstable modes, like the $m\neq 0$ 
co- or counter-rotating SARIs, in addition to the stable continuous Alfv\'en (and slow) 
singular eigenmodes. However, any frequency with a non-zero, but otherwise negligible 
complementary energy, should be easily excited, and is thus of equal relevance. This was 
shown to lead to entire 2D regions in the unstable eigenfrequency plane, where strictly 
speaking no modes exist but, for all practical purposes, a tiny perturbation suffices to get 
unlimited growth of localized wave packages that rotate with essentially the local 
Doppler frequency. These quasi-continua of SARIs are thus the central new concept 
elucidated in this work, which may bring a completely new view on turbulence, from the 
linear eigenmode perspective. 

\HG{}{Another viewpoint providing information beyond that obtained from the 
evolution of the individual eigenmodes comes from non-modal analysis. In typical 
hydrodynamical problems, like the solution of the Orr--Sommerfeld equation, it is 
usually stressed that the relevant linear operator is non-selfadjoint and, hence, that the 
eigenfunctions are non-orthogonal; see e.g.~\Ron{Schmid2007}. This permits the 
construction of linear combinations of eigenmodes that, for a limited time period, may 
grow faster than the fastest growing eigenmode proper. This is, rightly, considered to be 
extremely relevant for the excitation of turbulent motion. The view point of non-modal 
analysis is given additional impetus by the extension with the concept of 
pseudo-spectrum, where the linear operator is extended with an arbitrary operator of 
small norm $\epsilon$; see \Ron{Trefethen2005}. This yields contours in the complex 
$\omega$-plane that shrink to the spectrum proper in the limit $\epsilon \rightarrow 0$. 
Quite relevant for the excitation of turbulence: the pseudo-spectrum may even extend into 
the unstable (upper) part of the $\omega$-plane when the spectrum proper is restricted to 
the stable (lower) part. All this, evidently, evokes two questions: (1)~Can the present 
theory for the SARIs be extended with a non-modal analysis? (2)~Are the contours of the 
absolute value $|W_{\rm com}|$ of the complementary energy, delineating the 
quasi-continuum SARIs, just another way of representing a pseudo-spectrum?}

\HG{}{Before addressing these questions, it is useful to highlight the essential 
differences between the mentioned theory of the hydrodynamical problems and our 
analysis of the SARIs. The first one, in general, refers to non-conservative systems 
described by a non-selfadjoint linear operator having non-orthogonal eigenmodes. In our 
analysis, the system is conservative, the occurring operators $\bfG$ and $U$ are 
selfadjoint, nevertheless the eigenmodes are also non-orthogonal! Introducing 
eigenmodes $\bfxi_\alpha$ and $\bfxi_\beta$ corresponding to eigenvalues 
$\omega_\alpha$ and $\omega_\beta$ of the eigenvalue problem posed by the spectral 
differential equation~\E{6} + boundary conditions, this property has been demonstrated 
in \Ron{GKP2019}, Eqs.\ (12.97) and (12.98), which we here reproduce for the 
convenience of the reader:
\BEQ \langle \bfxi_\beta, \rho^{-1} \bfG(\bfxi_\alpha) \rangle = \omega_\beta^* 
\,\omega_\alpha \hs\langle \bfxi_\beta, \bfxi_\alpha \rangle \,,\qquad
\langle \bfxi_\beta, U \bfxi_\alpha \rangle = \half(\omega_\beta^*+ \omega_\alpha) 
\hs\langle \bfxi_\beta, \bfxi_\alpha \rangle \,. \En{92}\EEQ
Clearly for non-vanishing Doppler--Coriolis operator~$U$, the RHS inner products do 
not vanish so that the eigenmodes are non-orthogonal. However, the crucial feature of our 
approach is that {\em the eigenvalue problem is nonlinear:} the solutions are 
eigenfunctions neither of $\bfG$ nor of $U$, but of the specific nonlinear (with respect to 
the eigenvalue parameter $\omega$) combination occurring in Eq.~\E{6}. For the same 
reason, the question (usually dictated from quantum mechanical contexts) whether the 
operators $\bfG$ and $U$ commute or not is irrelevant: the two operators have different 
roles to play in the problem. In particular, the most significant distinction with all other 
approaches of the spectral problem is probably the focus it provides on the central 
importance of the Doppler--Coriolis operator $U$ describing the shear flow that occurs in 
almost all astrophysical plasmas [\,e.g.~\Ron{Zaqarashvili2007, Zaqarashvili2010}, and 
\Ron{Dikpati2020} on Rossby waves in the Sun, \Ron{Rudiger2007a, Rudiger2007b}, 
\Ron{Kitchatinov2010}, and \Ron{Hollerbach2010} on different extensions of the MRI, 
\Ron{Heifetz2015} on shear flow instabilities, to name just a few\,]. Having established 
the important difference with respect to selfadjointness, it remains to notice though that 
the non-modal analysis does not really depend on that, but only on the fact that the 
eigenmodes are non-orthogonal.  Hence, the answer to the first question is affirmative: in 
principle, it should be possible to extend the present theory of the SARIs also with a 
non-modal analysis. As a pertinent example, \Ron{Squire2014}, and also 
\Ron{Bhatia2016}, actually present a non-modal analysis of the MRIs, the spectral 
analysis of which we have shown to be one of the solutions of Eq.~\E{6}. However, 
these non-modal analyses crucially depend on WKB approximations and a local 
dispersion equation within a shearing box model, which we have shown in Sec.~\S{IVA} 
to be inadequate to describe the SARIs. To properly describe the discrete SARIs, at least 
the rapid oscillations approaching the continuum singularities should be incorporated, 
whereas for the quasi-continuum SARIs also the extreme connection between the two 
singularities, permitting continua of modes with large amplitudes at the Doppler 
frequencies, should be represented. Yet, the physics of the non-modal growth of the 
MRIs undoubtedly carries over to the SARIs and nature will know how to exploit that in 
the excitation of turbulence, irrespective of our inability to solve such a complicated 
problem.}

\HG{}{With respect to the second question, on the similarity of the contours of 
$|W_{\rm com}|$ and the pseudo-spectrum contours, it is clear that both methods are 
extensions to escape the narrow confines of the spectrum of discrete eigenvalues. They 
do that in entirely different directions though. The $|W_{\rm com}|$ contours enlarge the 
class of permitted solutions by relaxing the condition of continuity of the normal 
component of the displacement vector $\bfxi$, whereas the pseudo-spectrum contours 
enlarge the differential operator itself. Both $|W_{\rm com}|$ and $\epsilon$ represent a 
kind of measure of the distance to the actual eigenvalues, justified on physical grounds by 
the observation that the effects of external excitation and/or dissipation of the system (no 
matter how small) eventually are to be taken into account. This is actually what the 
non-modal analysis also is about, so that it is not surprising that the review on non-modal 
stability theory by \Ron{Schmid2007} contains an extensive discussion on external 
forcing, like time-dependent flows, stochastic forcing, flows in complex geometries, etc. 
For the quasi-continuum modes, here reported for the first time, it is not so clear that the 
pseudo-spectral method could also find those since then $W_{\rm com}$ is not literally a 
distance to a discrete mode (there are none in a wide neighborhood) but just an 
insignificant difference with such modes. On the other hand, it should also be pointed out 
that the present analysis, based on the Frieman--Rotenberg spectral equation~\E{6}, is 
strictly limited to ideal MHD, so that all the different dissipation mechanisms that occur 
in astrophysical plasmas cannot be incorporated. Fortunately, a multiplicity of such 
extensions has already been implemented in our finite element code 
Legolas~\R{CDK2021}. This code has verified most result on the SARIs reported here, 
including the fragmentation of the Spectral Web in the quasi-continuum range. We hope 
to report in the near future on the powerful broadening of scope by means of the ROC 
and Legolas codes, operated in tandem.}

In conclusion: Any small perturbation of the equilibrium will, in first instance, excite a 
huge collection of the quasi-continuum Super-Alfv\'enic Rotational Instabilities {\em at 
each point} of the accretion disk since the condition for instability is satisfied 
everywhere, whereas the modes are truly local in all three directions. Moreover, the 
breaking of the axisymmetry by these modes also implies possible dynamo action and 
excitation of magnetohydrodynamic shocks. Thus, the crucial presence of a magnetic 
field, as correctly highlighted in the MRI analysis of \Ron{BH91a}, opens up an 
enormous potential of additional new dynamical pathways to turbulence, necessary to 
explain why accretion occurs at all and why the accreted magnetized plasma is 
accelerated to powerful jets. Which of these pathways is chosen by nature is open to 
future research.

\begin{acknowledgements}
{\em Acknowledgements} \HG{}{We thank one of the referees for pointing out 
implications of our work that enlarge the focus considerably.} DIFFER is part of the 
institutes organisation of NWO. RK is supported by Internal funds KU Leuven through 
the project C14/19/089 \nobreak{TRACESpace,} an FWO project G0B452 and a joint 
FWO-NSFC grant G0E9619N. RK also received funding from the European Research 
Council (ERC) under the European Union Horizon 2020 research and innovation 
programme (grant agreement No. 833251 \nobreak{PROMINENT} ERC-ADG 2018).
\end{acknowledgements}

\appendix

\section{Spectral differential equations, quadratic forms, current density}{\Sn{A}}

\subsection{Coefficients of the spectral differential equations}{\Sn{A1}}

The singularity coefficients $N$ and $D$ of the cylindrical spectral equations~\E{10} 
and \E{12} were given in Eq.~\E{16}. For the convenience of the reader we here 
reproduce the definitions of the remaining coefficients from \Ron{Goed2018b}:
\BEQAR
A \,&\equiv&\, \frac{1}{r} \big( \widetilde{A} + \Delta \big) \,, \En{A1}\\[1mm]
B \,&\equiv&\, - \frac{4}{r} \hs\Big\{ (\rho \widetilde{\omega}^2 - k^2 \gamma p) P^2 - 
\rho \hs\widetilde{\omega}^2 R + \big(m \rho \hs\widetilde{\omega}^2 P - r^2 h^2 Q 
\big) \Lambda + \onefourth r^2 h^2 \widetilde{A} \hs\Lambda^2 \hs\Big\} \,, 
\En{A2}\\[0mm]
C \,&\equiv&\, \frac{2}{r^2} \hs\Big\{ m \widetilde{S} P
- r^2 \rho \hs\widetilde{\omega}^2 \big(Q - \half \widetilde{A} \hs\Lambda \big) \Big\} 
\,, \En{A3}\\[0mm]
E \,&\equiv&\, - \frac{1}{r^2} \Big\{ \widetilde{A}\hs\widetilde{S} 
\,\big(\widetilde{A} + \Delta\big) - 4 \widetilde{S} P^2 + 4 r^2 \big(Q - \half 
\widetilde{A} \hs\Lambda \big)^2 \Big\}  \,, \En{A4} \EEQAR
where $h^2 \equiv m^2/r^2 + k^2$ and abbreviations have been introduced for two 
equilibrium functions, $\Delta(r)$ and $\Lambda(r)$, and three perturbation functions, 
$P(r;\widetilde{\omega})$, $Q(r;\widetilde{\omega})$ and $R(r;\widetilde{\omega})$, 
as follows:
\BEQAR \Delta \,&\equiv&\, r(B_\theta^2/r^2 - \rho \Omega^2)' + \rho' G 
M_{\textstyle*}/r^2 \,, \qquad \Lambda \equiv \rho\hs(\Omega^2 - G 
M_{\textstyle*}/r^3) \,, \En{A5}\\[2mm]
P \,&\equiv&\, (B_\theta/r) F + \rho \hs\Omega \hs\widetilde{\omega} \,,\qquad Q \equiv  
(B_\theta/r) \big[F P + (B_\theta/r) \widetilde{A}\hs\big] \,,\qquad
R \equiv (B_\theta/r)^2 \big[2 m P - h^2 (B_\theta^2 + \rho{\hs}r^2\Omega^2)\big] 
\,.\En{A6} \EEQAR
All coefficients are complex through $\widetilde{\omega} \equiv \sigma - \Omega_0 + 
{\rm i}\hs\nu$, which is also a function of $r$ through the Doppler shift $\Omega_0(r)$.

For complex $\omega$, the system of first order ODEs~\E{12} splits into equations for 
the real and imaginary components:
\BEQ
\left(\! \begin{array}{c}
\chi_1' \\[1mm] 
\chi_2' \\[1mm] 
\Pi_1' \\[1mm] 
\Pi_2'
\end{array} \!\right)
+ \left(\! \begin{array}{rrrr}
\hat{C}_1 & - \hat{C}_2 & \hat{D}_1 & - \hat{D}_2 \\[1mm]
\hat{C}_2 & \hat{C}_1 & \hat{D}_2 & \hat{D}_1 \\[1mm]
 \hat{E}_1 & - \hat{E}_2 & - \hat{C}_1 & \hat{C}_2 \\[1mm]
\hat{E}_2 & \hat{E}_1 & -\hat{C}_2 & - \hat{C}_1
\end{array} \!\right) 
\left(\! \begin{array}{c}
\chi_1 \\[1mm] 
\chi_2 \\[1mm] 
\Pi_1 \\[1mm] 
\Pi_2
\end{array} \!\right) = 0 \,, \En{A7}\EEQ
where $\hat{C} \equiv C/N$, $\hat{D} \equiv D/N$, $\hat{E} \equiv E/N$. The real and 
imaginary components $\hat{C}_{1,2}$, $\hat{D}_{1,2}$, $\hat{E}_{1,2}$ follow by 
straightforward expansion of the expressions~\E{16}, \E{A3} and \E{A4}.

\subsection{Estimates of growth rates from the quadratic forms}{\Sn{A2}}

Estimates of the complex frequencies of the SARIs may be obtained by deriving a 
quadratic form corresponding to the accretion disk ODE~\E{23}, or the approximated 
form~\E{32},
\BEQ \int (\widetilde{\omega}^2 - \omega_{\rm A}^2) |\chi^\prime|^2 dr - k^2 \int 
\Big[\hs\kappa_{\rm e}^2 - (\widetilde{\omega}^2 - \omega_{\rm A}^2) + \frac{4 
\Omega^2 \omega_{\rm A}^2}{\widetilde{\omega}^2 - \omega_{\rm A}^2} \hs\Big] \, 
|\chi|^2 dr = 0 \,, \En{A8}\EEQ
where the boundary term resulting from integration by parts has been cancelled by 
applying the BCs. Writing $\widetilde{\omega} \equiv \tilde{\sigma}  + {\rm i}\hs\nu$, 
where $\tilde{\sigma} \equiv \sigma - \Omega_0$, and splitting this equation into real 
and imaginary parts yields
\BEQAR
&& \int (\tilde{\sigma}^2 - \nu^2 - \omega_{\rm A}^2) |\chi^\prime|^2 dr
- k^2 \int \Big[\hs\kappa_{\rm e}^2 - (\tilde{\sigma}^2 - \nu^2 - \omega_{\rm A}^2) + 
\frac{4 \Omega^2 \omega_{\rm A}^2 (\tilde{\sigma}^2 - \nu^2 - \omega_{\rm 
A}^2)}{(\tilde{\sigma}^2 - \nu^2 - \omega_{\rm A}^2)^2 + 4 \tilde{\sigma}^2 \nu^2} 
\hs\Big] \, |\chi|^2 dr = 0 \,, \En{A9}\\[0mm]
&& \nu \,\bigg\{\int \tilde{\sigma} |\chi^\prime|^2 dr + k^2 \int \tilde{\sigma} \Big[\hs 1
+ \frac{4 \Omega^2 \omega_{\rm A}^2}{(\tilde{\sigma}^2 - \nu^2 - \omega_{\rm 
A}^2)^2 + 4 \tilde{\sigma}^2 \nu^2} \hs\Big] \, |\chi|^2 dr \bigg\} = 0 \,. 
\En{A10}\EEQAR
From the latter equation it follows directly that, for instabilities ($\nu \ne 0$), the real 
Doppler  shifted frequency $\tilde{\sigma}$ should change sign on the interval, as 
discussed in Sec.~\S{IVA}. From the first equation it follows directly that there is no 
solution for $\nu\rightarrow \infty$, so that there should be a maximum growth rate 
$\nu_{\rm max}$. An estimate is obtained by noting that, from the previous argument, 
the approximation $\tilde{\sigma}^2 \ll \nu^2 +\omega_{\rm A}^2$ should hold for the 
integrands for the most global modes corresponding to $\nu_{\rm max}$. The first 
equation then simplifies to
\BEQ \int (\nu^2 + \omega_{\rm A}^2) |\chi^\prime|^2 dr + k^2 \int \Big[\hs\kappa_{\rm 
e}^2 + (\nu^2 + \omega_{\rm A}^2) - \frac{4 \Omega^2 \omega_{\rm A}^2}{\nu^2 + 
\omega_{\rm A}^2} \hs\Big] \, |\chi|^2 dr = 0 \,, \En{A11}\EEQ
where the integrand of the second integral should be negative, on average, to have 
solutions at all. This yields:
\BEQ \nu^2 \le \nu_{\rm max}^2 = \half \sqrt{\langle \kappa_{\rm e}^4 \rangle + 16 
\langle \Omega^2 \omega_{\rm A}^2 \rangle} - \half \langle \kappa_{\rm e}^2 \rangle - 
\langle \omega_{\rm A}^2 \rangle \,. \En{A12}\EEQ
For the explicit example of the SARIs discussed in Sec.~\S{IVA}, this gives $\nu_{\rm 
max} = 0.6307$ and $0.6277$, in agreement with the actual growth rates of the upper 
most modes of the Spectral Webs shown in  Fig.~\F{5} and \F{8} which are much lower 
($0.3913$ and $0.3246$). This is understood since the first integral of Eq.~\E{A11} is 
not incorporated in the expression~\E{A12}.

\subsection{Expressions for the current density perturbations}{\Sn{A3}}

Distinguishing perturbations with a tilde from background equilibrium quantities, the 
current density perturbation may be written as
\BEQ \tilde{\bfj} = \curl{\widetilde{\bfB}} \,,\qquad \widetilde{\bfB} = \curl{(\bfxi
\times \bfB)} \,. \En{A13}\EEQ
Projecting $\tilde{\bfj}$ onto the radial, the perpendicular and parallel (to $\bfB$) 
directions yields
\BEQAR 
\tilde{j}_r &=& - {\rm i\hs} ({m}/{r}) \hs(B_z \chi)' + {\rm i\hs} k 
\hs\big(({B_\theta}/{r}) \chi\big)' - {\rm i\hs} (m/r^2 + k^2) B \eta \,, \non\\[1mm]
\tilde{j}_\perp &=& \frac{1}{rB} \Big\{ - F^2 \chi + rB_z \Big[ ({1}/{r}) (B_z \chi)' + 
({mB}/{r}) \hs\eta \Big]' + B_\theta \Big[ r \big(({B_\theta}/{r}) \hs\chi\big)' - k r B 
\hs\eta \Big]' \Big\}  \,, \non\\[1mm]
\tilde{j}_\parallel &=& \frac{1}{rB} \Big\{ F G \chi + rB_\theta \Big[ ({1}/{r}) (B_z 
\chi)' + ({mB}/{r}) \hs\eta \Big]' - B_z\Big[ r \big(({B_\theta}/{r}) \hs\chi\big)' - k r B 
\hs\eta \Big]' \Big\}  \,, \En{A14}\EEQAR
where $\eta \equiv {\rm i}\hs(B_z \xi_\theta - B_\theta \xi_z)/B$ is the perpendicular 
component of the displacement vector~$\bfxi$, and $G \equiv m B_z/r - k B_\theta$ and 
$F \equiv m B_\theta + k B_z$ are proportional to the perpendicular and parallel 
components of the `wave vector'. One easily checks that the three current density 
components satisfy the constraint $\div{\tilde{\bfj}} = (1/r) (r \tilde{j}_r)' + {\rm i\hs} 
(G/B) \tilde{j}_\perp + {\rm i\hs} (F/B) \tilde{j}_\parallel = 0$. 

For our purpose, the radial and parallel current density perturbations  are not important 
since only $\tilde{j}_\perp$ produces a radially directed Lorentz force $\bfe_r \cdot 
(\tilde{\bfj} \times \bfB) \equiv \tilde{j}_\perp B$. Note, from Eq.~\E{A14}, that this 
expression involves the second derivative of $\chi$, which is dominant for the rapidly 
(spatially) oscillating solutions involved,  so that the contribution $\bfe_r \cdot (\bfj 
\times \tilde{\bfB})$ to the radial Lorentz force may be neglected. The expression for 
$\tilde{j}_\perp$ needs to be transformed to one in terms of the variables $\chi$ and 
$\Pi$ that are available from the ODE solver used in the program ROC. The expression 
of the variable $\eta$ in terms of $\chi$ and $\chi'$ may be found in Eq.~(12.93) of 
\Ron{GKP2019}. It is convenient though to transform it again into one involving $\Pi 
\equiv -(N/D) \chi' - (C/D) \chi$, exploiting Eq.~\E{20} for $(N/D)$ and Eqs.~\E{16} 
and \E{A3} for $(C/D)$ {\em in the incompressible approximation,} giving
\BEQ \eta \approx \frac{1}{mB} \hs\Big( - B_z \chi' + \frac{k r F}{\rho 
\hs\widetilde{\omega}^2 - F^2}\hs \Pi \Big) \,. \En{A15}\EEQ
This yields, after some algebra, the final expression for $\tilde{j}_\perp$:
\BEQAR
\tilde{j}_\perp &=&  \frac{1}{rB} \Big\{ a \chi'' + b \chi' + c \chi + 
\frac{d}{\widetilde{A}} \hs\Pi' + \Big( \hs\frac{e}{\widetilde{A}} + 
\frac{f}{\widetilde{A}^2} + \frac{g \hs\widetilde{\omega}}{\widetilde{A}^2} \hs\Big) 
\hs\Pi \Big\}\,, \qquad \widetilde{A} \equiv \rho \widetilde{\omega}^2 - F^2 
\,,\non\\[2mm]
&& a \equiv {r B_\theta F}/{m} \,,\qquad b \equiv B_z B_z' 
 + 2 B_\theta B_\theta' - {B_\theta^2}/{r} + ({k B_\theta}/{m})\hs(r B_z)' \,,\non\\[2mm]
&& c \equiv B_z B_z'' - {B_z B_z'}/{r} + B_\theta B_\theta'' - {B_\theta B_\theta'}/{r} + 
{B_\theta^2}/{r^2} - F^2 \,,\qquad d \equiv k r^2 F G/m \,,\non\\[2mm]
&& e \equiv [\hs k r^2 G \hs(F' - \rho' F/\rho) - 2 k^2 r B_\theta F \hs]/m \,,\qquad
 f \equiv - k r^2 F^2 G \hs(\rho' F/\rho - 2 F')/m \,,\qquad g \equiv 2 k r^2 F G \rho 
\hs\Omega'
\,. \En{A16}\EEQAR
The real coefficients $a,$ \dots $g$ are explicitly known through the equilibrium 
solutions~\E{4}. The complex functions $\chi'$, $\chi$, $\Pi'$ and $\Pi$ are directly 
provided by the solution of the ODEs~\E{12}, whereas $\chi''$ is produced from $\chi'$ 
by fourth order accurate interpolation on a fine grid. Note that the Alfv\'en factor 
$\widetilde{A}$ and the Doppler-shifted frequency $\widetilde{\omega}$ are complex, 
so that the contributions of  the real and imaginary components of $\Pi'$ and $\Pi$ to the 
expression~\E{A16} for the perpendicular current density $\tilde{j}_\perp$ are mixed in 
a complicated way. Because of the rapid spatial oscillations of the `eigenfunctions', those 
contributions are rather small compared to the first term involving the second derivative 
$\chi''$.

\bibliographystyle{apj}

\bibliography{SARI_refs}

\end{document}